\documentclass [prl,twocolumn,superscriptaddress,floatfix,amsmath,amssymb] {revtex4-2}
\pdfoutput=1
\usepackage{graphicx}
\usepackage{dcolumn}
\usepackage{bm}
\usepackage{color}
\usepackage[tight]{subfigure}
\usepackage{titlesec}

\usepackage[papersize={8.5in,11in}]{geometry}
\usepackage{xcolor}
\definecolor{darkblue}{rgb}{0.,0.,0.4}
\definecolor{darkred}{rgb}{0.5,0.,0.}
\definecolor{BlueViolet}{RGB}{138,43,226}
\definecolor{SkyBlue}{RGB}{30,144,255}
\definecolor{DarkGreen}{RGB}{0,100,0}
\usepackage[pdftex,colorlinks=true,linkcolor=darkblue,citecolor=red,urlcolor=darkred]{hyperref}
\usepackage{float}
\usepackage{physics}
\usepackage{cancel}
\usepackage[version=4]{mhchem}

\usepackage{bigints}
\newcommand{\bk}{\mathbf{k}}
\newcommand{\bp}{\mathbf{p}}
\newcommand{\bq}{\mathbf{q}}

\newcommand{\bl}{\mathbf{l}}

\def \nn{\nonumber \\}

\DeclareUnicodeCharacter{2212}{-}

\begin{document}

\title{Unconventional Superconductivity near a
Nematic Instability in a Multi-Orbital system}
\author{Kazi Ranjibul Islam}
\affiliation{School of Physics and Astronomy and William I. Fine Theoretical Physics Institute,
University of Minnesota, Minneapolis, MN 55455, USA}

\author{Andrey Chubukov}
\affiliation{School of Physics and Astronomy and William I. Fine Theoretical Physics Institute,
University of Minnesota, Minneapolis, MN 55455, USA}

\begin{abstract}
We analyze superconductivity in a multi-orbital fermionic system near the onset of a nematic order,
 using doped FeSe as an example.
We
associate the nematic order with spontaneous polarization between $d_{xz}$ and $d_{yz}$ orbitals.  We
 derive the pairing interaction, mediated by soft nematic fluctuations, and show that it is attractive,
  and that
    its strength depends on the position on the Fermi surface
       As the consequence, right at the nematic
        quantum-critical point (QCP),
          superconducting gap opens up at $T_c$ only at special points and extends into finite arcs at $T < T_c$.  In between the arcs the Fermi surface remains intact.   This gives rise to highly unconventional behavior of the specific heat, with no jump at $T_c$ and an apparent finite offset at $T=0$, when extrapolated from a finite $T$.   We argue that this behavior is consistent with the specific heat data for FeSe$_{1-x}$S$_x$ near critical $x$ for the onset of a nematic order.  We
          discuss the behavior of the gap away from a QCP and the pairing symmetry, and apply the results to FeSe$_{1-x}$S$_x$  and FeSe$_{1-x}$Te$_x$, which both show superconducting behavior near the QCP distinct from that in a pure FeSe.
\end{abstract}

\maketitle

\section{\textbf{INTRODUCTION.}}
It is widely believed that superconductivity in the cuprates, Fe-pnictides, heavy fermion, and other correlated electron systems is of electronic origin and at least in some portion of the phase diagram can be understood as mediated by soft fluctuations of a particle-hole order parameter, which is about to condense. The most studied scenario of this kind is pairing mediated by spin fluctuations. For the cuprates, it naturally leads to $d_{x^2-y^2}$ pairing. For Fe-pnictides,
 spin-mediated pairing interaction is attractive in both s-wave ($s^{+-}$) and $d_{x^2-y^2}$ channels. The argument, why pairing holds despite that the electron-electron interaction is repulsive, is the same in the two cases - antiferromagnetic spin fluctuations, peaked at momentum $Q$, increase the magnitude of a repulsive pairing interaction at the momentum transfer $Q$ (the pair hopping from $(k, -k)$ to $k+Q, -k-Q$). A repulsive pair hopping  allows for a solution for a gap function, which changes sign between Fermi points at $k_F$ and $k_F + Q$. There is still a repulsion at small momentum transfer, which is detrimental to any superconductivity,  and the bare Coulomb interaction is indeed larger at small momenta than at $Q$.  However, when spin fluctuations are strong, a repulsion at $Q$ gets stronger than at small momentum, and  sign-changing superconducting gap does develop.  This scenario has been verified by e.g., observation of a spin resonance peak below $T_c$\citep{christianson2008unconventional,lumsden2009two,chi2009inelastic,
 inosov2010normal,li2009spin,parshall2009spin}.
  Spin fluctuations were also identified as the source for spontaneous breaking of lattice rotational symmetry (nematicity)  in Fe-pnictides, as
  nematicity there develops in the immediate vicinity of the stripe magnetic order with momenta
 $Q =(\pi,0)$ or $(0,\pi)$.
   It has been argued multiple times \citep{fernandes2012preemptive,fernandes2014drives,fang2008theory,xu2008ising,fernandes2019intertwined}
    that spin fluctuations
    create an intermediate phase
  with a composite spin order,  which breaks symmetry between $(\pi,0)$ and  $(0,\pi)$, but reserves $O(3)$ spin-rotational symmetry.

 Situation is different, however, in bulk Fe-chalcogenide FeSe, which has been extensively studied in the last few years using various techniques.  A pure FeSe develops a nematic order at $T_p \sim 85K$, and becomes superconducting at $T_c \sim 9K$.   A nematic order decreases upon isovalent doping by either $S$ or $Te$ (FeSe$_{1-x}$S$_x$ and FeSe$_{1-x}$Te$_x$) and in both cases disappears at critical $x_c$ (0.17 for S doping and 0.53 for Te doping).  There is no magnetic order below $T_p$ for any $x$.

 The absence of magnetism lead to two conjectures: (i)  that nematicity  in FeSe is  a $d-$wave Pomeranchuk order, with order parameter bilinear in fermions, rather than a composite spin order, for which an order parameter in a 4-fermion operator, and (ii) that the origin of superconductivity may be different from the one in Fe-pnictides.  On (i), there is a consistency between the Pomeranchuk scenario for nematicity and the data already in pure  FeSe:
  a Pomeranchuk order parameter necessary changes sign between hole and electron pockets, consistent with the data\citep{suzuki2015momentum,chubukov2016magnetism,onari2016sign,benfatto2018nematic},
  and the temperature dependence of nematic susceptibility, measured by Raman, is in line with the Pomeranchuk scenario\citep{
  gallais2016charge,udina2020raman,klein2018dynamical}.
  On (ii), superconductivity in pure FeSe is  likely  still mediated by spin fluctuations\citep{imai2009does,glasbrenner2015effect,wang2016magnetic,liu2018orbital,hashimoto2018superconducting,classen2017interplay,xing2017competing}, as evidenced by the correlation between NMR $1/T_1$ and superconducting $T_c$, the consistency  between ARPES data on the gap anisotropy and calculations within spin fluctuation scenario,
   and the fact that a magnetic order does develop under pressure\citep{sun2016dome}. However near and above critical $x_c$, magnetic fluctuations are far weaker \citep{wiecki2018persistent,ayres2021decoupled}, e.g., a magnetic order does not develop until high enough pressure. It has been argued  \citep{hanaguri2018two,shibauchi2020exotic,sato2018abrupt,ishida2022pure,mukasa2023enhanced} based on a variety of data (see below) that superconductivity for such $x$ is qualitatively different from the one in pure FeSe. One argument here is that the gap anisotropy changes sign,
   another is that $T_c$ in FeSe$_{1-x}$Te$_x$ shows a clear dome-like behavior around $x_c$.

  In this communication we address the issue whether superconductivity in doped FeSe near $x_c$ can be mediated by nematic fluctuations.  It seems natural at a first glance to replace spin fluctuations by soft nematic fluctuations as a pairing glue. However, there are two obstacles, both related to the fact that soft nematic fluctuations are at small momentum transfer. First,  they do not affect the pair hopping term between hole and electron pockets, which is the key element for spin-mediated superconductivity. Second, the bare pairing interaction at small momentum transfer is repulsive, and dressing it by  nematic fluctuations only makes the repulsion stronger.

  We show that the pairing interaction $V_{eff} (k,-k;p,-p)$, mediated by nematic fluctuations (first two momenta are incoming, last two are outgoing), does become attractive near $x_c$, however for a rather special reason, related to the very origin of the Pomeranchuk order.  Namely, the driving force for a $d-$wave Pomeranchuk order is density-density interaction between hole and electron pockets. It does have a
   $d-$wave component $U^d_{he}$ because low-energy excitations in the band basis are constructed of $d_{xz}$ and $d_{yz}$ orbitals.  A  sign-changing nematic order (a spontaneous splitting of densities of $d_{xz}$ and $d_{yz}$ orbitals) develops~\citep{xing2018orbital} when $U^d_{he}$ exceeds d-wave intra-pocket repulsion, much like sign-changing $s^{+-}$ order develops when pair hopping  exceeds intra-pocket repulsion in the particle-particle channel.
   By itself, $U^d_{he}$  does not contribute to pairing, however taken at the second order, it produces an effective attractive interaction between fermion on the same pocket.
    We go beyond second order and collect all ladder and bubble diagrams which contain d-wave polarization bubbles at a  small momentum transfer. We show that this induced attraction is  proportional to the susceptibility for a $d-$wave Pomeranchuk order. Because a nematic susceptibility diverges at $x_c$,  the induced attraction necessary  exceeds the bare intra-pocket repulsion in some range around $x_c$, i.e., the full intra-pocket pairing interaction
      becomes attractive.

    This attractive interaction $V_{eff} (k-k;,p,-p) \propto A_{k,p} \chi_{nem} (|{\bf k}-{\bf p}|)$  is rather peculiar  because it inherits from $U^d_{he}$  the $d-$wave form-factor
     $A_{k,p} = \cos{(2\theta_k)} \cos{(2\theta_p)}$, where $\theta_k$ and $\theta_p$ specify the location of the fermions ((in our case, this holds on the hole pocket,  which is made equally out of $d_{xz}$ and $d_{yz}$  orbitals).
       A similar pairing interaction  has been earlier suggested for one-band models on
        phenomenological grounds \citep{lederer2015enhancement,schattner2016ising,lederer2017superconductivity,klein2018superconductivity} assuming that $d-$wave nematic coupling is attractive. We show that such an interaction emerges in the model with
         purely repulsive interactions, once we add the pairing component, induced by intra-pocket density-density  $U^d_{he}$.

 Because $\chi_{nem} (k-p)$ diverges at ${\bf k}={\bf p}$, the  presence of the form-factor $A_{k,p}$ in $V_{eff} (k,-k;p,-p)$ implies that the strength of the attraction depends on the position of a fermion on a Fermi surface
. As the consequence, the gap function on the hole pocket
  is the largest around hot points, specified by $\theta_h = n \pi/2, n=0-3$, and rapidly decreases in cold regions centered at $\theta_c = n \pi/2 + \pi/4, n=0-3$,
   This has been already emphasized in the phenomenological  study~\citep{lederer2015enhancement}. This behavior shows up most spectacularly  right at a nematic QCP, where the gap emerges at $T_c$ only at hot spots and extends at smaller $T$ into finite size arcs.
  The arcs length grows as $T$ decreases, but as long as $T$ is finite, there exist cold regions where the gap vanishes, i.e., the system preserves pieces of the original Fermi surface.
At $T=0$, the gap opens everywhere except at the cold spots $\theta_c$, where nematic form factor $\cos2\theta$ vanishes, but is still exponentially small near them, $\Delta (\theta) \propto \exp{-1/(\theta-\theta_c)^2}$.

  This, we argue, leads to highly unconventional behavior of  the specific heat coefficient $C_v/T$, which  does not display a jump at $T_c$ and instead increases as $(T_c-T)^{1/2}$, passes through a maximum at $T \sim 0.8 T_c$, and  behaves at smaller  $T$ like there is a non-zero residual $C_v/T$ at $T \to 0$  (see Fig.\ref{specific heat1}). In reality, $C_v/T$ vanishes at $T=0$, but nearly discontinuously, as $1/(\log(T_c/T))^{1/2}$.
   Also, because the regions, where the gap is non-zero, are disconnected,  the gap phases are uncorrelated, and $s-$wave, $d-$wave and two-component $p-$wave ($k_x + e^{i\alpha}k_y$) states are degenerate.

At a finite distance from a QCP and/or in the presence of non-singular pair-hopping between hole and electron pockets,   the gap function becomes continuous, but maxima at $\theta = n \pi/2$ remain.  The specific heat coefficient $C(T)/T$  acquires a finite jump at $T_c$, but holds the same behavior at intermediate $T$  as in Fig.\ref{sp heat away},  within some distance to a QCP.  The condensation energies for $s-$wave, $d-$wave and $p-$wave states split.  Which order develops depends on the interplay between the attractive pairing interaction, mediated by nematic fluctuations, and  non-singular repulsion. The letter is far stronger in $s-$wave and $d-$wave channels, which favors $p-$wave symmetry.  In this case, the most likely outcome is $k_x \pm i k_y$ state, which breaks time-reversal symmetry.
\section{\textbf{RESULTS}}
\subsection{\textbf{Model.}}~~~The electronic structure of pure/doped FeSe in the tetragonal phase consists of two non-equal hole pockets, centered at $\Gamma$, and two electron pockets centered at $X = (\pi, 0)$ and $Y = (0, \pi)$ in the $1$FeBZ.  The hole pockets are composed of $d_{xz}$ and $d_{yz}$ fermions, the X pocket is composed of
$d_{yz}$ and $d_{xy}$ fermions, and the $Y$ pocket is composed of $d_{xz}$ and $d_{xy}$ fermions.
The inner hole pocket is quite small and likely does not play much role for nematic order and superconductivity. We assume that heavy $d_{xy}$ fermions also do not play much role and  consider an effective two-orbital model with a single $d_{xz}/d_{yz}$ circular hole pocket, and mono-orbital electron pockets ($d_{yz}$ X-pocket and $d_{xz}$ Y-pocket). We define fermionic operators for mono-orbital Y and X pockets  as $f_1$ and $f_2$, respectively
 ($f_{1,k,\sigma} = d_{xz, k+Y, \sigma}$, $f_{2,k,\sigma} = d_{yz, k+X, \sigma}$).
 The band operator for the hole pocket is $h_{k,\sigma} = \cos{\theta_k} d_{yz, k, \sigma} + \sin{\theta_k} d_{xz, k, \sigma}$.  The kinetic energy is quadratic in fermionic densities and there are 11 distinct $C_4$-symmetric interactions \citep{fernandes2016low} involving low-energy fermions near the hole and the two electron pockets (see
 Ref. \citep{sm} for details).  We take  the absence of strong magnetic fluctuations in doped FeSe as an evidence that interactions at momentum transfer between $\Gamma$ and $X$ ($Y$) are far smaller than
  the interactions at small momentum transfer and neglect them. This leaves 6 interactions with small momentum transfer: 3  within hole or electron pockets and 3 between densities of fermions near different pockets.
 The single interaction between hole fermions contains an angle-independent term and terms proportional to $\cos{2\theta_k} \cos{2\theta_p}$ and $\sin{2\theta_k} \sin{2\theta_p}$, the two interactions between hole and electron pockets contain an angle-independent and a $\cos{2\theta_k}$ term, where $k$ belongs to the hole pocket and
 the three interactions between fermions on electron pockets contain only angle-independent terms.

\subsection{\textbf{Nematic susceptibility}} Like we said, we associate the nematic order with a d-wave Pomeranchuk order. In the orbital basis, this order is an orbital polarization (densities of $d_{xz}$ and $d_{yz}$ fermions split). In the band basis, we introduce two $d-$wave order parameters on hole and electron pockets: $\phi_h = \sum_{\bk,\sigma}\langle h_{\bk,\sigma}^\dagger h_{\bk,\sigma} \rangle\cos 2 \theta_{\bk}$ and
 $\phi_e= \sum_{\bk}\langle f_{2,\bk,\sigma}^\dagger f_{2,\bk,\sigma} \rangle-\langle f_{1,\bk\sigma}^\dagger f_{1,\bk,\sigma}\rangle$.  The set of two coupled self-consistent equations for $\phi_h$ and $\phi_e$ is obtained by summing ladder and bubble diagrams (see Ref. \citep{sm}) and is
 \begin{align}
\phi_h&= -\phi_h\, U_{h}^d \, \Pi_h^d -\phi_e \,U^d_{he}\, \Pi_e , \nonumber \\
\phi_e&=-\phi_e\, U_e^d \Pi_e-2 \phi_h \, U^d_{he} \Pi_h^d.
\label{phie equation1}
 \end{align}
 Here, $\Pi_h^d =-\bigintss G_p^h \, G_p^h \cos2\theta_\bp,$ and $ \Pi_e=-(1/2) \bigintss_p \left(G_p^X \, G_p^X + G_p^Y \, G_p^Y\right)$ are the polarization bubbles for the hole and the electron pockets, ($G^i_p = G^i (p, \omega_m)$ are the corresponding Green's functions and $\bigintss_p$ stands for $T \sum_{\omega_n} \bigintss \dfrac{d^2\bp}{(2\pi)^2}$). As defined,  $\Pi_h^d$ and $\Pi_e$ are positive. The couplings $U_{h}^d, U_e^d$ and $U^d_{he}$ are $d-$wave components of intra-pocket and inter-pocket density-density interactions.
   All interactions are positive (repulsive). The  analysis of (\ref{phie equation1})  shows that the
    nematic order with different signs of $\phi_h$ and $\phi_e$ develops when  $U^d_{he}$ is strong enough with the condition $2\, (U_{he}^d)^2\geq U_h^d\, U_e^d$.

The nematic susceptibility is inversely proportional to the determinant of (\ref{phie equation1}). Evaluating it at a small but finite  momentum $q$, we obtain $\chi_{\text{nem}} (q) \propto 1/Z$, where
 \begin{align}
Z =  \left(1+U_h^d \,\Pi_h^d (q) \right)\left(1+U_e^d\, \Pi_e (q) \right)-2(U^d_{he})^2 \Pi_h^d (q)\Pi_e(q).
\label{nematic susceptibility1}
\end{align}

\subsection{\textbf{Pairing interaction}} Our goal is to verify whether the pairing interaction near the onset of a nematic order is (i) attractive, (ii) scales with the nematic susceptibility, and (iii) contains  the d-wave form-factor $\cos^2 (2 \theta_k)$.   To do this, we use the fact that $\chi_{nem} (q)$ contains polarization bubbles $\Pi^d_h (q)$ and $\Pi_e(q)$, and obtain the fully dressed pairing interaction by collecting infinite series of renormalizations that contain $\Pi^d_h (q)$ and $\Pi_e (q)$ with small momentum $q$. This can be done analytically (see Refs. \citep{dong2022spin,sm} for detail). Because $q$ is small, the dressed pairing interactions are between fermions on only  hole pocket or only electron pockets: $V_{\text{eff}}^h(\bk, \bq)  = V_{eff}^h (\bk + \bq/2, -(\bk + \bq/2); \bk -\bq/2, -(\bk - \bq/2)$, $V_{\text{eff}}^e(\bk, \bq)  = V_{eff}^e (\bk + \bq/2, -(\bk + \bq/2); \bk -\bq/2, -(\bk - \bq/2)$.  We find
\begin{align}
\label{hole pairing interaction1}
V_{\text{eff}}^h(\bk, \bq)&= \frac{U^d_h}{1 + U^d_h \Pi^d_h (\bq)}  - A_h (U^d_{he})^2 \, \cos^2 2 \theta_\bk\,  \chi_{\text{nem}} (\bq) +...\\
V_{\text{eff}}^e(\bk,\bq)&= \frac{U^d_e}{1 + U^d_e \Pi_e (\bq)} -A_e (U^d_{he})^2\,  \chi_{\text{nem}}(\bq) + ...,
\label{electron pairing interaction1}
\end{align}
where $A_h=\dfrac{\Pi_e}{1+U_h^d\, \Pi_h^d (\bq)}$ and $A_e=\dfrac{1}{2}\dfrac{\Pi_h^d}{1+U_e^d\, \Pi_e (\bq)}$.
The dots stand for other terms which do not contain $\Pi_h^d (\bq)$ and $\Pi_e^d (\bq)$ and are terefore not sensitive to the nematic instability.

We see that each interaction contains two terms. The first is the dressed intra-pocket pairing interaction.
It does get renormalized, but remains repulsive and non-singular at the nematic instability.  The second term is the distinct interaction, induced by $U^d_{he}$.  It is (i) attractive, (ii) scales with the nematic susceptibility, and (iii) contains  the $d-$wave nematic form-factor  $\cos^2{2 \theta_\bk}$.   We emphasize that the attraction is induced by induced by inter-pocket density-density interaction, despite that relevant nematic fluctuations are with small momenta and the pairing interactions involve fermions from the same pocket.

  \begin{figure*}
  \centering
  \includegraphics[scale=1.5]{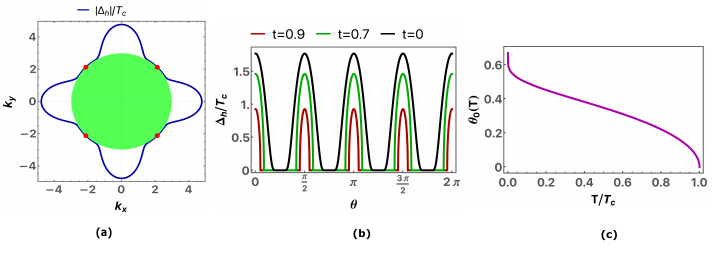}
  \caption{\textbf{Superconductivity at the nematic QCP($\mathbf{\delta=0)}$:} (a) We plot the absolute value of the SC gap scaled by the transition temperature $T_c$, $\Delta_h/T_c$(blue) on the hole pocket at zero temperature as a function of the angle on the Fermi surface(green disk) with radius $3$ unit. The gap is exponentially small near the cold spots(red dots) while maximum along the $k_x$ and $k_y$ axis. In (b) we plot the angular variation of the gap, $|\Delta_h|/T_c$ on the hole pocket for three different reduced temperatures, $t=0.9,0.7$ and $0$ below the transition point $T_c$. At finite temperature, the gap vanishes on the four patches of the Fermi surface arcs(flat segments of the gap). In (c) we plot the width of the gap, $\theta_0(T)$ as a function of the reduced temperature $t=T/T_c$. We keep the coupling strength $g=1$ here. }
\label{gap plots1}
  \end{figure*}
\subsection{\textbf{ Gap equation}} Near a nematic QCP, $\chi_{nem} (q)$ is enhanced and  the interaction, induced by $U^d_{he}$, is the dominant one.  In the absence of pair-hopping, the gap equation decouples between hole and electron pockets.  The most interesting case is  when the gap  develops first on the hole pocket (the case $A_h > A_e$).
We use Ornstein-Zernike form  $\chi_{\text{nem}}(\bq) = \chi_0/(\delta^2 + q^2)$, where $\delta$ is the distance to a nematic QCP in units of momentum.  At small $\delta$, relevant $q$ are of order $\delta$. To first approximation, the non-linear equation for $\Delta_h ({\bf k})$  then becomes local,  with angle-dependent coupling:
\begin{align}
1=g \, \cos^2 2\theta_{\bk} \bigintsss_0^\Lambda dx  \dfrac{\tanh\Bigr(\dfrac{\sqrt{x^2+|\Delta_h(\bk)|^2}}{2 T}\Bigr)}{\sqrt{x^2+\Delta_h(\bk)^2}}.
\label{int4}
\end{align}
where $g = m(U^d_{he})^2 \chi_0/(4\pi k_F \delta)$.
Because the coupling is larger at $\theta_k = n \pi/2, n=0-3$, the gap appears at  $T_c =1.13\Lambda \exp(-1/g)$ only at these points.  As $T$ decreases, the range, where the gap is non-zero, extends to four finite arcs with the width $\theta_0 (T) = 0.5 \arctan{\sqrt{g\log{T_c/T}}}$ (see Fig.\ref{gap plots1}b).  In the  areas between the arcs,
 the original Fermi surface survives. We emphasize that this is the original Fermi surface, not the Bogoliubov one, which could potentially develop inside the superconducting state\citep{setty2020topological,setty2020bogoliubov}.
We plot $|\Delta_h ({\bf k})|$  along the Fermi surface at $T=0$ and a finite $T$  in Fig.\ref{gap plots1}a,b, and plot $\theta_0(T)$  as a function of $T/T_c$ in Fig.\ref{gap plots1}c. The phases of the gap function in the four arcs are not correlated, hence $s-$wave, $d-$wave ($d_{x^2-y^2}$) and two-component $p-$wave ($k_x + e^{i\alpha} k_y$ with arbitrary $\alpha$) are all degenerate.
At $T=0$, the arcs ends merge at $\theta_k = n \pi/2 + \pi/4, n=0-3$ and the gap becomes non-zero everywhere except these cold spots(red dots in Fig.\ref{gap plots1}a). In explicit form, $|\Delta_h (k)| = 1.76 T_c  \exp{-\, \tan^2 2\theta_\bk/g}$.  The gap near cold spots becomes a bit smoother if we keep the Landau damping in $\chi_{nem}$ and solve the dynamical pairing problem, but still
$\Delta_h ({\bf k})$  remains highly anisotropic.

\subsection{\textbf{Specific heat}}  
\begin{figure}
  \centering
 \includegraphics[scale=.7]{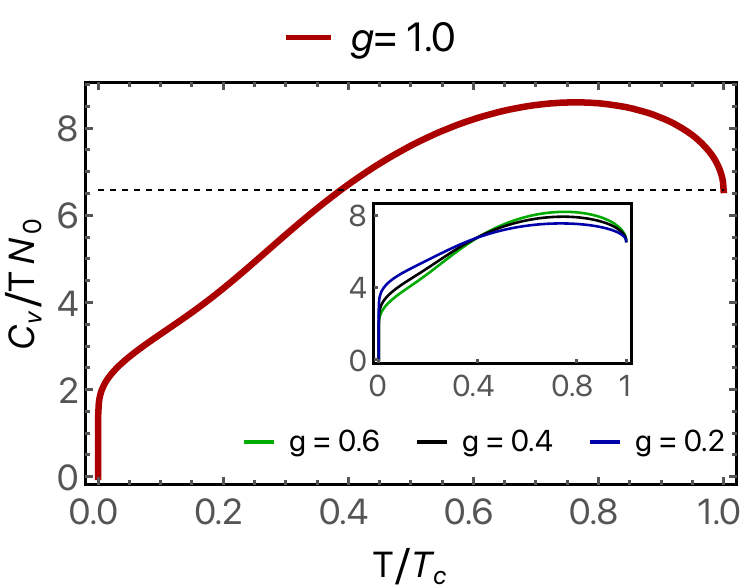}
  \caption{\textbf{Specific heat at the nematic QCP($\delta=0$):} We plot the specific heat coefficient scaled by the density of state on Fermi surface, $N_0$ as a function of the reduced temperature, $T/T_c$ for the coupling strength $g=1.0$ in the main frame and for $g=0.6,0.4$ and $0.2$ in the inset. The black dashed line represents the normal state contribution and is equal to $2\pi^2/3$.}
\label{specific heat1}
 \end{figure}

We split the specific heat coefficient $\gamma_c =C_v (T)/T$   into  contributions from the gapped and ungapped regions of the Fermi surface:
  $\gamma_c(T)=\gamma_c^n(T)+ \gamma_c^s(T)$. The first term,   $\gamma_c^n(T)=\dfrac{8 N_0 \pi}{3} \Big[\dfrac{\pi}{4}-\theta_0(T)\Bigr]$ which at small $T$ becomes: $\gamma^n_c (T) \approx \dfrac{4 \,N_0 \,\pi}{3\, \sqrt{\bar{g}\, \log T_c/T}}$. It evolves almost discontinuously:  vanishes at $T=0$, but reaches $1/3$ of the normal state value already at $T = 0.01 T_c$. We obtained $\gamma^s_c (T)$ at higher $T$ numerically and show the result for the full $\gamma_c (T)$  in Fig.\ref{specific heat1}.
  We see that
  $\gamma_c (T)$ does not jump at $T_c$. Instead it increases from its normal state value as
 $\sqrt{T_c-T}$, passes through  maximum at $T \approx 0.8 T_c$ and nearly linearly  decreases at smaller $T$,
  apparently with  a finite offset at $T=0$.  It eventually drops to zero at $T=0$, but only at extremely small $T$, as $1/(\log{T_c/T})^{1/2}$.  We emphasize that  $\gamma_c (T)$ is a function of a single parameter $T/T_c$, i.e., the smallness of the range, where $\gamma_c (T)$ drops, is purely numerical.

\begin{figure*}[]
  \centering
  \includegraphics[scale=1.5]{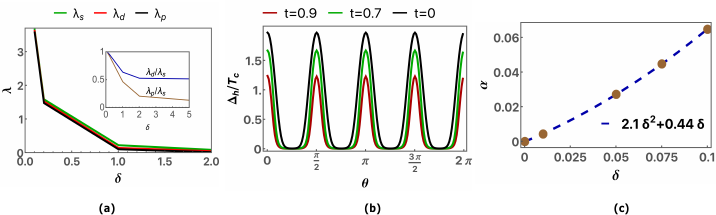}
   \caption{\textbf{Superconductivity away from the nematic QCP($\mathbf{\delta\neq0)}$:} In (a) we plot the largest eigenvalue $\lambda$ of the linearized gap equation defined in \citep{sm} in different angular momentum channels labeled as $\lambda_s,\lambda_d$ and $\lambda_p$ for the $s,d$ and $p-$ wave respectively as a function of the nematic mass parameter $\delta$. We show how the ratio of these eigenvalues vary with $\delta$ in the inset and find $\lambda_p/\lambda_s< \lambda_d/\lambda_s<1$ which indicates that $s-$ wave is the leading instability. With $\delta$ going to zero, the ratios become closer to each other, indicating possible degeneracy among different pairing channels. In (b) we plot the angular variation($\theta$) of the gap function, $\Delta_h(\theta)/T_c$ on the hole pocket for a set of reduced temperatures, $t=T/T_c$ below the transition point $T_c$ for $\delta=0.01$. In (c) we plot the gap anisotropy $\alpha=\Delta_h(\theta=\pi/4)/\Delta_h(\theta=0)$ as a function of the nematic mass $\delta$ and fit (blue dashed line) our result upto second order in $\delta$ with the fitting parameters $\alpha(\delta)=2.12 \, \delta^2+0.44 \, \delta$.}
\label{numerical solution1}
  \end{figure*}
 \subsection{\textbf{Away from a nematic QCP}}
\begin{figure}
 \centering
 \includegraphics[scale=.7]{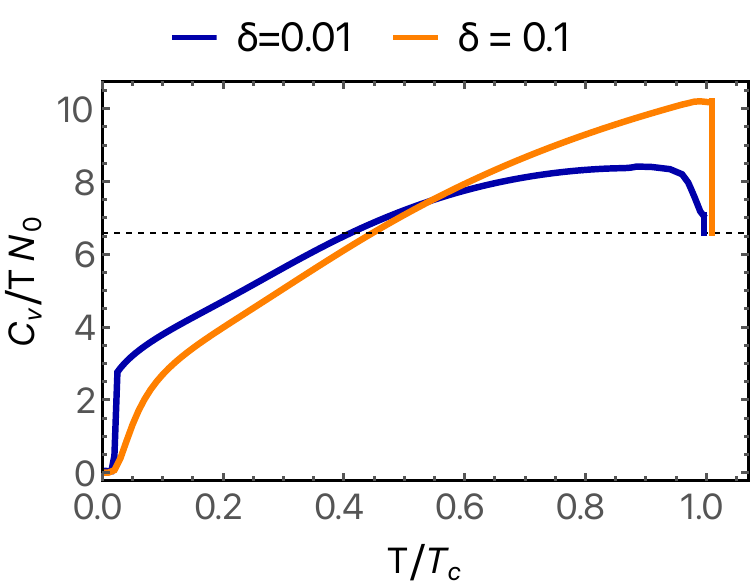}
 \caption{\textbf{Specific heat away from the nematic QCP($\delta \neq 0$):} We plot the variation of the specific heat coefficient scaled by the density of state on Fermi surface, $N_0$ with the reduced temperature $T/T_c$ for a set of values of the nematic mass $\delta=0.01$ and $0.1$. Here, $T_c$ is the superconducting transition temperature. The black dashed line represents the normal state contribution and is equal to $2\pi^2/3$. There is a finite specific heat jump at the transition point represented by the vertical line for both values of $\delta$. }
 \label{sp heat away}
 \end{figure}

~~~  At a finite $\delta$, $s-$wave, $d-$wave, and $p-$wave solutions for the gap function are no longer degenerate.  If we  keep  only the interaction induced by $U^d_{he}$ (the second term in
 \ref{electron pairing interaction1}), we find that s-wave solution has the lowest condensation energy. We show the  eigenvalues $\lambda_{s,p,d}$ and the gap functions in Fig.\ref{numerical solution1}a,b.  The gap function is smooth and finite for all angles, but remains strongly anisotropic up to sizable $\delta/k_F$. We define the gap anisotropy $\alpha$ as the ratio of the gap function on the hole fermi surface at $\theta=\pi/4$($k_x- k_y$ axis) to $\theta=0$($k_x$ axis): $\alpha=\Delta_h(\pi/4)/\Delta_h(0)$ and show its variation with the nematic mass parameter $\delta$ in Fig.\ref{numerical solution1}c. The specific heat coefficient $\gamma_c (T)$ has a finite jump at $T_c$,  whose magnitude increases with $\delta$, yet the low temperature behavior remains nearly the same as at a QCP up to sizable $\delta/k_F$ (Fig.\ref{sp heat away}).
 If we consider the full pairing interaction in (\ref{hole pairing interaction1}), situation may change as  the first term in (\ref{hole pairing interaction1}) has comparable repulsive $s-$wave and $d-$wave harmonics,
 but a much smaller $p-$wave harmonic.  As the consequence, $p-$wave may become the leading instability.  The condensation energy for a p-wave state is the lowest for $k_x  + i k_y$ and $k_x - i k_y$ gap functions. A selection of one of these states breaks time-reversal symmetry.

\subsection{\textbf{Comparison with experiments}}
We argued in this work is that  pairing in  doped FeSe
   near a nematic QCP is mediated by nematic fluctuations rather than by spin fluctuations.  This is generally consistent with the observations in Ref. \citep{shibauchi2020exotic,ishida2022pure,hanaguri2018two} of two distinct pairing states in pure FeSe and in doped FeSe$_{1-x}$S$_x$ and FeSe$_{1-x}$Te$_x$ at $x \geq x_c$.
     More specifically, one can distinguish between magnetic and nematic pairing scenarios by measuring the angular dependence of the gap along the hole $d_{xz}/d_{yz}$ pocket.
   We argued that a nematic-mediated pairing gives rise to an anisotropic gap, with maxima along $k_x$ and $k_y$ directions.  Within spin-fluctuation scenario,  the gap  $\Delta_h (k) = a + b \cos{4 \theta}$
     is the largest along the diagonal  directions  $k_x \pm k_y$ ($b <0$, see e.g., Ref.~\citep{graser2010spin}).
      The angular dependence of the gap in pure and doped FeSe has been extracted from ARPES and STM data in  Ref. \citep{xu2016highly,liu2018orbital,sprau2017discovery,nagashima2022discovery,
      walker2023electronic}.
       For pure and weakly doped FeSe, an extraction of $\cos{4\theta}$ dependence is complicated because superconductivity co-exists with long-range nematic order, in which case the gap additionally has $\cos{2 \theta}$ term due to nematicity-induced mixing of $s-$wave and $d-$wave components~\citep{hashimoto2018superconducting,kushnirenko2018three}.  Still, the fits of the ARPES data in Refs.~\citep{xu2016highly,liu2018orbital} yielded a negative $b$, consistent with spin-fluctuation scenario. A negative $b$  is also consistent with the flattening of the gap on the  hole pocket near $\theta =\pi$, observed in the STM study~\citep{sprau2017discovery}. A negative prefactor for $\cos{4 \theta}$ term was also reported for Fe-pnictides, e.g.,  Ba$_0.24$K$_{0.76}$Fe$_2$As$_2$, Ref. \citep{ota2014evidence}.   In contrast, STM data for tetragonal FeSe$_{0.45}$Te$_{0.55}$ (Ref.\citep{sarkar2017orbital})
        found the maximal gap along $k_x$ and $k_y$ directions, consistent with the pairing by nematic fluctuations. The gap maximum along $k_y$ has also been reported in a recent laser ARPES study of
   FeSe$_{0.78}$S$_{0.22}$ (Ref. \citep{nagashima2022discovery}). Further, recent  STM data on FeSe$_{1-x}$S$_{x}$  (Ref. \citep{walker2023electronic} detected a shift of the   gap maxima from $k_x = \pm k_y$ for $x <0.17$ to $k_x$ and $k_y$ for $x > 0.17$, and STM data for FeSe$_{0.81}$S$_{0.19}$ (Ref. \citep{silva_neto_last}) showed clear gap maxima along  $k_x$ and $k_y$. Taken together, these data strongly support the idea about different pairing mechanisms in pure FeSe and in doped ones at $x \geq x_c$, and are consistent with the change of the pairing glue from spin fluctuations at $x <x_c$
       to nematic fluctuations at $ x \geq x_c$.

    Next, we argued that right at a nematic QCP,  the gap vanishes in the cold regions on the Fermi surface, and this leads to highly unconventional behavior of the specific heat coefficient $\gamma_c (T)$
   \footnote{This holds when we neglect pair hopping between hole and electron pockets.  In the presence of pair hopping, the gap becomes non-zero everywhere except, possibly, special symmetry-related points. Still,  in the absence of magnetism nearby, pair-hopping is a weak perturbation, and the gap in cold regions is small.}
   The specific heat of FeSe$_{1-x}$S$_x$ has been measured in Refs.~\citep{sato2018abrupt,mizukami2021thermodynamics}. The data clearly indicate that the jump of $\gamma_c (T)$ at $T_c$ decreases with increasing $x$ and vanishes at around $x_c$.  At smaller $T$,  $\gamma_c (T)$  passes through a maximum at around $0.8 T_c$ and then decreases nearly linearly towards  apparently  a finite value at $T=0$.
     The  authors of Ref.\citep{hanaguri2018two} argued that this behavior  is not  caused by fluctuations,
     because
residual resistivity does not exhibit a noticeable increase around $x_c$ (Ref. \citep{hosoi2016nematic}). Other experiments~\citep{coldea2019evolution} also indicated that correlations only get weaker with increasing $x$.
       The behavior of $\gamma_c (T)$ around $x_c$ was first interpreted first as potential BCS-BEC crossover~\citep{shibauchi2020exotic} and later as a potential evidence of an exotic pairing that creates a Bogolubov Fermi surface in the superconducting state~\citep{agterberg2017bogoliubov,setty2020topological,nagashima2022discovery}. We argue that the specific heat data are consistent with the nematic-mediated pairing, in which near $x_c$  the gap develops in the arcs near $k_x$ and $k_y$  and  nearly  vanishes in between the arcs.   This explanation is also consistent with recent observation~\citep{matsuura2023two} that superfluid density in FeSe$_{1-x}$S$_x$ drops at $x \geq x_c$, indicating that some fermions remain unpaired.

  Finally, recent $\mu$SR experiments~\citep{matsuura2023two,roppongi2023} presented evidence for time-reversal symmetry breaking in FeSe.  The $\mu$SR signal is present below $T_c$  for all $x$, however in FeSe$_{1-x}$Te$_x$ it clearly increases above $x_c$.  This raises a possibility that the superconducting state at $x > x_c$ breaks time-reversal symmetry, at least in FeSe$_{1-x}$Te$_x$.   Within our nematic scenario, this would indicate a $p-$wave pairing with $k_x \pm i k_y$ gap structure. We argued that $p-$wave pairing, mediated by nematic fluctuations, is  a strong competitor to $s^{+-}$ pairing.

 There is one recent data set, which we cannot explain at the moment.  Laser ARPES study of FeSe$_{0.78}$S$_{0.22}$ (Ref. \citep{nagashima2022discovery}) detected superconducting gap in the polarizarion of light, which covers momenta near  the $X$ direction, but no gap in polarization selecting momenta near $Y$.  Taken at a face value, this data  implies that superconducting order strongly breaks $C_4$ symmetry.  In our nematic scenario, pure $k_x$ (or $k_y$) order is possible, but has smaller condensation energy than $k_x \pm i k_y$.  More analysis  is needed to resolve this issue.

\section{DISCUSSION} In this paper we derived an effective pairing interaction near the onset of a nematic order in a 2D two-orbital/three band system of fermions nd applied the results to doped FeSe.   The model consists of a hole band, centered at $\Gamma$ and made equally of $d_{xz}$ and $d_{yz}$  fermions, and two electron bands, centered at  $X$ and $Y$ and made out $d_{yz}$ and $d_{xz}$ fermions, respectively. The nematic order is a spontaneous polarization between $d_{xz}$ and $d_{yz}$ orbitals, which changes sign between hole and electron pockets. We found the pairing interaction as the sum of two terms: a dressed bare interaction, which remains non-singular and repulsive, and the term, induced by intra-pocket density-density interaction $U^d_{he}$. This last term contains the square of the nematic form-factor and scales with  the nematic susceptibility,  and is the dominant pairing interaction near the onset of a nematic order. We obtained the gap function and found that it is highly anisotropic with gap maxima along $k_x$ and $k_y$ directions.  This is in variance with pairing by spin fluctuations, for which the gap has maxima along diagonal directions $k_x \pm k_y$.  Right at the nematic QCP, the
 gap develops in four finite arcs around $k_x$ and $k_y$, while in between the arcs the original Fermi surface survives.  Such a gap function, degenerate between $s$-wave, $d-$wave, and $p-$wave,  gives rise to highly unconventional behavior of the specific heat coefficient with no jump at $T_c$ and seemingly finite value at $T=0$
  (the actual $C_v (T)/T$ vanishes at $T=0$, but drops only at extremely low $T \sim 10^{-2} T_c$).  In the tetragonal phase away from a QCP, the degeneracy is lifted, and there is a competition between $s-$wave and $k_x \pm i k_y$, the latter breaks time-reversal symmetry.  In both cases, the gap remains strongly anisotropic, with maxima along $X$ and $Y$ directions.   We compared our theory with existing experiments in some details.
\section{\textbf{ACKNOWLEDGMENTS}}
We acknowledge with thanks useful conversations with D. Agterberg, E. Berg, P. Canfield, P. Coleman, Z. Dong, R. Fernandes, Y. Gallais, E. Gati, T. Hanaguri, P. Hirschfeld, B. Keimer,  A. Klein, H. Kontani, L. Levitov, A. Pasupathi, I. Paul, A. Sacuto, J. Schmalian, T. Shibauchi, and R. Valenti.
This work  was supported by  U.S. Department of Energy, Office of Science, Basic Energy Sciences, under Award No. DE-SC0014402.
\bibliographystyle{naturemag}
\bibliography{biblio}

\clearpage
\widetext
\begin{center}
\textbf{\large SUPPLEMENTAL MATERIALS}
\end{center}
\setcounter{equation}{0}
\setcounter{figure}{0}
\setcounter{table}{0}
\setcounter{page}{1}
\setcounter{section}{0}
\makeatletter
\renewcommand{\theequation}{S\arabic{equation}}
\renewcommand{\thefigure}{S\arabic{figure}}
\renewcommand{\thesection}{A\arabic{section}}

\section{ Model Hamiltonian}
  We consider an effective two dimensional $2$-orbital model Hamiltonian with two hole pockets, centered at the $\Gamma$ point of the
Brillouin zone(BZ) and two electron pockets, centered at $X = (0, \pi)$ and $Y = (\pi, 0)$ points of the BZ, respectively. The
hole pockets and the corresponding hole bands are composed of $d_{xz}$ and $d_{yz}$ orbitals. The $X$ pocket/band is composed of $d_{yz}$
and the $Y$ pocket/band is composed of $d_{xz}$ orbitals respectively. For simplicity, we neglect the $d_{xy}$ orbital contribution to
the $X$ and $Y$ electron pockets. We also neglect the spin-orbit coupling on the band dispersion. The kinetic energy of the model Hamiltonian near the $
\Gamma$, $Y$ and $X$ points are captured by 
\begin{align}
H_{\text{Kin}}=H_\Gamma+H_Y+H_X,
\end{align}
with
\begin{align}
H_\Gamma&=\sum_{\bk,\sigma}\psi_{\Gamma,\bk,\sigma}^\dagger \Big[\Big(\mu_h-\dfrac{\bk^2}{2\, m_h }\Big)\tau_0-\dfrac{b}{2}\bk^2\, \cos2\theta_\bk\, \tau_3+ \dfrac{b}{2}\bk^2\, \sin2\theta_\bk\, \tau_1\Big]\psi_{\Gamma,\bk,\sigma},\label{hole band}\\
H_{Y}&=\sum_{\bk,\sigma} \xi_{Y,\bk} f_{1,\bk,\sigma}^\dagger f_{1,\bk,\sigma},\\
H_{X}&=\sum_{\bk,\sigma} \xi_{X,\bk} f_{2,\bk,\sigma}^\dagger f_{2,\bk,\sigma}.
\end{align}
Here, $\psi_{\Gamma,\bk,\sigma}^\dagger=\Big(d^\dagger_{xz,\bk,\sigma},d^\dagger_{yz,\bk,\sigma}\Big)$ where $d^\dagger_{i,\bk,\sigma}$ creates a fermion with the orbital index $i(xz,yz)$, momentum $\bk$(measured from $\Gamma$) and spin index $\sigma(\pm 1)$. $\theta_\bk$ is the polar angle for momentum $\bk$, measured from the $\Gamma-X$-direction in the anti-clockwise direction. $\tau_i$'s are the conventional Pauli matrices with $\tau_0$ being the $2\times 2$ identity matrix. The parameters $\mu_h,m_h$ and $b>0$ are taken from the ARPES data for FeSe at $k_z=\pi$ \cite{fernandes2016low}. Diagonalizing Eq.~\eqref{hole band}, we find two bands called outer($h$) and inner($H$) hole pockets with the dispersion relation  $
\xi_{h,H}=\mu_h- \bk^2/2 \, m_h\pm b\, \bk^2.$ The fermionic band operators $h,H$ are the linear combination of the orbital operators,
\begin{align}
\label{hole orbital to band 1}
h_{\bk,\sigma}&=\cos\theta_\bk\, d_{yz,\bk,\sigma}+\sin\theta_\bk\, d_{xz,\bk,\sigma}\\
H_{\bk,\sigma}&=\sin\theta_\bk\, d_{yz,\bk,\sigma}-\cos\theta_\bk\, d_{xz,\bk,\sigma}.
\label{hole orbital to band 2}
\end{align}  
 
On the other hand,  $f^\dagger_{1,\bk}$ and $f^\dagger_{2,\bk}$ creates fermion with orbital $d_{xz}$ and $d_{yz}$ and momentum $\bk$ near $Y-$ and $X-$ points respectively. The corresponding energy dispersions for the electron pockets are taken of the form
\begin{align}
\xi_Y(\bk)=\dfrac{k_{x}^2}{2 m_x}+\dfrac{k_y^2}{2 m_y}-\mu_e \hspace{2 cm} \xi_X(\bk)=\dfrac{k_{x}^2}{2 m_y}+\dfrac{k_y^2}{2 m_x}-\mu_e,
\end{align}
where $m_x$ and $m_y$ are fitting parameters from the ARPES data. In the orbital basis,
the interaction part of the Hamiltonian involves $14$ distinct $C_4$ symmetric interactions between the low energy fermions near the $\Gamma, X$ and $Y$ pockets,
\begin{align}
H_{\text{int}}&=\dfrac{U_4}{2} \sum  \left[d_{xz,\sigma}^\dagger d_{xz,\sigma}d_{xz,\sigma'}^\dagger d_{xz,\sigma'}+d_{yz,\sigma}^\dagger d_{yz,\sigma}d_{yz,\sigma'}^\dagger d_{yz,\sigma'}\right]+\tilde{U}_4 \sum \, d_{xz,\sigma}^\dagger d_{xz,\sigma}d_{yz,\sigma'}^\dagger d_{yz,\sigma'} 
 +\tilde{\tilde{U}}_4 \sum  \, d_{xz,\sigma}^\dagger d_{yz,\sigma}d_{yz,\sigma'}^\dagger d_{xz,\sigma'}\nn & +\dfrac{\bar{U}_4}{2} \sum \left[d_{xz,\sigma}^\dagger d_{yz,\sigma}d_{xz,\sigma'}^\dagger d_{yz,\sigma'}+d_{yz,\sigma}^\dagger d_{xz,\sigma}d_{yz,\sigma'}^\dagger d_{xz,\sigma'}\right]+U_1 \sum\left[f_{1,\sigma}^\dagger f_{1,\sigma} d_{xz,\sigma'}^\dagger d_{xz,\sigma'}+f_{2,\sigma}^\dagger f_{2,\sigma} d_{yz,\sigma'}^\dagger d_{yz,\sigma'}\right]\nn & +\bar{U}_1 \sum\left[f_{1,\sigma}^\dagger f_{1,\sigma} d_{yz,\sigma'}^\dagger d_{yz,\sigma'}+f_{2,\sigma}^\dagger f_{2,\sigma} d_{xz,\sigma'}^\dagger d_{xz,\sigma'}\right]+ U_2 \sum\left[ f_{1,\sigma}^\dagger d_{xz,\sigma} d_{xz,\sigma'}^\dagger f_{1,\sigma'}+ f_{2,\sigma}^\dagger d_{yz,\sigma} d_{yz,\sigma'}^\dagger f_{2,\sigma'}\right]\nn &+ \bar{U}_2 \sum\left[ f_{1,\sigma}^\dagger d_{yz,\sigma} d_{yz,\sigma'}^\dagger f_{1,\sigma'}+ f_{2,\sigma}^\dagger d_{xz,\sigma} d_{xz,\sigma'}^\dagger f_{2,\sigma'}\right]+\dfrac{U_3}{2}\sum \left[ f_{1,\sigma}^\dagger d_{xz,\sigma} f_{1,\sigma'}^\dagger d_{xz,\sigma'}+f_{2,\sigma}^\dagger d_{yz,\sigma} f_{2,\sigma'}^\dagger d_{yz,\sigma'}\right]\nn & +\dfrac{\bar{U}_3}{2} \sum \left[ f_{1,\sigma}^\dagger d_{yz,\sigma} f_{1,\sigma'}^\dagger d_{yz,\sigma'}+ f_{2,\sigma}^\dagger d_{xz,\sigma} f_{2,\sigma'}^\dagger d_{xz,\sigma'}\right]+\dfrac{U_5}{2} \sum  \left[f_{1,\sigma}^\dagger f_{1,\sigma}f_{1,\sigma'}^\dagger f_{1,\sigma'}+f_{2,\sigma}^\dagger f_{2,\sigma}f_{2,\sigma'}^\dagger f_{2,\sigma'}\right]\nn &+ \tilde{U}_5 \sum \, f_{1,\sigma}^\dagger f_{1,\sigma}f_{2,\sigma'}^\dagger f_{2,\sigma'}+
 +\tilde{\tilde{U}}_5 \sum  \, f_{1,\sigma}^\dagger f_{2,\sigma}f_{2,\sigma'}^\dagger f_{1,\sigma'} +\dfrac{\bar{U}_5}{2} \sum \left[f_{1,\sigma}^\dagger f_{2,\sigma}f_{1,\sigma'}^\dagger f_{2,\sigma'}+f_{2,\sigma}^\dagger f_{1,\sigma}f_{2,\sigma'}^\dagger f_{1,\sigma'}\right].
 \label{Interaction in orbital basis}
\end{align}
Here, we omit the momentum index in the fermionic operators for the simplicity of the notation and $"\sum"$ represents summation over the momenta with momentum conservation. In terms of bare Hubbard-Hund interactions, these interactions are 
\begin{align}
\label{bare interaction 1}
U_1&=U_2=U_3=U_4=U_5=U,\\
\bar{U}_1&=\tilde{U}_4=\tilde{U}_5=U',\\
\bar{U}_2&=\tilde{\tilde{U}}_4=\tilde{\tilde{U}}_5=J,\\
\bar{U}_3&=\bar{U_4}=\bar{U}_5=J',
\label{bare interaction 4}
\end{align}
with $U,U',J,J'$ are intra-orbital Hubbard repulsion, inter-orbital Hubbard repulsion, inter-orbital Hund exchange and inter-orbita Hund exchange term known as pair hopping term respectively. 
We convert these interactions in the band basis(for $X-, Y-$ pockets Eq.~\eqref{Interaction in orbital basis} is already in the band basis) using the Eqs.~(\ref{hole orbital to band 1}-\ref{hole orbital to band 2}) and write the interaction Hamiltonian as

\begin{align}
H_{\text{int}}=\sum_{i,j\in \{h,X,Y\}}H_{i,j}+H_\text{other}
\end{align}
Here, $H_{i,j}$ is the interaction between the pockets $i$ and $j\epsilon\{h,Y,X\}$ and takes the form
\begin{align}
 \label{inth}
H_{h,h}&=\dfrac{1}{2} \sum_{\bk,\bp,\bq} h_{\bk,\sigma}^\dagger h_{\bk+\bq,\sigma}h_{\bp,\sigma'}^\dagger h_{\bp-\bq,\sigma'} V_{h,h}^{\text{den}}(\bk,\bk+\bq;\bp,\bp-\bq),\\ 
H_{h,Y}&= \sum_{\bk,\bp,\bq} f_{1,\bk,\sigma}^\dagger f_{1,\bk+\bq,\sigma} h_{\bp,\sigma'}^\dagger h_{\bp-\bq,\sigma'} V_{h,Y}^{\text{den}}(\bp,\bp-\bq)+\sum_{\bk,\bp,\bq}f_{1,\bk,\sigma}^\dagger h_{\bk+\bq,\sigma} f_{1,\bp,\sigma'}^\dagger h_{\bp-\bq,\sigma'} V_{h,Y}^{\text{ph}}(\bk+\bq,\bp-\bq)\nn & +\sum_{\bk,\bp,\bq}f_{1,\bk,\sigma}^\dagger h_{\bk+\bq,\sigma} h_{\bp,\sigma'}^\dagger f_{1,\bp-\bq,\sigma'} V_{h,Y}^{\text{ex}}(\bk+\bq,\bp),\\ 
H_{h,X}&= \sum_{\bk,\bp,\bq} f_{2,\bk,\sigma}^\dagger f_{2,\bk+\bq,\sigma} h_{\bp,\sigma'}^\dagger h_{\bp-\bq,\sigma'} V_{h,X}^{\text{den}}(\bp,\bp-\bq)+\sum_{\bk,\bp,\bq}f_{2,\bk,\sigma}^\dagger h_{\bk+\bq,\sigma} f_{2,\bp,\sigma'}^\dagger h_{\bp-\bq,\sigma'} V_{h,X}^{\text{ph}}(\bk+\bq,\bp-\bq)\nn & +\sum_{\bk,\bp,\bq}f_{2,\bk,\sigma}^\dagger h_{\bk+\bq,\sigma} h_{\bp,\sigma'}^\dagger f_{2,\bp-\bq,\sigma'} V_{h,X}^{\text{ex}}(\bk+\bq,\bp).
\label{interaction in band basis}
\end{align}
The interaction within the electron pockets labeled as $H_{X,X}, H_{Y,Y}$ and $H_{X,Y}$ are already in the band basis and captured by the  last $4$ terms of Eq.(\ref{Interaction in orbital basis}).
We group interactions into density-density, exchange and pair-hopping by labeling them "den", "ex" and "ph" respectively. Since the electron pockets are assumed to be mono-orbital, in Eqs.(\ref{inth}-\ref{interaction in band basis}), interaction strengths are momentum dependent only for the hole pockets in the follwoing way \begin{align}
 \label{Vhh}
V_{h,h}^{\text{den}}=& U_4  \Big\{\sin\theta_\bk \sin\theta_{\bk+\bq}\sin\theta_\bp\sin\theta_{\bp-\bq}+\cos\theta_\bk \cos\theta_{\bk+\bq}\cos\theta_\bp\cos\theta_{\bp-\bq}\Big\}\nn & +\tilde{U}_4\Big\{\sin\theta_\bk \sin\theta_{\bk+\bq}\cos\theta_\bp\cos\theta_{\bp-\bq}+\cos\theta_\bk \cos\theta_{\bk+\bq}\sin\theta_\bp\sin\theta_{\bp-\bq}\Big\}\nn & +\tilde{\tilde{U}}_4\Big\{\sin\theta_\bk \cos\theta_{\bk+\bq}\cos\theta_\bp\sin\theta_{\bp-\bq}+\cos\theta_\bk \sin\theta_{\bk+\bq}\sin\theta_\bp\cos\theta_{\bp-\bq}\Big\}\nn & + \bar{U}_4\Big\{\sin\theta_\bk \cos\theta_{\bk+\bq}\sin\theta_\bp\cos\theta_{\bp-\bq}+\cos\theta_\bk \sin\theta_{\bk+\bq}\cos\theta_\bp\sin\theta_{\bp-\bq}\Big\} ,\\
V_{h,Y}^{\text{den}}&= U_1 \, \sin\theta_\bp \sin\theta_{\bp-\bq}+\bar{U}_1\, \cos\theta_\bp \cos\theta_{\bp-\bq},\\ V^{h,X}_{\text{den}}&= U_1 \, \cos\theta_\bp \cos\theta_{\bp-\bq}+\bar{U}_1\, \sin\theta_\bp \sin\theta_{\bp-\bq},  \\
V_{h,Y}^{\text{ph}}&= \dfrac{U_3}{2} \, \sin\theta_{\bk+\bq} \sin\theta_{\bp-\bq}+\dfrac{\bar{U}_3}{2}\, \cos\theta_{\bk+\bq} \cos\theta_{\bp-\bq},\\  V^{h,X}_{\text{ph}}&= \dfrac{U_3}{2} \, \cos\theta_{\bk+\bq} \cos\theta_{\bp-\bq}+\dfrac{\bar{U}_3}{2}\, \sin\theta_{\bk+\bq} \sin\theta_{\bp-\bq}, \\
V_{h,Y}^{\text{ex}}&=U_2 \, \sin\theta_{\bk+\bq} \sin\theta_{\bp}+\bar{U}_2\, \cos\theta_{\bk+\bq} \cos\theta_{\bp},\\ V_{h,X}^{\text{ex}}&=U_2 \, \cos\theta_{\bk+\bq} \cos\theta_{\bp}+\bar{U}_2\, \sin\theta_{\bk+\bq} \sin\theta_{\bp}.
\label{Vhx}
\end{align}
On the other hand, all other interactions which involve the inner hole band fermions are contained in $H_{\text{other}}$ term. We ignore the effects of this part on the nematic criticality and the superconductivity in the rest of the paper because of the two reasons: (a) unlike the outer hole pocket, inner hole pocket is quite small in size and can even disappear below the Fermi energy in the presence of a small spin-orbit coupling or external strain, and (b) the presence of the inner hole pocket does not change the main results of our paper qualitatively, but complicates the calculation unnecessarily.   
 \section{Orbital Order Instability}
 We define the orbital/nematic order as the density difference between the $d_{xz}$ and $d_{yz}$ fermions. A non-zero value for this order parameter breaks the $Z_2$ symmetry of the system. In the band language, the order parameter translates to two zero momentum d-wave components: one for the hole pocket
 \begin{align}
 \phi_h=&\sum_{\bk,\sigma}\langle d_{yz,\bk,\sigma}^\dagger d_{yz,\bk,\sigma}\rangle-\langle d_{xz,\bk\sigma}^\dagger d_{xz,\bk,\sigma}\rangle \\
 &=\sum_{\bk,\sigma}\langle h_{\bk,\sigma}^\dagger h_{\bk,\sigma} \rangle\cos 2 \theta_{\bk}
 \end{align}
 and another for the electron pockets
 \begin{align}
 \phi_e=&\sum_{\bk}\langle f_{2,\bk,\sigma}^\dagger f_{2,\bk,\sigma} \rangle-\langle f_{1,\bk\sigma}^\dagger f_{1,\bk,\sigma}\rangle \\
 =& \phi_x-\phi_y.
 \end{align}
 Here, $\phi_{x}$ and $\phi_{y}$ are the fermionic density in the $X-$ and $Y-$ pockets respectively. 
 \begin{figure}
\includegraphics[scale=.8]{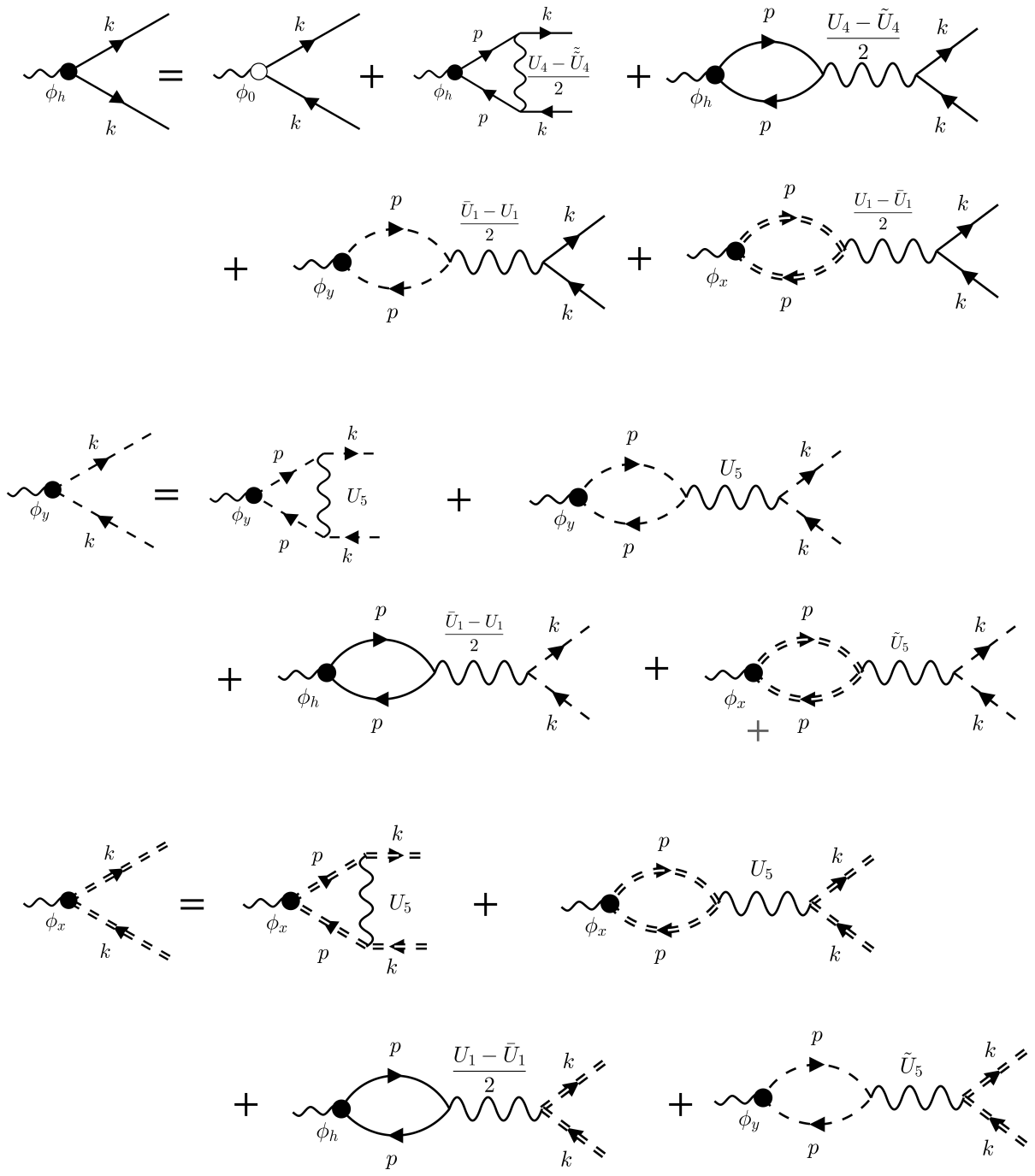}
\centering
\caption{Set of ladder equations for the nematic order vertex for hole pocket($\phi_h)$, $X-$pocket($\phi_x$) and $Y-$ pocket($\phi_y$). Here we label only the d-wave component of the interactions. The  solid line, dashed line and double dashed line represents Green's function for the hole pocket, $X-$ pocket and $Y-$ pocket fermion. The curvy line represents interaction between fermions. Black dot represent the nematic vertex while the void dot represent the bare vertex for the hole pocket only. We keep both the external fermion momentum $\bk$ same. }
\label{nematic diagram}
\end{figure}

 To discuss the nematic instability, we have introduced an infinitesimally small $Z_2$ symmetry breaking external perturbation to the interaction Hamiltonian in the form of $\phi_0 \, \cos2\theta_\bk\, h^\dagger_\bk h_\bk$ and define nematic susceptibility as $\phi_h=\chi_{\text{nem}}\, \phi_0$. The onset of the nematic order is signaled by the divergence of its susceptibility, $\chi_{\text{nem}}$. We represent the set of self-consistent equations for $\phi_h, \phi_x$ and $\phi_y$ in Fig.~\ref{nematic diagram}.    They are obtained by adding up the Hartree and Fock diagrams for different bands and turn out to be
 \begin{align}
 \phi_h \cos 2\theta_\bk  =\phi_0\, \cos 2\theta_\bk+ \phi_h &\bigintss_p G_p^h G_{p}^h\,
\cos 2\theta_\bp  \,V_{h,h}^{\text{den}}\left(\bk,\bp;\bp,\bk\right) -2 \phi_h \bigintss_p G_p^h G_{p}^h
\,\cos2\theta_\bp \,V_{h,h}^{\text{den}}\left(\bk,\bk;\bp,\bp\right) \nn & -2 \phi_y \bigintss_p G_p^Y G_{p}^Y
\,V_{h,Y}^{\text{den}}\left(\bk,\bk;\bp,\bp\right)-2 \phi_x \bigintss_p G_p^X G_{p}^X
\,V_{h,X}^{\text{den}}\left(\bk,\bk;\bp,\bp\right),
\label{hole order eq}
 \end{align}
 \begin{align}
 \phi_y&=\phi_y \bigintss_p G^Y_p G^Y_p\, U_5-2\, \phi_y\bigintss_p G^Y_p G^Y_p \, U_5-2\,\phi_h \bigintss_p G^h_p G^h_p \, \cos2\theta\bp\, V_{h,Y}^{\text{den}}(\bk,\bk;\bp,\bp)-2 \phi_x\,\bigintss_p G^X_p G^X_p \, \tilde{U}_5,
 \label{y order eq}
 \end{align}
  \begin{align}
 \phi_x&=\phi_x \bigintss_p G^X_p G^X_p\, U_5-2\, \phi_x\bigintss_p G^Y_p G^Y_p \,U_5 -2\,\phi_h \bigintss_p G^h_p G^h_p \, \cos2\theta\bp\, V_{h,X}^{\text{den}}(\bk,\bk;\bp,\bp)-2 \phi_y\,\bigintss_p G^Y_p G^Y_p \,  \tilde{U}_5.
 \label{x order eq}
 \end{align}
Here $G^i(\bp,\omega)=1/i \omega-\xi_i(\bp)$ is the Green's function for the pocket $i$ with $4-$ momentum $p=(\bp,\omega)$. $\bigintss_p$ stands for $T \sum_{\Omega_n} \bigintss \dfrac{d^2\bp}{(2\pi)^2}$, where $\Omega_n$ is the fermionic Matsubara frequency($\Omega_n=(2n+1)\pi T$), $\bp$ is the lattice momentum and $T$ is the temperature. Because orbital order is a zero momentum order, we only keep the low momentum transfer interaction like density interaction and ignore large momentum transfer interaction like exchange interaction in Eqs.(\ref{hole order eq}-\ref{x order eq}). Since nematic orders $\phi_h$ and $\phi_e$ are d-wave order, only the d-wave component of the interactions( proportional to $\cos2\theta$ terms) will contribute in Eqs.~(\ref{hole order eq}-\ref{x order eq}).
 Using Eqs.~(\ref{Vhh}-\ref{Vhx}), one finds the $s-$ and $d-$ wave components of the interactions  
 \begin{align}
  V_{h,h}^{\text{den}}\left(\bk,\bp;\bp,\bk\right)&=\dfrac{U_4+\tilde{\tilde{U}}_4}{2}  +\dfrac{U_4-\tilde{\tilde{U}}_4}{2} \cos2\theta_\bk\cos2\theta_\bp+\dfrac{\tilde{U}_4+\bar{U}_4}{2}\sin2\theta_\bk \, \sin2\theta_\bp,
 \end{align}
 \begin{align}
  V_{h,h}^{\text{den}}\left(\bk,\bk;\bp,\bp\right)&=\dfrac{U_4+\tilde{U}_4}{2} +\dfrac{U_4-\tilde{U}_4}{2} \cos2\theta_\bk\cos2\theta_\bp+\dfrac{\tilde{\tilde{U}}_4+\bar{U}_4}{2}\sin2\theta_\bk \, \sin2\theta_\bp,
 \end{align}
 \begin{align}
  V_{h,Y}^{\text{den}}\left(\bk,\bk;\bp,\bp\right)&=\dfrac{U_1+\bar{U}_1}{2}-\dfrac{U_1-\bar{U}_1}{2} \cos2\theta_\bp,
 \end{align}
 \begin{align}
  V_{h,X}^{\text{den}}\left(\bk,\bk;\bp,\bp\right)&=\dfrac{U_1+\bar{U}_1}{2}+\dfrac{U_1-\bar{U}_1}{2} \cos2\theta_\bp.
 \end{align}
 Combining all these terms for Eqs.~(\ref{hole order eq}-\ref{x order eq}), we get
 \begin{align}
 \label{phih equation}
\phi_h&= \phi_0-\phi_h\, U_{h}^d \, \Pi_h^d-\phi_e \,U_{he}\, \Pi_e,\\
\phi_e&=-\phi_e\, U_e^d \Pi_e-2 \phi_h \, U_{he} \Pi_h^d.
\label{phie equation}
 \end{align}
  Here,
 $U_h^d=\dfrac{U_4-2\tilde{U}_4+\tilde{\tilde{U}}_4}{2} 
 $ and 
 $U_e^d =U_5-2 \tilde{U}_5$ are the effective d-wave intra-pockets interactions for hole and electron pockets respectively. $U_{he}= (U_1-\bar{U}_1)$ is the effective d-wave inter-pocket interaction for the nematic order. We define the effective polarization bubbles with vertex factor $\cos2\theta$ for hole pocket and $1$ for electron pockets as
\begin{align}
\label{hole polarization}
\Pi_h^d&=-\bigintss_p G^h_p G^h_p \cos^22\theta_\bp,\\
\Pi_e&=-\bigintss_p G^e_p G^e_p 
\label{electron polarization} 
\end{align}
respectively. Within our convention, the polarization bubbles are positive $\Pi_h^d,\Pi_e>0$. Eq.~\eqref{phie equation} is found from two other equations, $\phi_y=-\phi_y \, U_5 \, \Pi_e+\phi_h \, U_{he} \, \Pi_h^d-2\, \phi_x\,\tilde{U}_5 \, \Pi_e$ and 
$\phi_x=-\phi_x \, U_5 \, \Pi_e-\phi_h \, U_{he} \, \Pi_h^d-2\, \phi_y\, \tilde{U}_5 \, \Pi_e$. Combining Eqs.~(\ref{phih equation}) and (\ref{phie equation})), we find the expression for $\chi_{\text{nem}}$,
\begin{align}
\chi_{\text{nem}}=\dfrac{1}{\left(1+U_h^d\,\Pi_h^d\right)\left(1+U_e^d\, \Pi_e\right)-2U_{he}^2 \Pi_h^d \Pi_e}=\dfrac{1}{Z}.
\label{nematic susceptibility}
\end{align}
 As stated before the onset of the orbital order is set when $Z=0$. To see what kind of orbital order (measured by the relative sign between $\phi_h$ and $\phi_e$) is produced at the transition point, we need to solve Eqs.~(\ref{phih equation}) and (\ref{phie equation})) without the source term $\phi_0$. At the bare interaction level(\ref{bare interaction 1}-\ref{bare interaction 4}), we find no solution exists for $J>U/5$ while only sign preserving solution(called $d^{++}$ with sgn$(\phi_h)$=sgn$(\phi_e)$) exists when $J<U/5$. To see this we need to include the effect of the large momentum transfer interaction($U_2,\bar{U}_2,\tilde{\tilde{U}}_5$) as at the bare level they are equal in magnitude to the low momentum transfer density interaction and can't be ignored. The calculation is straightforward \cite{xing2018orbital} and makes both the effective intra-pocket and inter-pocket interactions comparable: $U_h^d=U_{he}=U_e^d/2=(5\,J-U)/2$. A possibility of another solution called $d^{\pm}$ orbital order with sgn$(\phi_h)$=-sgn$(\phi_e)$ can exist if we consider $U_h^d, U_e^d$ and $U_{he}$ differ from the bare values such that $U_h^d \, U_e^d\neq 2 U^2_{he}$ and $U_{he}>0$. This can happen as a consequence of the renormalization coming from the high energy fermions. In experiments, $d^\pm$ orbital order is found \cite{sprau2017discovery}. 
  
\section{Bare Pairing Interaction}\label{Bare interaction}

   In the superconducting channel, the bare pairing interaction from Eq.~\eqref{Interaction in orbital basis} reads
\begin{align}
H_{\text{pair}}= &\dfrac{ V_0^{h}}{2} h_{\bk,\uparrow}^\dagger h_{-\bk,\downarrow}^\dagger h_{-\bp,\downarrow} h_{\bp,\uparrow}+\dfrac{V_0^{e}}{2} \sum_{i=1}^2 f_{i,\bk,\uparrow}^\dagger f_{i,-\bk,\downarrow}^\dagger f_{i,-\bp,\downarrow} f_{i\bp,\uparrow}+V^{h,e}_s h_{\bk,\uparrow}^\dagger h_{-\bk,\downarrow}^\dagger \left(f_{1,-\bp,\downarrow} f_{1\bp,\uparrow}+f_{2,-\bp,\downarrow} f_{2\bp,\uparrow}\right)\\
& +V^{h,e}_d h_{\bk,\uparrow}^\dagger h_{-\bk,\downarrow}^\dagger \left(f_{2,-\bp,\downarrow} f_{2\bp,\uparrow}-f_{1,-\bp,\downarrow} f_{1\bp,\uparrow}\right)\, \cos 2\theta_{\bk}.
\end{align}
Here,$V^{h}_0$ and $V_0^e$ are the intra-pocket pairing interaction for the hole and electron pockets(same for $X-$ and $Y-$) respectively with the expression $V^{h}_0=\dfrac{U_4+\bar{U}_4}{2}+\dfrac{U_4-\bar{U}_4}{2}\cos 2\theta_\bk \cos 2 \theta_\bp+\dfrac{\tilde{U}_4+\tilde{\tilde{U}}_4}{2}\sin 2\theta_\bk \sin 2 \theta_\bp$ and $V^{e}_0=U_5$. On the other hand, $V^{h,e}_s$ and $V^{h,e}_d$ are the $s-$ and $d-$ wave components of the inter-pocket pairing interactions respectively with the expression $V^{h,e}_s=\dfrac{U_3+\bar{U}_3}{2}$ and $V^{h,e}_d=\dfrac{U_3-\bar{U}_3}{2}$. We neglect the pairing interaction between the electron pockets. Using Eqs.~(\ref{bare interaction 1}-\ref{bare interaction 4}), one finds the pairing interactions are repulsive and can only lead to a superconducting instability if the inter-pocket repulsion overcomes intra-pocket repulsion. We define the superconducting gap on the hole pocket as $\Delta_h$, on $X-$ pocket as $\Delta_x$ and on $Y-$pocket as $\Delta_y$. In the tetragonal phase, $s-$ and $d-$ wave components of the superconducting gap equations decouple. For the $s-$ wave component(labeled by a  suffix $s$) the set of linearized equations for the gap is obtained straightforwardly and reads as
\begin{align}
\begin{pmatrix}
\Delta^s_h \\
\Delta^s_e
\end{pmatrix}=L \begin{pmatrix}
-u^s_h & -u_{he}^s\\
-2 u_{he}^s & -u_e
\end{pmatrix} \begin{pmatrix}
\Delta^s_h\\
\Delta^s_e
\end{pmatrix}.
\label{matrix equ}
\end{align}
Here, $\Delta^s_e=\Delta_x+\Delta_y$, $u_h^s=N_0 \dfrac{U_4+\bar{U}_4}{2}$, $u_{he}^s=N_0 \dfrac{U_3+\bar{U}_3}{2}$, $u_e=N_0 \, U_5$, $N_0$ is the density of states, $L=\log\dfrac{1.13 \Lambda}{T_c}$ and $\Lambda$ is the upper cut-off for the pairing. Eq.~\eqref{matrix equ} has a solution only if the largest eigen-value $\lambda_s=\dfrac{-(u^e+u_h^s)+\sqrt{(u^e-u_h^s)^2+8 (u_{he}^s)^2}}{2}$ is positive. For simplicity, if we assume all intra-pocket pairing interactions are same $u_s^h=u^e_s=u$, then this criteria translates to $u_{he}^s>\dfrac{u}{\sqrt{2}}$. For $u_{he}^s<\dfrac{u}{\sqrt{2}}$, there is no $s-$ wave superconducting instability. For the d-wave, the set of linearized equation become
\begin{align}
\begin{pmatrix}
\Delta^d_h \\
\Delta^d_e
\end{pmatrix}=L \begin{pmatrix}
-u_h^d/2 & -u_{he}^d\\
- u_{he}^d & -u_e
\end{pmatrix} \begin{pmatrix}
\Delta^s_h\\
\Delta^s_e
\end{pmatrix}.
\label{matrix equ d}
\end{align}
Here, $\Delta^d_e=\Delta_x-\Delta_y$, $u^d_h=N_0 \dfrac{U_4-\bar{U}_4}{2}$, $u_{he}^d=N_0 \dfrac{U_3-\bar{U}_3}{2}$. Eq.~\eqref{matrix equ d} has a solution only if the largest eigen-value $\lambda_d=\dfrac{-(u_e+u_h^d/2)+\sqrt{(u_e-u_h^d/2)^2+4 (u_{he}^d)^2}}{2}$ is positive. For simplicity, if we assume $u_h^d=u_h^s=u_e=u$ and $u_{he}^s=u_{he}^d=v$, then $\lambda_s=-u+\sqrt{2\, v^2}>\lambda_d=-3\, u+\sqrt{u^2+16\, v^2}/4>0$ when $v>u/\sqrt{2}$: $s-$ wave will be the leading instability. In the next section, we show that near a nematic instability order parameter fluctuations renormalize the intra-pocket pairing interactions $u_s^h, u^e_s$ attractive and scale with the nematic susceptibility $\chi_{\text{nem}}$ which makes $\lambda$ always positive. 
\section{Renormalization of the Pairing Interaction near the Nematic Transition Point}
\label{section $4$}
In Sec.\ref{Bare interaction}, We find that the bare intra-pocket pairing interactions for both the hole and electron pockets, $V^h_0$ and $V^e_0$ respectively are repulsive, and in order to have superconductivity, inter-pocket repulsion, $V^{h,e}_{s,d}$ has to overcome the intra-pocket repulsion. In this section, we derive the corrections to the intra-pocket and inter-pocket pairing interaction due to the proximity of a nematic order. We show that the dressed intra-pocket pairing interactions labeled as $V^h_\text{eff}$(for hole) and  $V^e_\text{eff}$ (for electron pockets) get an attractive singular component proportional to the nematic susceptibility $\chi_{\text{nem}}$ besides its regular repulsive part near the nematic order. On the other hand, the large momentum transfer($\pi$) type interaction pair hopping/inter-pocket pairing interaction is mostly unaffected by the low momentum nematic fluctuations.
\subsection{Intra-pocket pairing interaction} 
 We first focus on the pairing interaction on the hole pocket describing scattering between fermion pair states ($\bk,\alpha
 ;-\bk,\gamma) \rightarrow (\bp,\beta;-\bp,\delta)$ where $\bk,\bp$ are the momentum and $\alpha,\beta,\gamma,\delta$ are the spin components of the incoming and outgoing fermions. Because of the orbital degrees of freedom for the hole pocket fermions, it is convenient to cast Eqs. (\ref{inth}-\ref{interaction in band basis}) in a slightly different format
\begin{align}
H_{hh}=\dfrac{1}{2}\sum_{\bk,\bp,\bq} b^T(\bk,\bk+\bq) \,\Gamma_{hh} \,b(\bp,\bp-\bq) h_{\bk,\sigma}^\dagger h_{\bk+\bq,\sigma}h_{\bp,\sigma'}^\dagger h_{\bp-\bq,\sigma'}
\label{Hhh}
\end{align}
\begin{align}
H_{h,X}=\sum_{\bk,\bp,\bq} d^T \,  \Gamma_{hx} \, b(\bp,\bp-\bq) f_{2,\bk,\sigma}^\dagger f_{2,\bk+\bq,\sigma} h_{\bp,\sigma'}^\dagger h_{\bp-\bq,\sigma'}
\end{align}
\begin{align}
H_{h,Y}=\sum_{\bk,\bp,\bq} d^T \,  \Gamma_{hy} \, b(\bp,\bp-\bq) f_{1,\bk,\sigma}^\dagger f_{1,\bk+\bq,\sigma} h_{\bp,\sigma'}^\dagger h_{\bp-\bq,\sigma'}.
\end{align}
 
Here, $b$ and $d$ are the bare vertex for the intra-hole($\Gamma_{hh}$) and inter-pocket($\Gamma_{hx}$ and $\Gamma_{hy}$) density-density interactions defined as \begin{align}
\label{b vertex}
b^T(\bk,\bk+\bq)&=\begin{pmatrix}
\cos\theta_{\bk} \cos\theta_{\bk+\bq} & \sin\theta_{\bk} \sin\theta_{\bk+\bq} & \cos\theta_{\bk} \sin\theta_{\bk+\bq} & \sin\theta_{\bk} \cos\theta_{\bk+\bq}
\end{pmatrix},\\ d^T&=\begin{pmatrix}
1 & 1 & 0 & 0
\end{pmatrix} ,
\end{align} and 
\begin{align}
\label{gamma interaction}
\Gamma_{hh}=  \begin{pmatrix}
 u_4 & \tilde{u_4} & 0 & 0 \\
 \tilde{u_4} & u_4 &0 & 0\\
 0 & 0 & \bar{u_4} & \tilde{\tilde{u_4}}\\
 0 & 0 & \tilde{\tilde{u_4}} & \bar{u_4}
 \end{pmatrix} 
 \hspace{1 cm}
 \Gamma_{hx}= \begin{pmatrix}
 U_1 & 0 & 0 & 0 \\
0 & \bar{U}_1 &0 & 0\\
 0 & 0 & 0 & 0\\
 0 & 0 & 0 & 0
 \end{pmatrix} \hspace{1 cm}
 \Gamma_{hy}= \begin{pmatrix}
 \bar{U}_1 & 0 & 0 & 0 \\
0 & U_1 &0 & 0\\
 0 & 0 & 0 & 0\\
 0 & 0 & 0 & 0
 \end{pmatrix} .
\end{align}
The lower $2\times 2$ diagonal block of $\Gamma_{hx}$ and $\Gamma_{hy}$ is zero as a consequence of the orbital preserving interaction between hole and electron pockets.
To simplify the presentation, we first ignore the inter-electron pocket density interaction $\tilde{U}_5$ in our microscopic derivation of the dressed interaction, and later give our full result when one includes it. 
\subsubsection{Singular part of the pairing interaction} 
\label{singular part}

\begin{figure}[]
\includegraphics[scale=.3]{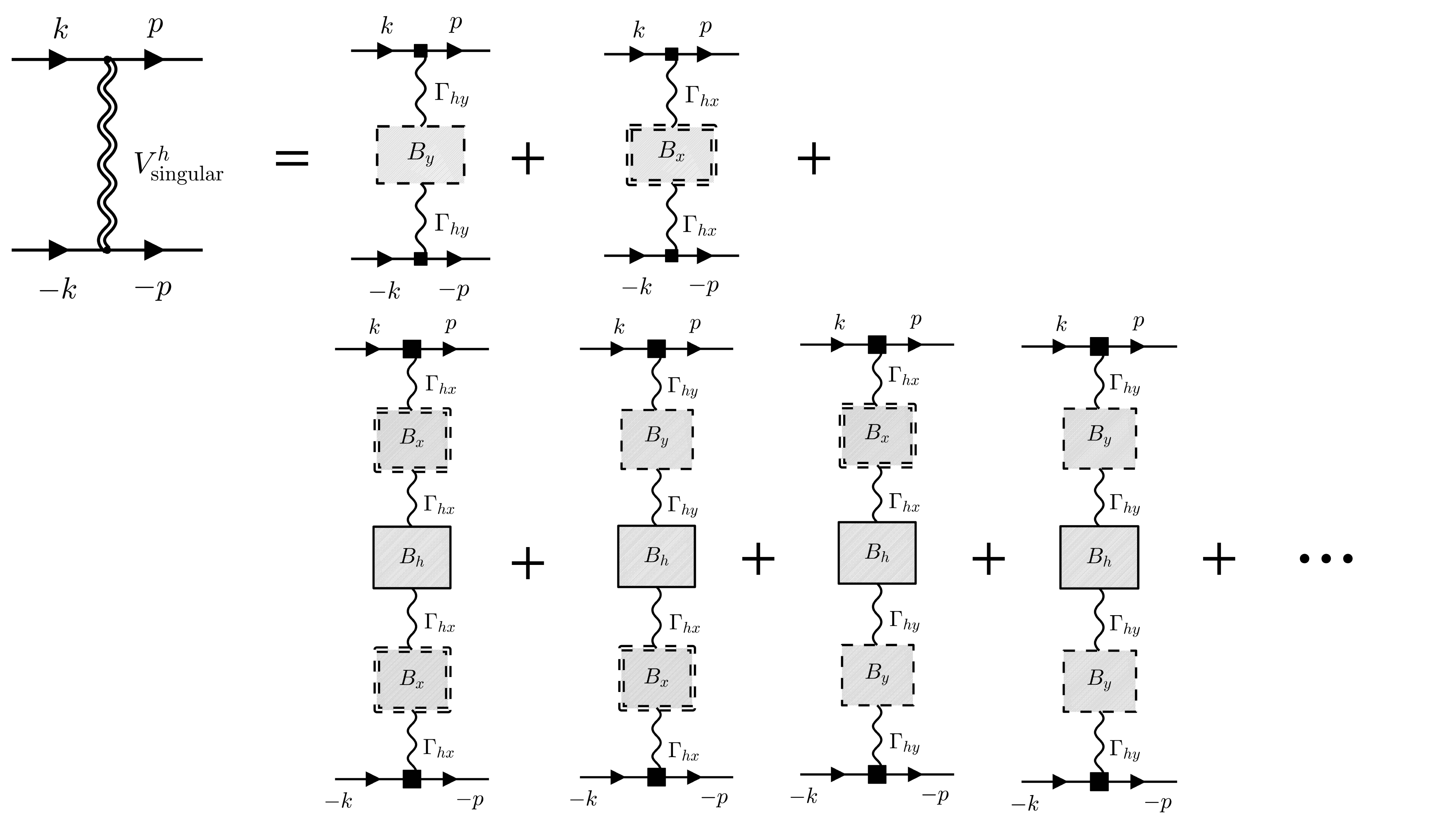}
\centering
\caption{Dyson equation for the singular part of the dressed pairing interaction labeled as $V^h_\text{singular}$ represented by the double curvy line between the fermions with paired state $(\bk,-\bk)$ scattering to ($\bp,-\bp$) within the hole pocket. The black straight arrow represents hole fermionic Green's function.The shaded boxes labeled as $B_h,B_x$ and $B_y$ are defined in Fig.\ref{shaded box}. The black rectangular vertex is defined in  Fig.\ref{external vertex}. Single curvy line represents the bare interaction between the hole and electron pocket. "$\cdots$" represents higher order terms of series. }
\label{effective interaction diagram}
\end{figure}

\begin{figure}
\includegraphics[scale=.3]{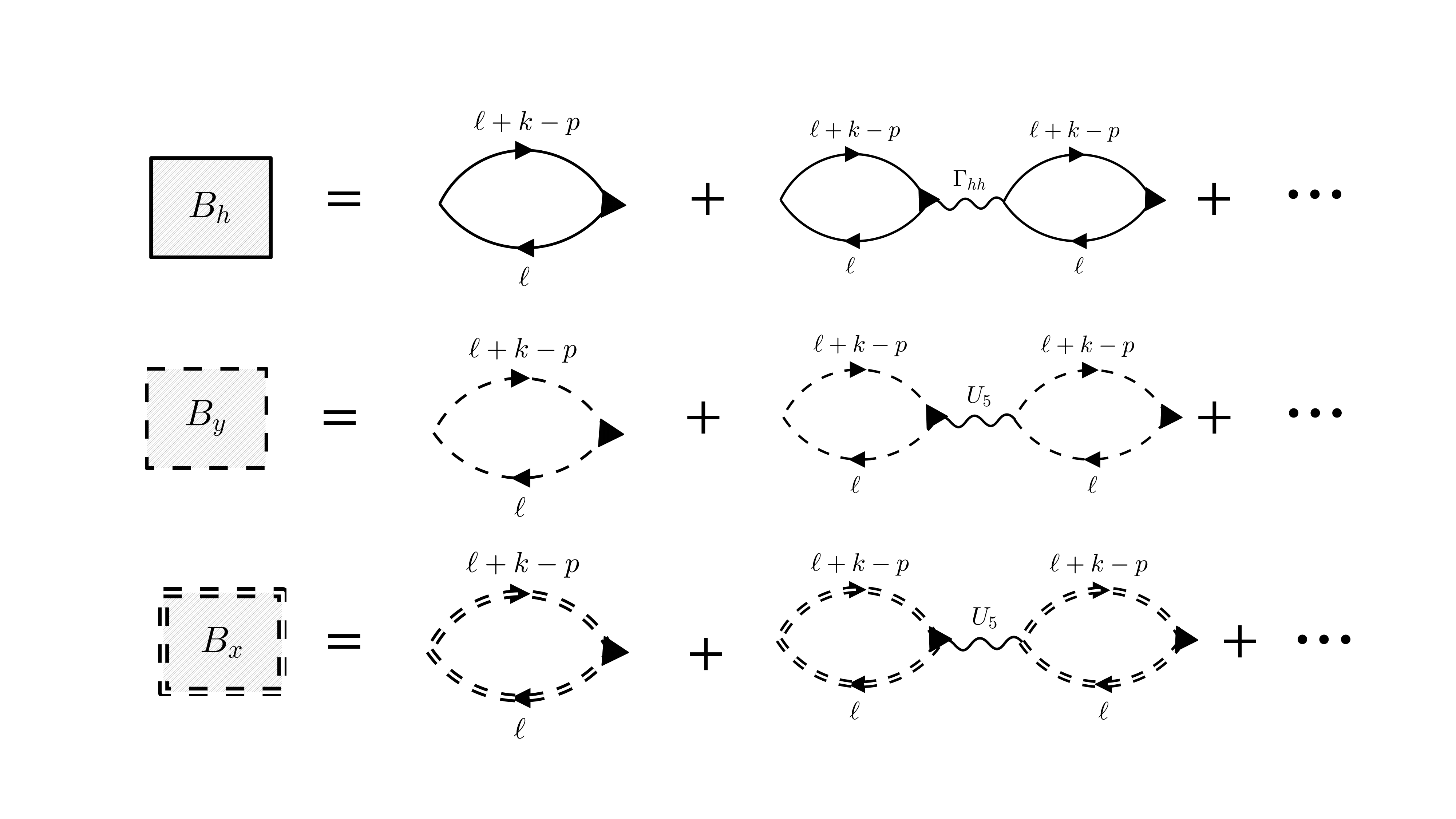}
\centering
\caption{Dyson equation for the shaded boxes $B_h, B_y$ and $B_x$ used in Fig.\ref{effective interaction diagram} are represented here. The solid, dashed and double-dashed line represent the Green's functions for the hole, $Y-$ and $X-$ pocket fermions. The Black triangle represents the dressed vertex $\gamma$ for different pockets depicted in Fig.\ref{vertex correction in bubble}. The curvy lines represent the interactions defined in the main text."$\cdots$" captures the higher order terms of the corresponding series.}
\label{shaded box}
\end{figure}

\begin{figure}
\includegraphics[scale=.3]{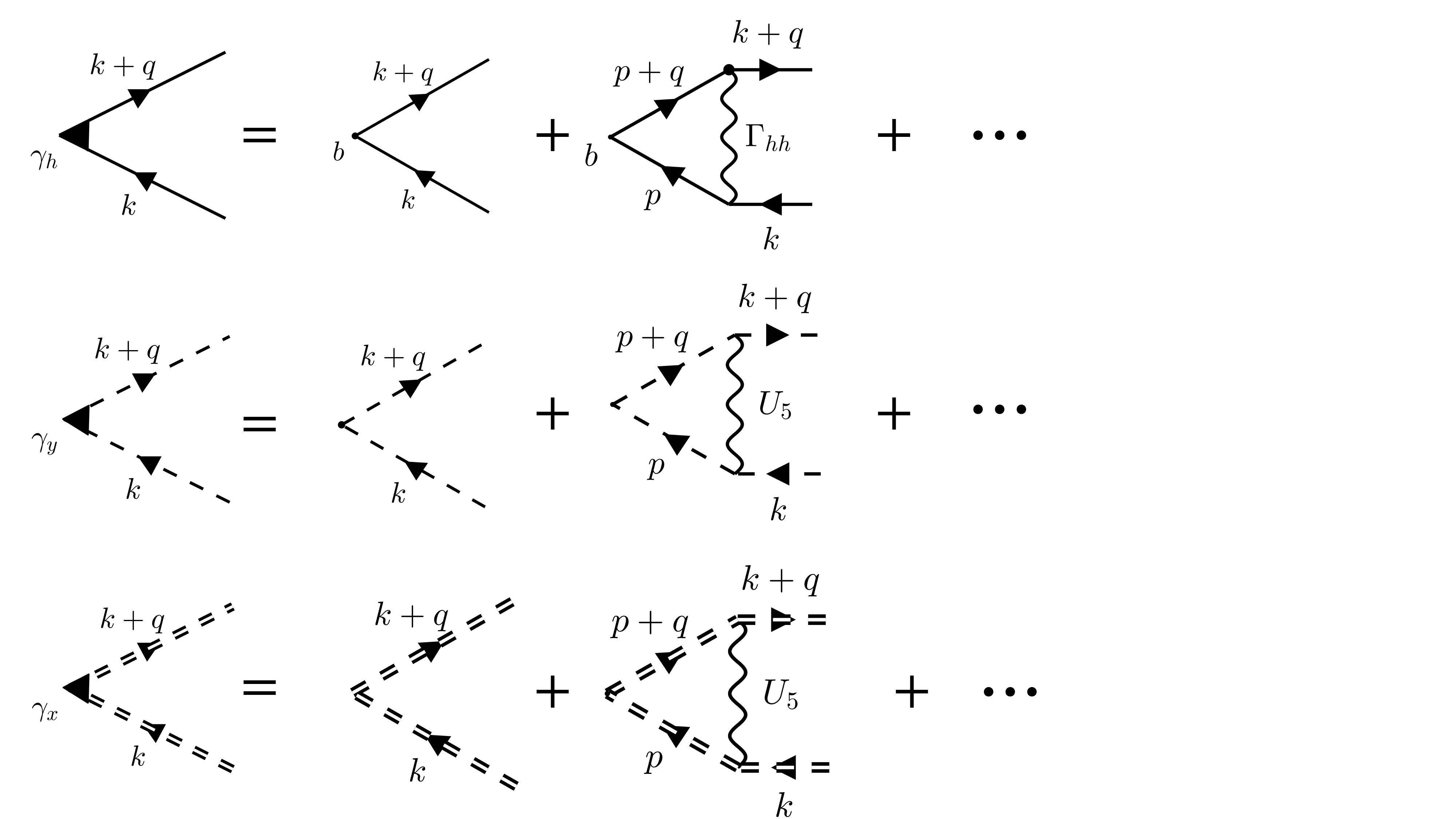}
\centering
\caption{Ladder series for the polarization bubble vertex correction defined in Fig.\ref{shaded box}. We label the dressed vertex $\gamma_h,\gamma_y$ and $\gamma_x$ for the hole, $Y-$ and $X-$ pocket fermions. }
\label{vertex correction in bubble}
\end{figure}

\begin{figure}
\includegraphics[scale=.3]{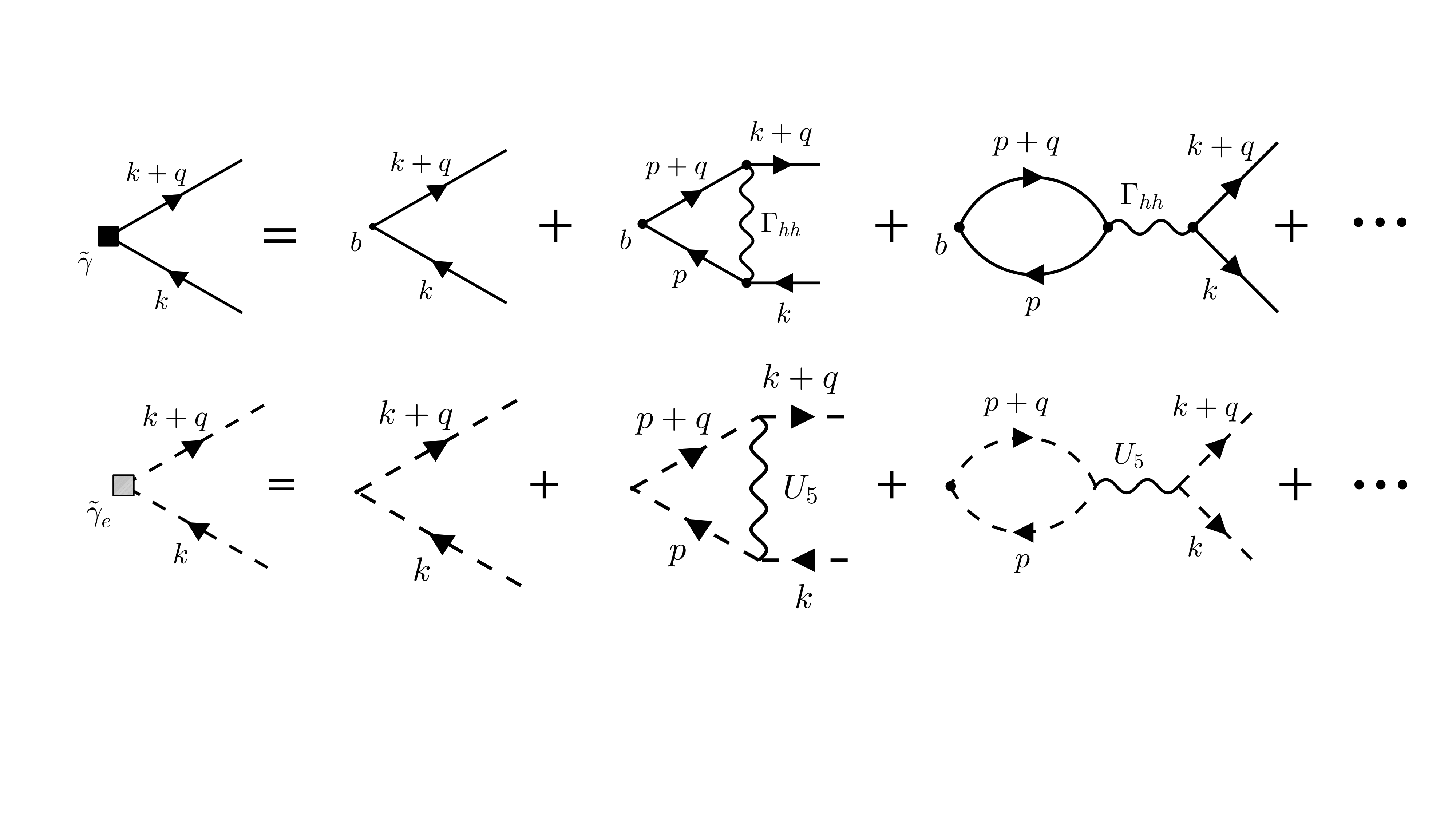}
\centering
\caption{Dyson equation for the dressed vertex between the external fermions on the hole pocket labeled as $\tilde{\gamma}_h$(black square) and on the electron pockets labeled as $\tilde{\gamma}_e$(gray square). $b$ represents the bare vertex for the intra-hole interaction defined in Eq.~\eqref{Hhh} for hole pocket. The black and dashed straight arrow represent the Green's function for the hole and electron pocket(in this case same for both $X-$ and $Y-$) fermions. Wavy lines represent the interactions defined in the main text. "$\cdots$" represent the higher order terms in the series.}
\label{external vertex}
\end{figure}

We argue below that the singular part of the renormalized intra-pocket pairing interaction, $V^h_\text{eff}$ near the onset of the nematic order is captured by the Dyson equation illustrated in Fig.\ref{effective interaction diagram} and we call it $V^h_\text{singular}$. Each shaded boxes in Fig.\ref{effective interaction diagram} labeled as $B_h$, $B_x$ and $B_y$ are an infinite series of bubble diagrams(defined below) made of hole, $X-$ and $Y-$ pocket fermions respectively as illustrated in Fig.\ref{shaded box}. Each bubble is dressed by an insertion of a ladder series of interactions at one of the vertices labeled as $\gamma_h, \gamma_x$ and $\gamma_y$ for the hole, $X-$ and $Y-$ pocket bubbles respectively shown in Fig.\ref{vertex correction in bubble}. We also dress the bare external vertex $b$ by including a ladder series of wine-glass (second diagram of the right hand side of the equation of Fig.\ref{external vertex}) and bubble diagram (third diagram of the right hand side of the equation of Fig.\ref{external vertex}) shown in Fig.\ref{external vertex} and label it as $\tilde{\gamma}_h$. With these preliminary definitions, the computation of the singular part of the effective pairing interaction for the hole pocket fermions as depicted in Fig.\ref{effective interaction diagram} goes as
\begin{align}
\label{veff1}
-V^h_{\text{singular}}(\bk,\bp)& =\tilde{\gamma}^T_h(\bk,\bp)\cdotp \Bigr[S+S \circ B_h \circ S+\cdots \Bigr]\cdot\tilde{\gamma}_h(-\bk,-\bp), \\
&= \tilde{\gamma}^T_h(\bk,\bp)\cdot S \circ \dfrac{1}{1-B_h\circ S}\cdot \tilde{\gamma}_h(-\bk,-\bp),
\label{veff2}
\end{align}
where "$\cdots$" represents the  higher order terms of the series and the spin component of each diagram is $\delta_{\alpha \beta}\delta_{\gamma \delta}$. $S$ is the sum of the first two diagrams of the right hand side of Fig.\ref{effective interaction diagram}(without the external vertex corrections represented by the black square) which translates to
\begin{align}
\label{veff3}
S=\Gamma_{hy}\cdot d\cdot d^T\cdot \Gamma_{hy} B_{y}+\Gamma_{hx}\cdot d\cdot d^T\cdot \Gamma_{hx} B_{x}.
\end{align} In Eqs.~(\ref{veff1}-\ref{veff3}), for convenience we omit the momentum($q=k-p$) dependence of $S$ and $B_h$, and use "$\cdot$" to denote the product between a matrix and vector and "$\circ$" to denote the product between two matrices. Because of the orbital structure of the hole pocket, $\tilde{\gamma}_h, \gamma_h, B_h$ and $S$ are matrix in nature. Below we compute the analytical expression for $S,\gamma_h, \tilde{\gamma}_h, B_h, B_x$ and $B_y$ to give the final form of Eq.~\eqref{veff2}. We first calculate $\gamma_h$, the vertex correction in the bubble with an insertion of the ladder series as shown in Fig.\ref{vertex correction in bubble}. A straightforward calculation for $\gamma_h$ gives 

 \begin{align}
\gamma_h(\bk,\bk+\bq)=b(\bk,\bk+\bq)- \bigintss_p b(\bp, \bp+\bq)G^h_p G^h_{p+q} b^T(\bk,\bp)\cdot\Gamma_{hh}\cdot b(\bp+\bq,\bk+\bq)+\cdots
\label{hole ladder 1}
\end{align}
where $\cdots$ represents higher order terms in the ladder series. We define \begin{align}
\bigintss_p b(\bp, \bp+\bq)\, G^h_p\,  G^h_{p+q} \, b^T(\bk,\bp)\cdot\Gamma_{hh}\cdot b(\bp+\bq,\bk+\bq)=-\alpha(q)\cdot b(\bk,\bk+\bq).
\label{alphat}
\end{align}
Here $\alpha(q)$ is a $4\times 4$ matrix defined below
\begin{align}
\alpha= \begin{pmatrix}
U_4\Pi_{cc}+\tilde{\tilde{U}}_4 \Pi_m & U_4\Pi_{m}+\tilde{\tilde{U}}_4 \Pi_{ss} & 0 & 0\\
U_4\Pi_{m}+\tilde{\tilde{U}}_4 \Pi_{cc} & U_4\Pi_{ss}+\tilde{\tilde{U}}_4 \Pi_m & 0 & 0\\
 0 & 0 & \tilde{U}_4\Pi_{cs}+\bar{U}_4\Pi_m & \tilde{U}_4\Pi_{m}+\bar{U}_4\Pi_{cs}\\
 0 & 0 &  \tilde{U}_4\Pi_{m}+\bar{U}_4\Pi_{cs} & \tilde{U}_4\Pi_{cs}+\bar{U}_4\Pi_m
\end{pmatrix}
\label{alpha form}
\end{align} with 
\begin{align}
\Pi_{cc}(q)&=-\bigintss_l G^h_l G^h_{l+q} \cos^2\theta_\bl \cos^2\theta_{\bl+\bq}, \quad  \Pi_{ss}(q)=-\bigintss_l G^h_l G^h_{l+q} \sin^2\theta_\bl \sin^2\theta_{\bl+\bq}  \\
\Pi_m(q)& =-\bigintss_l G^h_l G^h_{l+q}  \cos \theta_\bl \sin\theta_\bl \cos\theta_{\bl+\bq}\sin\theta_{\bl+\bq} \quad \Pi_{cs}(q)=-\bigintss_l G^h_l G^h_{l+q} \cos^2 \theta_\bl \sin^2\theta_{\bl+\bq}
\end{align} In Eq.~\eqref{alpha form} we omit the $q$ dependence of $\alpha$ and $\Pi$'s for the simplicity of the representation. We put a negative sign in Eq.~\eqref{alphat} to make $\alpha$ numerically positive. For $q=0$, $\Pi_{cc}=\Pi_{ss}$ and $\Pi_{m}=\Pi_{cs}$. After summing up the ladder series in Eq.~\eqref{hole ladder 1} we get
\begin{align}
\gamma_h(\bk,\bk+\bq)&=\dfrac{1}{1-\alpha}\cdot b(\bk,\bk+\bq).
\label{only gamma}
\end{align}
 
 For the electron pockets, the vertex corrections for $X-$ and $Y-$ pockets are same $\gamma_x=\gamma_y=\gamma_e$ and equal to
 \begin{align}
 \gamma_e=\dfrac{1}{1-\Pi_e U_5},
\end{align}  
where $\Pi_e$ is defined in Eq.~\eqref{electron polarization}. Next we find the expression for the shaded boxes $B_h, B_x$ and $B_y$ which are the ladder series of dressed bubbles, shown in Fig.\ref{shaded box}. For the electron pockets, the insertion of one bubble without the vertex correction makes a contribution of $2 \Pi_e$. A factor of $2$ is coming from the spin. Because we define the polarization bubble ($\Pi_e$) with an overall minus sign, it makes the overall expression for $B_i$ positive by canceling the effect of $-1$ coming from the fermionic loop. Summing up the bubbles with the vertex correction $\gamma_e$ upto infinite order gives 
\begin{align}
B_e&=2 \Pi_e \gamma_e-(2\Pi_e \gamma_e)^2 U_5+\cdots
=\dfrac{2\Pi_e}{1+U_5 \Pi_e},
\label{Be}
\end{align} where $B_x=B_y=B_e$. 
For the hole pocket, the inclusion of a bubble without the vertex correction($\gamma_h$) makes a contribution of(see first diagram for $B_h$ in Fig.\ref{shaded box}) \begin{align}
M(q)= -2\bigintss_l b(\bl,\bl+\bq) G_l G_{l+q}b^T(\bl+\bq,\bl),
\label{M}
\end{align} where $M(q)$ is a $4\times 4$ matrix defined below 
\begin{align}
M=-2 \bigintss_l & G_l \, G_{l+q} & \nn  \times &\begin{pmatrix}
\cos^2\theta_\bl \cos^2\theta_{\bl+\bq} & \cos \theta_\bl \sin\theta_\bl \cos\theta_{\bl+\bq}\sin\theta_{\bl+\bq} & \cos \theta_\bl \sin\theta_\bl \cos^2\theta_{\bl+\bq} & \cos^2 \theta_\bl \cos\theta_{\bl+\bq}\sin\theta_{\bl+\bq}\\
\cos \theta_\bl \sin\theta_\bl \cos\theta_{\bl+\bq}\sin\theta_{\bl+\bq} & \sin^2\theta_\bl \sin^2\theta_{\bl+\bq} & \sin^2\theta_\bl \cos\theta_{\bl+\bq}\sin\theta_{\bl+\bq} & \cos\theta_{\bl}\sin\theta_\bl \sin^2\theta_{\bl+\bq}\\
\cos^2 \theta_\bl  \cos\theta_{\bl+\bq}\sin\theta_{\bl+\bq} & \cos \theta_\bl \sin\theta_\bl \sin^2\theta_{\bl+\bq} & \cos \theta_\bl \sin\theta_\bl \cos\theta_{\bl+\bq}\sin\theta_{\bl+\bq} & \cos^2 \theta_\bl \sin^2\theta_{\bl+\bq}\\
\cos \theta_\bl \sin\theta_\bl \cos^2\theta_{\bl+\bq} & \sin^2\theta_\bl \cos\theta_{\bl+\bq}\sin\theta_{\bl+\bq} & \sin^2\theta_\bl \cos^2\theta_{\bl+\bq} & \cos \theta_\bl \sin\theta_\bl \cos\theta_{\bl+\bq}\sin\theta_{\bl+\bq}
\end{pmatrix}. 
\label{M matrix}
\end{align}

\begin{figure}
\includegraphics[scale=.3]{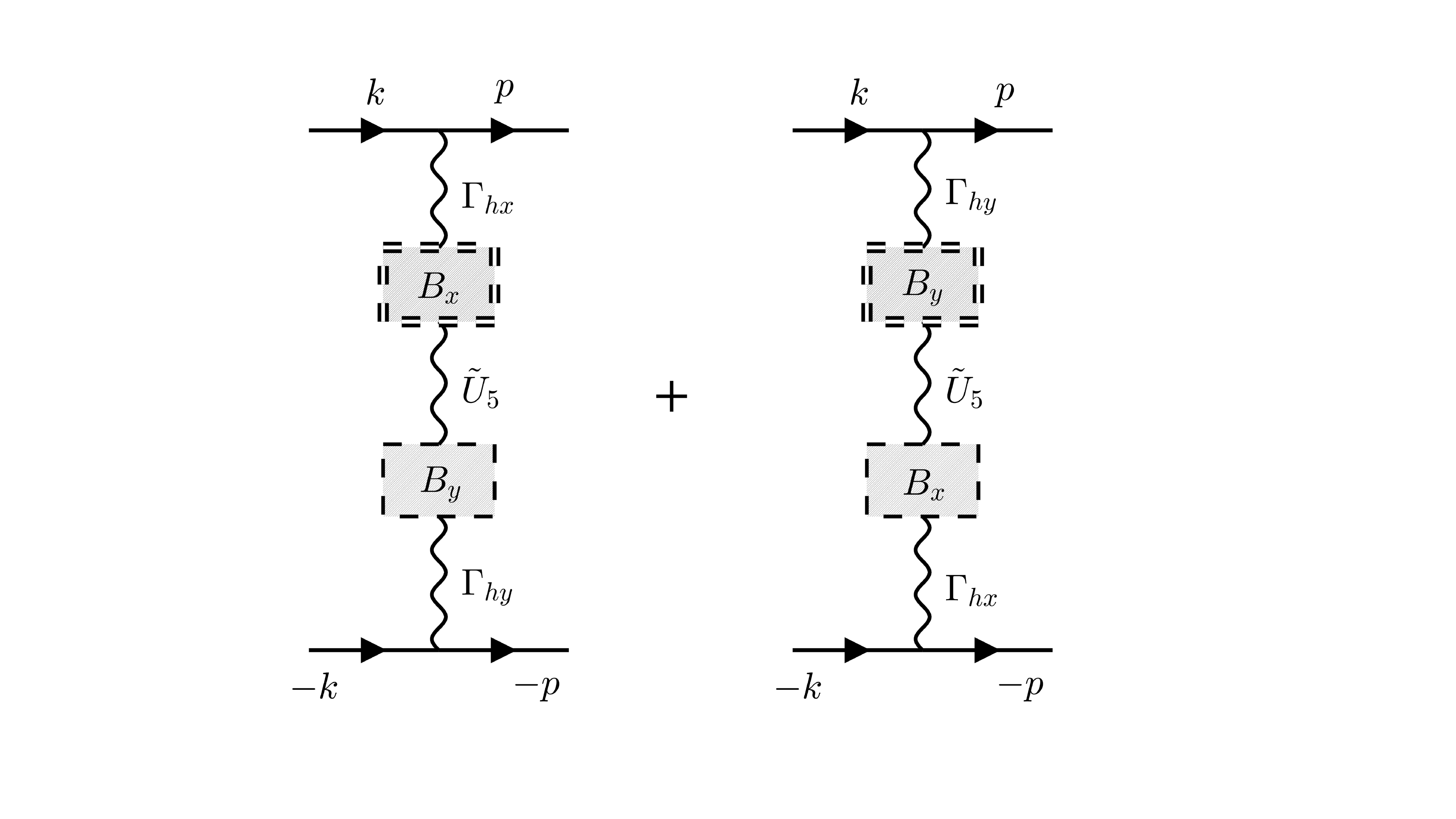}
\centering
\caption{In the presence of the density interaction between the $X-$ and $Y-$ electron pockets, the zeroth order diagram for $V^h_\text{singular}$ defined as $S$ in the main text gets corrected with these two new sets of diagrams.}
\label{U5 tilde}
\end{figure}

It is easy to check that in the presence of any odd number of "sine" term, the integration in Eq.~\eqref{M matrix} will give zero because of the parity symmetry of the underlying Hamiltonian. As a result, $M(q)$ becomes
\begin{align}
M=2 \begin{pmatrix}
\Pi_{cc} & \Pi_m & 0 & 0\\
\Pi_m & \Pi_{ss} & 0 & 0\\
0 & 0 & \Pi_m & \Pi_{cs}\\
0 & 0 & \Pi_{cs} & \Pi_m
\end{pmatrix}.
\label{M form}
\end{align}
Using Eq.~\eqref{only gamma} for the vertex correction $\gamma_h^T=b^T \cdot\dfrac{1}{1-\alpha^T}$, we sum up the bubbles upto infinite orders and get the expression for $B_h$ as

\begin{align}
B_h&=M\circ \dfrac{1}{1-\alpha^T}-M\circ \dfrac{1}{1-\alpha^T}\circ \Gamma_{hh}\circ M\circ \dfrac{1}{1-\alpha^T}+\cdots\\
&=M\circ \dfrac{1}{1-\alpha^T}\circ \dfrac{1}{1+\Gamma_{hh}\circ M\circ \dfrac{1}{1-\alpha^T}}.
\label{Bh}
\end{align}
Next we calculate the correction in the external vertex defined as $\tilde{\gamma}_h$ and depicted in Fig.\ref{external vertex}. Using Eqs. \eqref{alphat} and \eqref{M}, we find the expression for $\tilde{\gamma}_h$

\begin{align}
\tilde{\gamma}_h(\bk,\bk+\bq)&=b(\bk,\bk+\bq)+\alpha \circ b(\bk,\bk+\bq)-M\circ \Gamma_{hh}\cdot  b(\bk,\bk+\bq)+\cdots\\
&=\dfrac{1}{1-\alpha+M \circ \Gamma_{hh}}\cdot b(\bk,\bk+\bq) 
\label{gammah tilde}
\end{align}
 Combining Eqs.(\ref{veff2},\ref{Bh},\ref{gammah tilde}), we find and further simplify the expression for $V^h_{\text{singular}}$
 \begin{align}
-V^h_{\text{singular}}&=b^T\cdot\dfrac{1}{1-\hat{\alpha}^T+\Gamma_{hh} \circ M}\circ S \circ \dfrac{1}{1-B_h\circ S}\circ \dfrac{1}{1-\alpha \circ +M \circ\Gamma_{hh}}\cdot b\\
&=b^T\cdot\dfrac{1}{1-\hat{\alpha}^T+\Gamma_{hh} \circ M} \circ \left[S^{-1}-B_h\right]^{-1}\circ \dfrac{1}{1-\alpha \circ +M \circ\Gamma_{hh}}\cdot b\\
&=b^T\cdot \left[1-\hat{\alpha}^T+\Gamma_{hh} \circ M\right]^{-1}\left[S^{-1}-M \circ \left(1-\alpha^T+\Gamma_{hh} \circ M\right)^{-1}\right]^{-1} \circ\dfrac{1}{1-\alpha \circ +M \circ\Gamma_{hh}}\cdot b\\
&=b^T\cdot \left[S^{-1}\left(1-\alpha^T+ \Gamma_{hh}\circ M \right)-M \right]^{-1}\circ \dfrac{1}{1-\alpha \circ +M \circ\Gamma_{hh}}\cdot b\\
&=b^T\cdot  \dfrac{1}{1-\alpha^T+ \Gamma_{hh}\circ M - S \circ M}\circ S  \circ\dfrac{1}{1-\alpha \circ +M \circ\Gamma_{hh}}\cdot b
\label{Veff simple expression}
 \end{align}
Here, for simplicity of the presentation we suppress the momentum dependence of $V^h_{\text{singular}}, \,\alpha, \, M$ and the vertices $b$ and $b^T$.
Using Eqs.~(\ref{Be}, \ref{veff3}) we find the expression for $S$ as
\begin{align}
S=\dfrac{2 \Pi_e}{1+U_5 \Pi_e} \begin{pmatrix}
U_1^2+\bar{U}_1^2 & 2 U_1 \bar{U}_1 &0 &0\\
2 U_1 \bar{U}_1 & U_1^2+\bar{U}_1^2 & 0 & 0\\
0 & 0 & 0 & 0 \\
0 & 0 & 0 & 0
\end{pmatrix}
\label{S form}
\end{align}
To understand what qualitative results Eq.~\eqref{Veff simple expression} yields, we can calculate order by order in inter-pocket interaction $S$ while ignoring all intra-pocket interaction. To simplify the results we assume the external momenta are same $p=k$. To first order in $S$, direct computation of Eq.~\eqref{Veff simple expression} produce $\Pi_e \{(U_1+\bar{U}_1)^2+(U_1-\bar{U}_1)^2 \cos^22\theta_k\}$ while the second order produces $2 \Pi_e^2 \{(U_1+\bar{U}_1)^4  \Pi_h^s+ (U_1-\bar{U}_1)^4\Pi_h^d\cos^22\theta_k\}.$ $\Pi_h^s$ is the polarization bubble defined in the same way as $\Pi_h^d$, but with the vertex form factor $1$. \begin{align}
\Pi_h^s=-\bigintss_p G^h_p G^h_p.
\end{align}
Finding the pattern, we can add all the higher order terms and find
\begin{align}
-V^h_{\text{singular}}= \dfrac{\Pi_e (U_1-\bar{U}_1)^2}{1-2 \Pi_h^d\, \Pi_e (U_1-\bar{U}_1)^2}\cos^22\theta_{\bk}+\dfrac{\Pi_e (U_1+\bar{U}_1)^2}{1-2 \Pi_h^s\, \Pi_e (U_1+\bar{U}_1)^2}.
\label{simple effective int}
\end{align}
 Identifying with Eq.~\eqref{nematic susceptibility}, we find the first term of Eq.~\eqref{simple effective int} is nothing but the nematic susceptibility $\chi_{\text{nem}}$ with a prefactor proportional to $(U_1-\bar{U}_1)^2\, \cos^22\theta_{\bk}$ when one ignores the intra-pocket density interactions. The second term is non-singular as  it is not related to any instability and can be ignored. Next we will include intra-pocket interactions and calculate Eq.\eqref{Veff simple expression} exactly. To make the flow of the calculation simple, we introduce a matrix function $Q(x)$ as
 \begin{align}
 Q(x)=1-\alpha^T + \Gamma_{hh} \circ M- x\, S \circ M,
 \label{Q form}
 \end{align}  
such that Eq.~\eqref{Veff simple expression} can be written as(again suppressing the momentum dependence to simplify the presentation)
\begin{align}
-V^h_{\text{singular}}= b^T \cdot \left[Q(1)\right]^{-1} \circ S \circ \left[Q^T(0)\right]^{-1}\cdot b.
\end{align} 
Since, the lower $2\times 2 $ diagonal block of $S$ is zero, to compute the effective interaction we need to compute only the upper block of $Q(x)$. Using Eqs.~(\ref{Q form},\ref{S form},\ref{M form},\ref{gamma interaction}), we find the exact expression for $Q(x)$,
\begin{align}
Q(x)=\begin{pmatrix}
A_{11}(x) & A_{12}(x)\\
A_{21}(x) & A_{22}(x)
\end{pmatrix},
\end{align} with 
\begin{align}
\label{A11}
A_{11}(x)&=1+U_4 \Pi_{cc}+\left(2 \tilde{U}_4-\tilde{\tilde{U}}_4\right) \Pi_{m}-x\dfrac{2 \Pi_e}{B} \Bigr[(U_1+\bar{U}_1)^2 (\Pi_{cc}+\Pi_m)+(U_1-\bar{U}_1)^2 (\Pi_{cc}-\Pi_m)\Bigr]\\
A_{22}(x)&=1+U_4 \Pi_{ss}+\left(2 \tilde{U}_4-\tilde{\tilde{U}}_4\right) \Pi_{m}-x\dfrac{2 \Pi_e}{B} \Bigr[U_1+\bar{U}_1)^2 (\Pi_{ss}+\Pi_m)+(U_1-\bar{U}_1)^2 (\Pi_{ss}-\Pi_m)\Bigr]\\
A_{12}(x)&=U_4 \Pi_{m}+\left(2 \tilde{U}_4-\tilde{\tilde{U}}_4 \right) \Pi_{ss}-x\dfrac{2 \Pi_e}{B} \Bigr[(U_1+\bar{U}_1)^2 (\Pi_{ss}+\Pi_m)+(U_1-\bar{U}_1)^2 (\Pi_{ss}-\Pi_m)\Bigr]\\
A_{21}(x)&=U_4 \Pi_{m}+\left(2 \tilde{U}_4-\tilde{\tilde{U}}_4 \right) \Pi_{cc}-x\dfrac{2 \Pi_e}{B} \Bigr[(U_1+\bar{U}_1)^2 (\Pi_{cc}+\Pi_m)+(U_1-\bar{U}_1)^2 (\Pi_{cc}-\Pi_m)\Bigr]
\label{A22}
\end{align}
 where $B=1+U_5 \Pi_e$. Using the expression of $Q(x)$, the effective pairing interaction in terms of $A_{i,j}$ becomes
 \begin{align}
 -V^h_{\text{singular}}= \dfrac{1}{\text{Det}[Q(1)]} \dfrac{1}{\text{Det}[Q(0)}  b^T \cdot \begin{pmatrix}
 A_{22}(1) & -A_{12}(1)\\
 -A_{21}(1) & A_{11}(1)
 \end{pmatrix} \circ S \circ  \begin{pmatrix}
 A_{22}(0) & -A_{12}(0)\\
 -A_{21}(0) & A_{11}(0)
 \end{pmatrix}\cdot b,
 \label{veff simple 2}
 \end{align} where $"\text{Det}"$ represents the determinant. We express the form of determinant for $Q(x)$ using Eqs.~(\ref{A11}-\ref{A22}) as a power series in $B$ with coefficients $Z_i$,
\begin{align}
\text{Det}[Q(x)]= Z_1+x \left(\dfrac{Z_2}{B}+\dfrac{Z_3}{B^2}\right)
\label{Det Q}
\end{align}
with \begin{align}
\label{z1}
Z_1&=(1+U_h^d\, \Pi_h^d)(1+U_h^s\, \Pi_h^s)-(\Pi_{cc}-\Pi_{ss})^2 U_h^d\, U_h^s,\\
Z_2 &= -2 \Pi_e (U_1+\bar{U}_1)^2 \Pi_h^s \left(1+U_h^d\, \Pi_h^d\right)-2 \Pi_e (U_1-\bar{U}_1)^2 \Pi_h^d \left(1+U_h^s\, \Pi_h^s\right)+ 2 \Pi_e \left(\Pi_{cc}-\Pi_{ss}\right)^2 \times \nn & \hspace{2 cm} \left( (U_1+\bar{U}_1)^2 U_h^d+ (U_1-\bar{U}_1)^2 U_h^s\right)\\
Z_3&=4 \Pi_e^2  (U_1+\bar{U}_1)^2\, (U_1-\bar{U}_1)^2\ \left[\Pi_h^d\, \Pi_h^s-(\Pi_{cc}-\Pi_{ss})^2\right].
\label{z3}
\end{align}
 Here, $U_h^s=\dfrac{U_4+2\tilde{U}_4-\tilde{\tilde{U}}_4}{2}$ and we have used the following relations(suppressing the momentum dependence of the polarization bubble below)
 \begin{align}
 \Pi_h^s&=\Pi_{cc}+\Pi_{ss}+2\Pi_m,\\
 \Pi_h^d&=\Pi_{cc}+\Pi_{ss}-2\Pi_m.
\end{align}  In general, $\Pi_{cc}$ and $ \Pi_{ss}$ differ from each other for $q\neq 0$. Since nematic order is $q=0$ momentum order with strong fluctuations around it, the leading singularity in the effective pairing interaction comes form $q=0$. As a result, we can ignore the term $\Pi_{cc}-\Pi_{ss}$ in the Eqs.~(\ref{z1}-\ref{z3}) as the difference for $q\rightarrow 0$ is vanishingly small. With a little bit of algebraic manipulation Eq.~\eqref{Det Q} becomes $\text{Det}[Q(x)]^{-1}=(1+U_5 \, \Pi_e)^2 K_d(x) \times K_s(x),$ where $K_d$ and $K_s$ are defined as
 
 \begin{align}
 \label{Kd}
 K_d(x)=&\dfrac{1}{\left[\left(1+U_5 \Pi_e\right)\left(1+U_h^d \, \Pi_h^d\right)-2 \, x\, \Pi_h^d\, \Pi_e (U_1-\bar{U}_1)^2\right]}\\
 K_s(x)=&\dfrac{1}{\left[\left(1+U_5 \Pi_e\right)\left(1+U_h^s \, \Pi_h^s\right)-2 \, x\, \Pi_h^s\, \Pi_e (U_1+\bar{U}_1)^2\right]}.
\end{align}  
 Next, we compute the matrix multiplication in Eq.~\eqref{veff simple 2}. we assert that the bare vertex $b(\bk,\bp)$ has effectively $2$ components relevant to our calculation with
\begin{align}
b(\bk,\bp)=\begin{pmatrix}
\cos\theta_{\bk} \cos\theta_{\bp} & \sin\theta_{\bk} \sin\theta_{\bp}
\end{pmatrix}. 
\end{align} 
A straight forward calculation shows
\begin{align}
 &b^T(\bk,\bp) \cdot \begin{pmatrix}
 A_{22}(1) & -A_{12}(1)\\
 -A_{21}(1) & A_{11}(1)
 \end{pmatrix} \circ S \circ  \begin{pmatrix}
 A_{22}(0) & -A_{12}(0)\\
 -A_{21}(0) & A_{11}(0)
 \end{pmatrix}\cdot b(\bk,\bp)\nn &=\dfrac{2\Pi_e}{1+U_5\, \Pi_e}\Biggr[\dfrac{V_1(\bq)}{2}\left(1+\cos 2\theta_{\bk}\cos2\theta_{\bp}\right)+\dfrac{V_2(\bq)}{2} \sin 2\theta_{\bk}\sin2\theta_{\bp}\Biggr],
\end{align}
with \begin{align}
V_1&=\dfrac{(U_1+\bar{U}_1)^2}{2}\Bigr(A_{11}(1)-A_{12}(1)\Bigr)\Bigr(A_{11}(0)-A_{12}(0)\Bigr)+\dfrac{(U_1-\bar{U}_1)^2}{2}\Bigr(A_{11}(1)+A_{12}(1)\Bigr)\Bigr(A_{11}(0)+A_{12}(0)\Bigr)\\
V_2&=\dfrac{(U_1+\bar{U}_1)^2}{2}\Bigr(A_{11}(1)-A_{12}(1)\Bigr)\Bigr(A_{11}(0)-A_{12}(0)\Bigr)-\dfrac{(U_1-\bar{U}_1)^2}{2}\Bigr(A_{11}(1)+A_{12}(1)\Bigr)\Bigr(A_{11}(0)+A_{12}(0)\Bigr).
\end{align}
All the momentum dependence of $V_{1,2}$ are hidden in the polarization bubbles used in $A_{i,j}$ functions. Using Eqs.~(\ref{A11}-\ref{A22}), we find the following expression for
\begin{align}
\label{difference}
A_{11}(1)-A_{12}(1)&=1+U_h^d \, \Pi_h^d -\dfrac{2 \Pi_e\, \Pi_h^d}{1+U_5\, \Pi_e} (U_1-\bar{U}_1)^2=\dfrac{1}{K_d(1) (1+U_5 \Pi_e)}\\
A_{11}(0)-A_{12}(0)&=1+U_h^d \, \Pi_h^d = \dfrac{1}{K_d(0) (1+U_5 \Pi_e)}\\
A_{11}(1)+A_{12}(1)&=1+U_h^s \, \Pi_h^s -\dfrac{2 \Pi_e\, \Pi_h^s}{1+U_5\, \Pi_e} (U_1+\bar{U}_1)^2 = \dfrac{1}{K_s(1) (1+U_5 \Pi_e)}\\
A_{11}(0)+A_{12}(0)&=1+U_h^s \, \Pi_h^s = \dfrac{1}{K_s(0) (1+U_5 \Pi_e)}.
\label{addition}
\end{align}
Combining all the terms Eqs.(\ref{Kd}-\ref{addition}), we find the expression for the singular pairing interaction 
\begin{align}
-V^h_{\text{singular}}(\bk,\bp)=\Pi_e\, (U_1-\bar{U}_1)^2 \cos^2\left(\theta_{\bk}+\theta_{\bp}\right)(1+U_5 \, \Pi_e) K_d(1)\, K_d(0) + \Pi_e\, (U_1+\bar{U}_1)^2 \cos^2\left(\theta_{\bk}-\theta_{\bp}\right)(1+U_5 \, \Pi_e) K_s(1)\, K_s(0).
\label{final Veff 0} 
\end{align}
In the presence of the inter-electron pocket density interaction $\tilde{U}_5$, one needs to add two more diagrams to $S$ depicted in Fig.\ref{U5 tilde}. This changes the expression for $S$ from Eq.~\eqref{veff3}) to 
\begin{align}
S=\Gamma_{hy}\cdot d\cdot d^T\cdot \Gamma_{hy} \, B_{y}+\Gamma_{hx}\cdot d\cdot d^T\cdot \Gamma_{hx} \, B_{x}+2\, \Gamma_{hx}\cdot d\cdot d^T\cdot \Gamma_{hy} \, B_{y} \, B_x\, \tilde{U}_5 .
\label{S full form}
\end{align} Using Eq.~\eqref{S full form} to perform the subsequent calculations outlined before, Eq.~(\ref{Veff simple expression}-\ref{final Veff 0}), one arrives at the the final form for the $V^h_{\text{singular}}$ as 
\begin{align}
-V^h_{\text{singular}}(\bk,\bp)=\dfrac{\Pi_e}{1+U_h^d\, \Pi_h^d}\, (U_1-\bar{U}_1)^2 \cos^2\left(\theta_{\bk}+\theta_{\bp}\right)\, \chi_\text{nem} + \dfrac{\Pi_e}{1+U_h^s\, \Pi_h^s}\, (U_1+\bar{U}_1)^2 \cos^2\left(\theta_{\bk}-\theta_{\bp}\right)\, \chi^s_{\text{nem}},
\label{final expression for the singular part}
\end{align} where $\chi^s_\text{nem}=1/\left[\left(1+U_e^s \, \Pi_e\right)\left(1+U_h^s \, \Pi_h^s\right)-2 \, \Pi_h^s\, \Pi_e (U_1+\bar{U}_1)^2\right]$ with $U_e^s=U_5+2\, \tilde{U}_5$. $\chi^s_\text{nem}$ represent the susceptibility of the total fermionic density in the $A_{1g}$ channel and does not diverge at any temperature. On the other hand, near the onset of the d-wave nematic instability, the susceptibility $\chi_{\text{nem}}$ diverges. As a result, we can ignore the non-singular second term of the Eq.~\eqref{final expression for the singular part} and write the effective interaction near the nematic order as
\begin{align}
V_\text{eff}(\bk,\bp)=-A_h\, (U_1-\bar{U}_1)^2 \cos^2\left(\theta_{\bk}+\theta_{\bp}\right)\chi_{\text{nem}}(\bk-\bp),
\label{hole pairing interaction}
\end{align}
where $V_\text{eff}$ is the singular part of $V^h_\text{singular}$ and $A_h=\dfrac{\Pi_e}{1+U_h^d\, \Pi_h^d}.$ 

\subsubsection{Non-singular part of the pairing interaction} 
\begin{figure}[]
  \centering
 \subfigure[]{\includegraphics[width =  .7\textwidth]{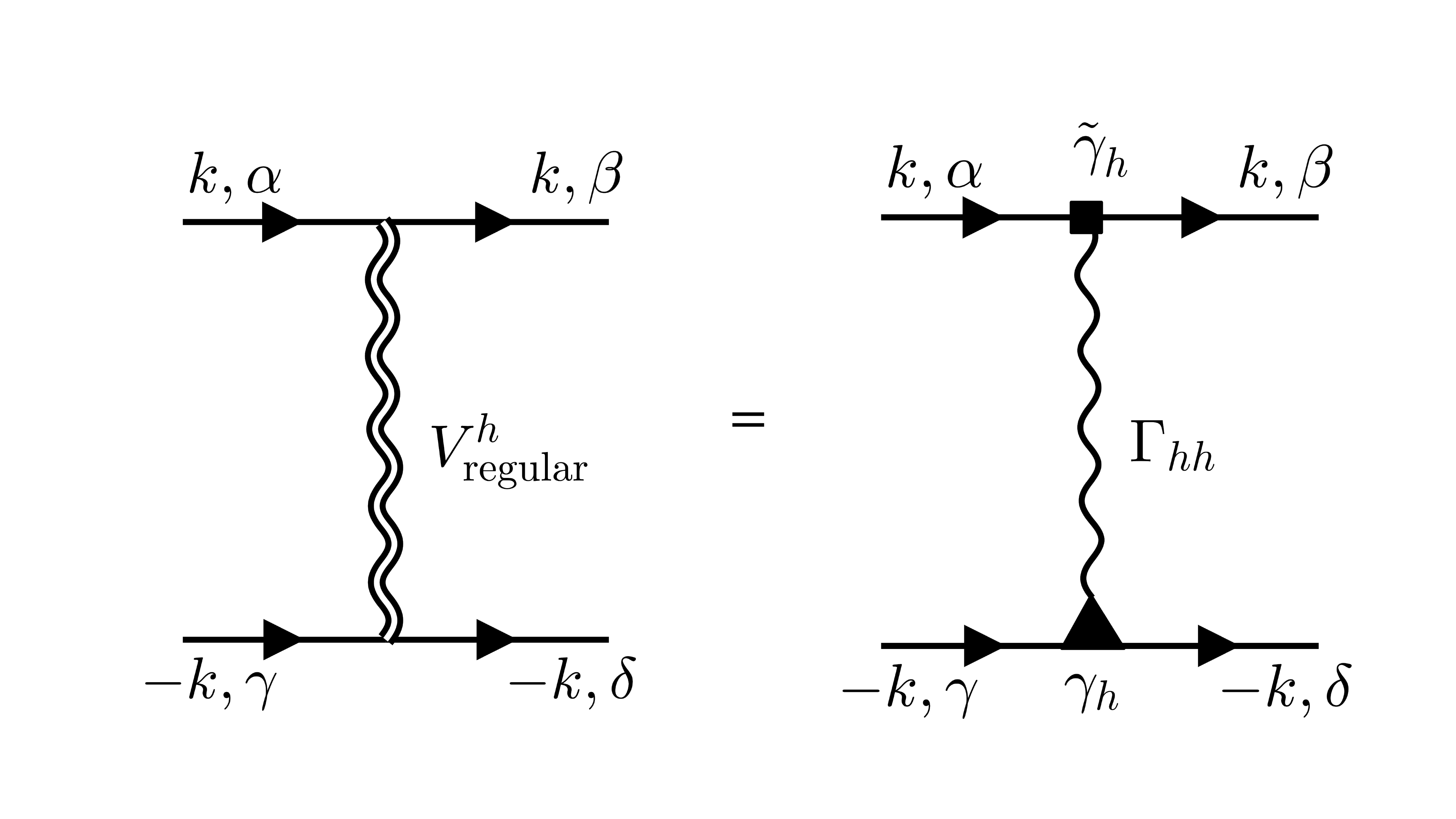}} 
 \subfigure[]{\includegraphics[width =  \textwidth]{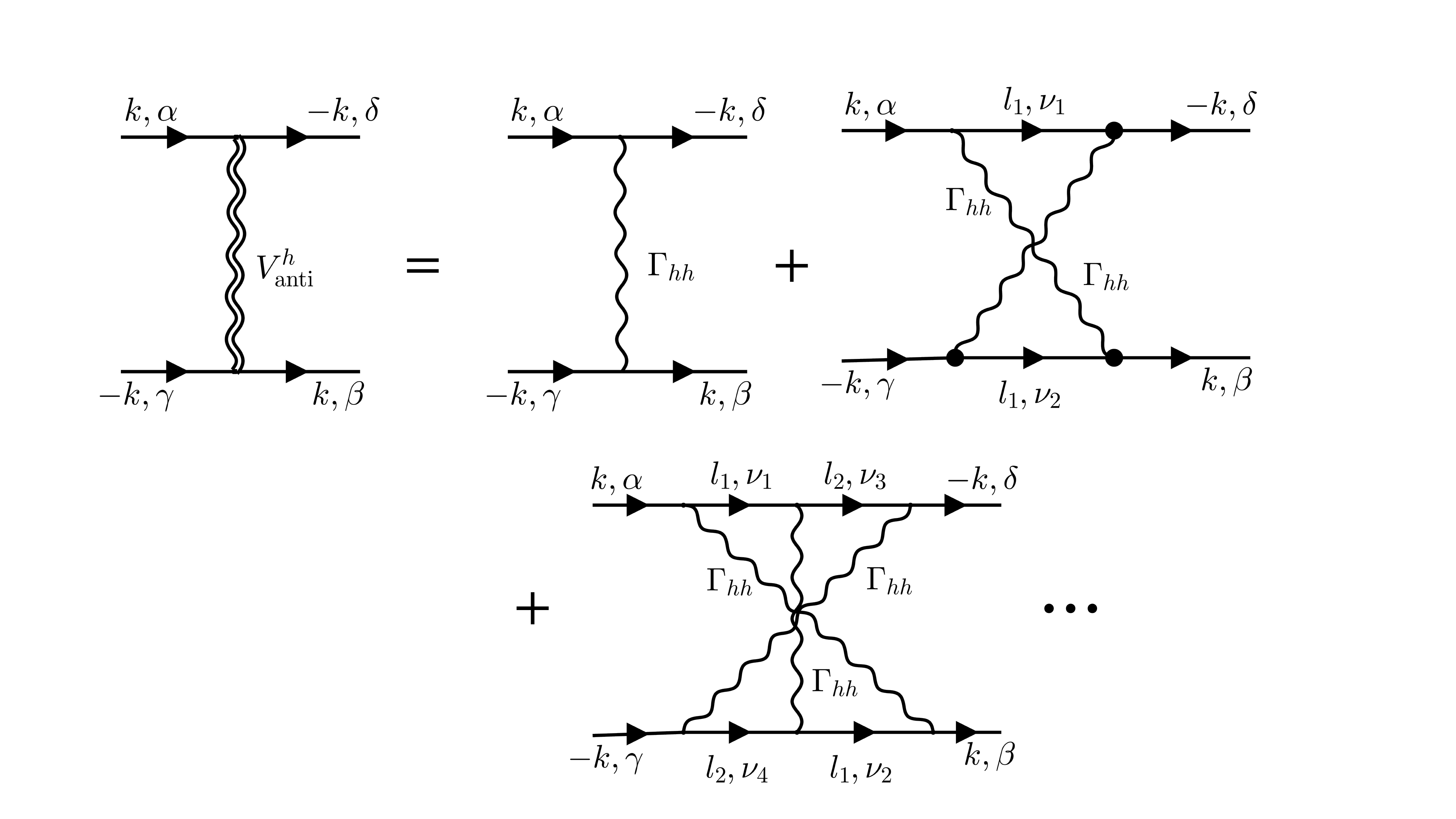}}
  \caption{ Dyson equation for the the regular part of the intra-pocket pairing interaction, $V^h_\text{regular}$ is shown in (a).The solid black line, double curvy line, single curvy line represent the hole fermion Green's function, dressed interaction and bare interaction $\Gamma_{hh}$ respectively. The black square and the triangle represent $\tilde{\gamma}_h$ and $\gamma$ defined in the main text. In (b) dyson equation for the anti-symmetric part of the pairing interaction is shown. }.
\label{non singular part}
 \end{figure}

In Sec.\ref{singular part}, we find the singular component of the dressed pairing interaction on the hole pocket. It scales with the nematic susceptibility $\chi_{\text{nem}}$ with an angular dependence of $\cos^22\theta_\bk$ and exist as a consequence of the inter-pocket density interaction $U_{he}$. On the other hand, the regular part of the dressed pairing interaction comes purely from the intra-pocket(in this case hole pocket) density interaction as we will show in this section. For the simplicity of the calculation, we keep the momentum of the external fermions same as the internal ones and argue that the regular component termed as $V^h_\text{regular}$ of the intra-pocket pairing interaction for the hole pocket is depicted in Fig.\ref{non singular part}

\begin{figure}
\includegraphics[scale=.3]{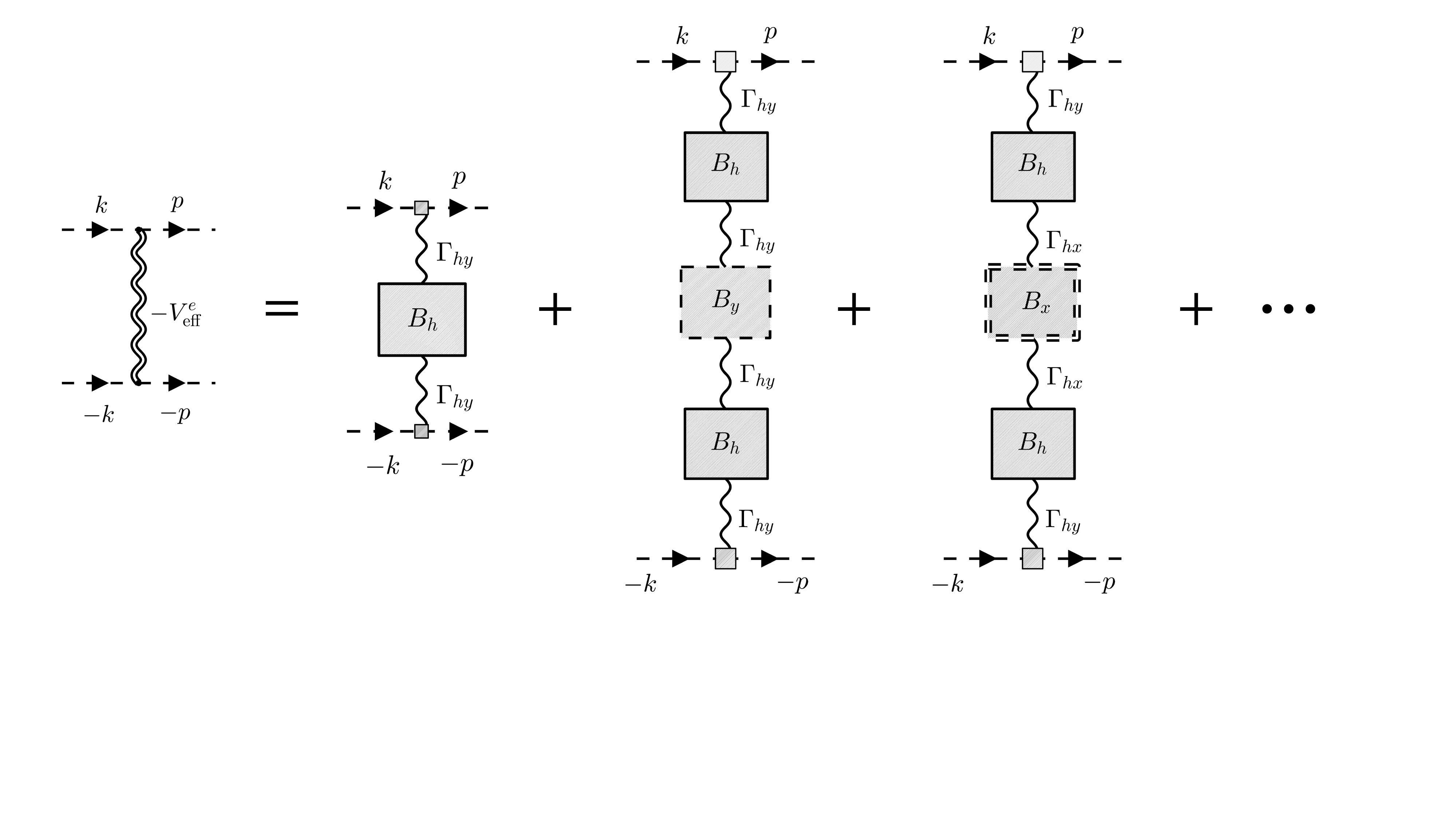}
\centering
\caption{Ladder series for the dressed pairing interaction represented by the double curvy line between the fermions with paired state $(\bk,-\bk)$ scattering to ($\bp,-\bp$) within the electron pocket. Here the shaded boxes labeled as $B_h,B_x$ and $B_y$ are defined in Fig.\ref{shaded box}. The black rectangular vertex is defined in  Fig.\ref{external vertex}. Single curvy line represents the bare interaction between the hole and electron pocket.}
\label{electron pairing}
\end{figure}

 From Fig.\ref{non singular part}a, $V^{h}_\text{regular}=\tilde{\gamma}_h.\Gamma_{hh}.\gamma_h$ where $\tilde{\gamma}_h$ and $\gamma_h$ are defined in Fig.\ref{vertex correction in bubble} and \ref{external vertex}.  A straight forward calculation using Eqs.\eqref{only gamma},\eqref{gammah tilde} gives
\begin{align}
V^{h}_\text{regular}=\dfrac{U_4+\tilde{U}_4}{2} \chi^s_{\text{spin}}\, \chi^s_{\text{den}}
+\dfrac{U_4-\tilde{U}_4}{2}\chi^d_{\text{spin}}\, \chi^d_{\text{den}}\, \cos^22\theta_\bk+\dfrac{\bar{U}_4+\tilde{\tilde{U}}_4}{2} \chi^{\text{mix}}_{\text{spin}}\, \chi^{\text{mix}}_{\text{den}}\, \sin^22\theta_\bk,
\label{direct 1}
\end{align}
with the spin component $\delta_{\alpha\beta}\delta_{\gamma \delta}$. Here, $\chi^s_{\text{spin}},\chi^d_{\text{spin}}$ and $\chi^\text{mix}_{\text{spin}}$ are the spin susceptibilities with the form factor $1, \cos2\theta$ and $\sin2\theta$ respectively, while $\chi^s_{\text{den}},\chi^d_{\text{den}}$ and $\chi^\text{mix}_{\text{den}}$ are the density susceptibilities with the form factor $1, \cos2\theta$ and $\sin2\theta$ respectively for a one band model(in our case it is the hole pocket). They are defined as
\begin{align}
\chi^s_{\text{spin}}&=\dfrac{1}{1-\dfrac{U_4+\tilde{\tilde{U}}_4}{2}\, \Pi_h^s,}\hspace{2 cm} \chi^s_{\text{den}}=\dfrac{1}{1+\dfrac{U_4+2\tilde{U}_4-\tilde{\tilde{U}}_4}{2}\, \Pi_h^s},\\
\chi^d_{\text{spin}}&=\dfrac{1}{1-\dfrac{U_4-\tilde{\tilde{U}}_4}{2}\, \Pi_h^s,}\hspace{2 cm} \chi^d_{\text{den}}=\dfrac{1}{1+\dfrac{U_4-2\tilde{U}_4\tilde{\tilde{U}}_4}{2}\, \Pi_h^s},\\
\chi^\text{mix}_{\text{spin}}&=\dfrac{1}{1-2 \, \Pi_m \, (\bar{U}_4+\tilde{U}_4)},\hspace{2 cm} \chi^\text{mix}_{\text{den}}=\dfrac{1}{1-2 \, \Pi_m \, (\bar{U}_4+2\tilde{\tilde{U}}_4-\tilde{U}_4)}.
\end{align}
To complete our calculation of the dressed pairing interaction, we also compute the  fully anti-symmetrized pairing interaction. The anti-symmetric part of the fully anti-symmetrized pairing interaction describes scattering between fermion pair states $(\bk,\alpha;-\bk,\gamma)\rightarrow(-\bk,\delta;\bk,\beta)$. We depict the relevant diagrams contributing to $V^{\text{anti}}_{hh}$ in Fig.\ref{non singular part}b. Using Eq.~\eqref{Vhh} one finds
\begin{align}
V^\text{anti}_{hh}=\dfrac{U_4+\bar{U}_4}{2}\dfrac{1}{1-\dfrac{U_4+\bar{U}_4}{2}\, \Pi_h^s}+\dfrac{U_4-\bar{U}_4}{2}\dfrac{\cos^22\theta_\bk}{1-\dfrac{U_4-\bar{U}_4}{2}\, \Pi_h^d}+\dfrac{\tilde{U}_4+\tilde{\tilde{U}}_4}{2}\dfrac{\sin^22\theta_\bk}{1-\dfrac{\tilde{U}_4+\tilde{\tilde{U}}_4}{2}\, \Pi_h^d}
\label{anti part}
\end{align}
with the spin component $\delta_{\alpha\delta}\delta_{\gamma \beta}$. To understand what Eqs.~(\ref{direct 1}-\ref{anti part}) produce, we combine both the terms and write the non-singular part of the fully anti-symmetrized pairing vertex $\Gamma^h_\text{regular}$ using the bare Hubbard-Hund interaction, Eq.~(\ref{bare interaction 1}-\ref{bare interaction 4})as
\begin{align}
\label{regular part}
\Gamma^h_\text{regular}=& \delta_{\alpha\beta}\delta_{\gamma \delta}\Bigg[ \dfrac{U-J}{\Big(1+\dfrac{3U-5J}{2}\, \Pi_h^s\Big)\Big(1-\dfrac{U+J}{2}\, \Pi_h^s\Big)}+\dfrac{J}{\Big(1+\dfrac{5J-U}{2}\, \Pi_h^d\Big)\Big(1-\dfrac{U-J}{2}\, \Pi_h^d\Big)}\Bigg]-\nn  &\delta_{\alpha\delta}\delta_{\gamma \beta} \Bigg[ \dfrac{U+J}{2\Big(1-\dfrac{U+J}{2}\, \Pi_h^s\Big)}+\dfrac{U-J}{2\Big(1-\dfrac{U-J}{2}\, \Pi_h^d\Big)}\Bigg]\\=&\Gamma_c \, \delta_{\alpha\beta}\delta_{\gamma \delta}+\Gamma_s\, \hat{\sigma}_{\alpha\beta}.\hat{\sigma}_{\gamma \delta}.
\label{regular piece}
\end{align}
Here, $\Gamma_c$ and $\Gamma_s$  are the spin and charge components of the antisymmetrized vertex with the form
\begin{align}
\Gamma_c&=\dfrac{3U-5J}{4}\dfrac{1}{1+\dfrac{3U-5J}{2}\,\Pi_h^s}+\dfrac{5J-U}{4}\dfrac{1}{1+\dfrac{5J-U}{2}\,\Pi_h^d}\\
\Gamma_s&=-\dfrac{U+J}{4}\dfrac{1}{1-\dfrac{U+J}{2}\, \Pi_h^s}-\dfrac{U-J}{4}\dfrac{1}{1-\dfrac{U-J}{2}\, \Pi_h^d}.
\end{align}
From Eq.~\eqref{regular part}, one finds that the regular part does not scale with the nematic instability $\chi_{\text{nem}}$ and is angle independent unlike the singular part $V^h_{\text{singular}}$ which has an angular dependence as $\cos^22\theta_\bk$. We calculate the effective intra-pocket interaction for the electron pockets depicted in Fig.\ref{electron pairing} in the similar way and find
\begin{align}
V_{\text{eff}}^e(\bk,\bp)=-A_e (U_1-\bar{U}_1)^2 \chi_{\text{nem}}(\bk-\bp),
\label{electron pairing interaction}
\end{align}
where $A_e=\dfrac{1}{2}\dfrac{\Pi_h^d}{1+U_e^d\, \Pi_e}$. Because of the extra factor of $1/2$ and $\Pi_h^d<\Pi_e$, $A_h>A_e.$ 
\subsection{Inter-pocket pairing interaction}
We argue that the inter-pocket pairing interaction a.k.a pair hopping interaction does not get affected by the strong nematic fluctuation. Pair hopping interaction is a large momentum($Q$) transfer interaction with $Q=(0,\pi)$ or $(\pi,0)$. On the other hand, nematic fluctuations are peaked near zero momentum and can't influence pair hopping  term much.   
\section{Solving the Superconducting Gap Equation at the QCP: $\delta=0$ }
\label{gap solution}
\begin{figure}[]
  \centering
  \includegraphics[width =.8\textwidth]{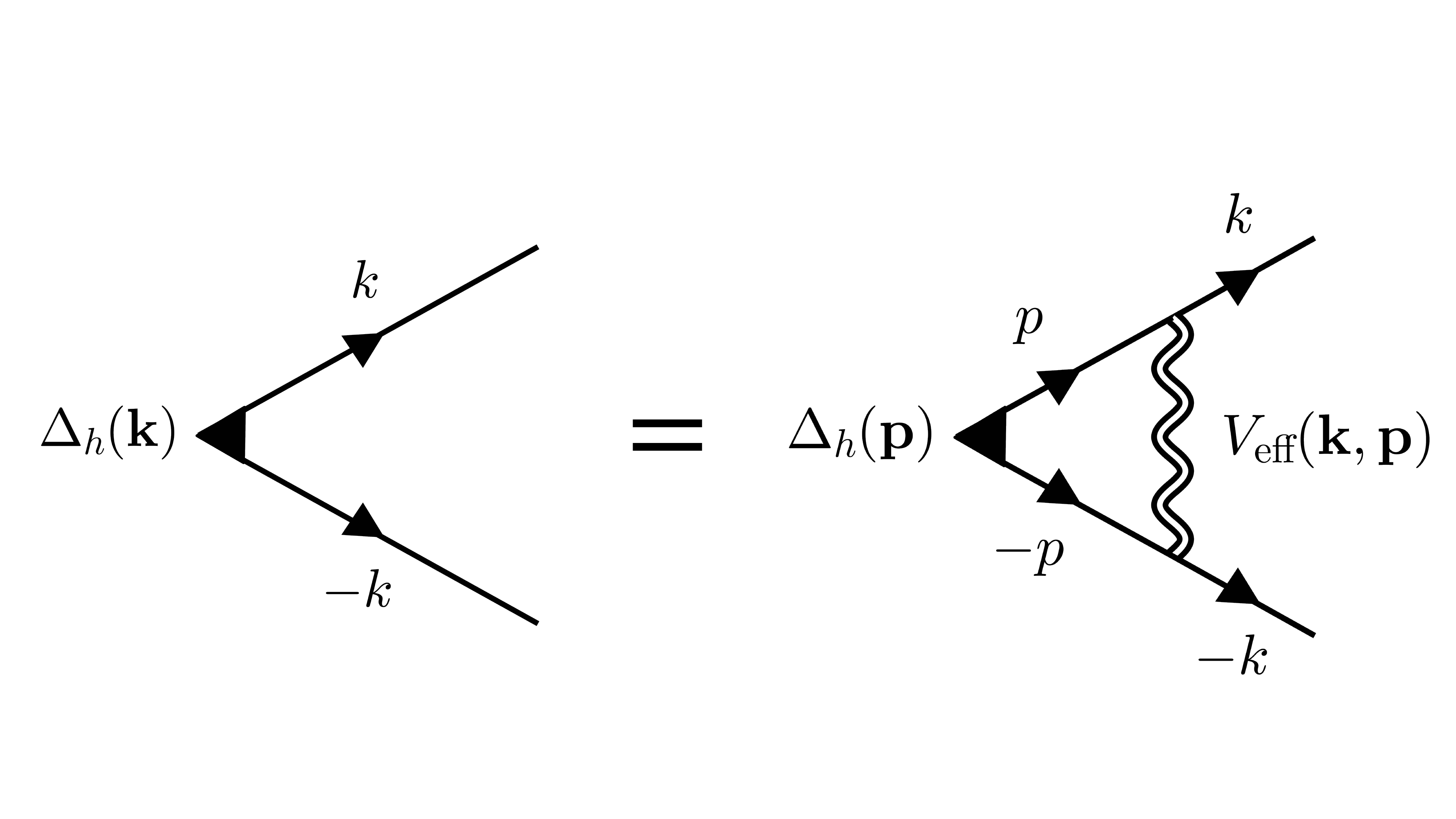} 
  \caption{Feynman diagram for the non-linear gap equation for the hole pocket. $\Delta_h$ represent the gap function. The double wavy line is the effective pairing interaction defined in Eq.~\eqref{pairing interaction} for the hole pocket fermions between momentum $k$ and $p$. }
\label{superconducting vertex}
 \end{figure}

\begin{figure}[]
  \centering
 \subfigure[]{\includegraphics[width = 0.3 \textwidth]{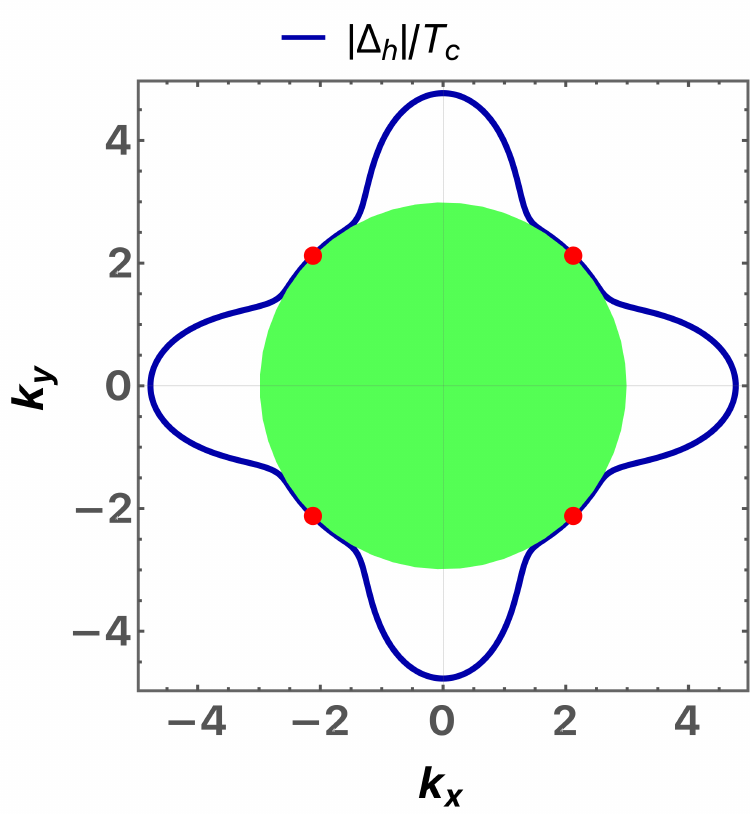}} 
 \subfigure[]{\includegraphics[width = 0.3 \textwidth]{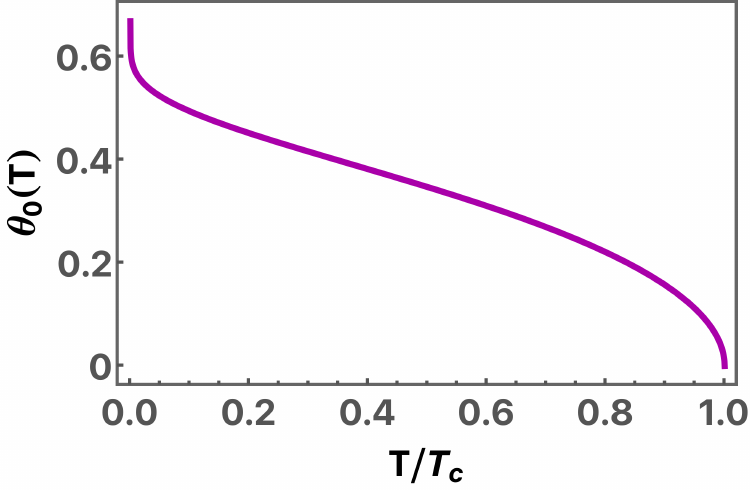}} 
 \subfigure[]{\includegraphics[width = 0.3 \textwidth]{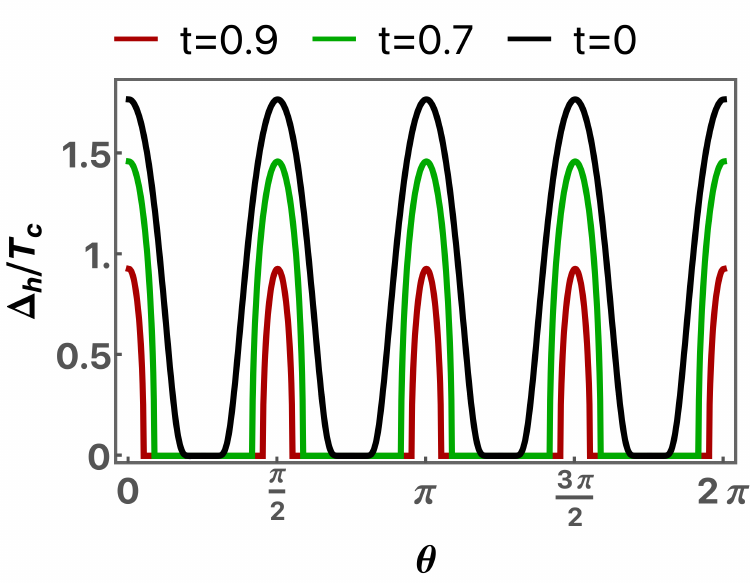}} 
  \caption{(a) We plot the absolute value of the SC gap scaled by the transition temperature $T_c$, $\Delta_h/T_c$(blue) on the hole pocket at zero temperature as a function of the angle on the Fermi surface(green disk) with radius $3$ unit. The gap is exponentially small near the cold spots(red dots) while maximum along the $k_x$ and $k_y$ axis. In (b) we plot the angular variation of the gap, $|\Delta_h|/T_c$ on the hole pocket for three different reduced temperatures, $t=0.9,0.7$ and $0$ below the transition point $T_c$. At finite temperature, the gap vanishes on the four patches of the Fermi surface arcs(flat segments of the gap). In (c) we plot the width of the gap, $\theta_0(T)$ as a function of the reduced temperature $t=T/T_c$. We keep the coupling strength $g=1$ here.}
\label{gap plots}
 \end{figure}

 In Section.\ref{section $4$} we find that near the nematic instability the intra-pocket pairing interaction becomes attractive and divergent as it scales with the nematic susceptibility $\chi_{\text{nem}}$ compared to the finite repulsive inter-pocket interaction. As a result, we can ignore the effect of the inter-pocket pairing interaction and study the superconducting gap equation for individual pockets  led by the intra-pocket pairing interaction. The pairing interaction within the hole and the electron pockets presented in Eqs.~(\ref{hole pairing interaction},\ref{electron pairing interaction}) differ by the prefactor $A_{h,e}$. Since $A_h> A_e$, pairing interaction is larger on the hole pocket and superconductivity will first develop on it. In our further analysis, we keep the pairing interaction(Eq.~\eqref{hole pairing interaction}) static, predominantly between fermions near the Fermi surface and used the regular Ornstein-Zernike form for the $\chi_{\text{nem}}(\bq)\left(= \chi_0/(\xi^{-2}+\bq^2)\right)$ and find
 \begin{align}
 V_{\text{eff}}(\bk,\bp)&=-A_h\, \chi_0 (U_1-\bar{U}_1)^2 \dfrac{\cos^2(\theta_{\bk}+\theta_{\bp})}{\xi^{-2}+(\bk-\bp)^2}\\
 &=-\bar{g} \dfrac{\cos^2(\theta_{\bk}+\theta_{\bp})}{\delta^2+4 \sin^2\Bigr(\dfrac{\theta_{\bk}-\theta_{\bp}}{2}\Bigr)}.
 \label{pairing interaction}
\end{align} Here $\chi_0$ is the bare susceptibility, $\delta=1/k_f \xi$ is the inverse of the nematic correlation length $\xi$, $\bar{g}=A_h\, \dfrac{(U_1-\bar{U}_1)^2}{k_F^2}$ is the effective coupling constant and $k_F$ is the Fermi momentum. With the pairing interaction listed in Eq.~\eqref{pairing interaction} we write the non-linear gap equation $\Delta_h(\bk)$ on hole pocket depicted in Fig.\ref{superconducting vertex} as
 \begin{align}
 \Delta_h(\bk)=\bar{g}\, \bigintsss \dfrac{d^2\bp}{(2\pi)^2} \dfrac{\Delta_h(\bp) \, \cos^2\Bigr(\theta_{\bk}+\theta_\bp\Bigr)}{\delta^2+4 \sin^2\Bigr(\theta_{\bk}-\theta_{\bp}/2 \Bigr)}\dfrac{\tanh\Bigr(\beta E_h(\bp)/2\Bigr)}{E_h(\bp)},
 \label{int1}
 \end{align}
 with $\beta=1/T$ and the excitation energy $E_h(\bp=\sqrt{\xi_h(\bp)^2+\Delta_h(\bp)^2})$. We shift the momentum $\bp=\bk+\tilde{\bp}$ and rewrite the integration in Eq.~\eqref{int1} as
 \begin{align}
  \Delta_h(\bk)=\bar{g}\, \bigintsss \dfrac{d^2\tilde{\bp}}{(2\pi)^2} \dfrac{\Delta_h(\bk+\tilde{\bp}) \, \cos^2\Bigr(2 \theta_{\bk}+\theta_{\tilde{\bp}}\Bigr)}{\delta^2+4 \sin^2\Bigr(\theta_{\tilde{\bp}}/2 \Bigr)}\dfrac{\tanh\Bigr(\beta E_h(\bk+\tilde{\bp})/2\Bigr)}{E_h(\bk+\tilde{\bp})}.
 \label{int2}
 \end{align}
 Eq.~\eqref{int2} is analytically tractable in the limit $\delta\rightarrow 0$. In this limit, the integration is peaked near $\theta_{\tilde{\bp}}=0$. As a result, we can ignore the $\tilde{\bp}$ dependence of the gap and the cosine factor in Eq.~\eqref{int2} and transform it from an integral equation to an algebraic equation
 \begin{align}
  \Delta_h(\bk)&=\bar{g} \, N_0 \,\Delta_h(\bk)\, \cos^2 2\theta_{\bk} \bigintsss_0^{2\pi} \dfrac{d \theta_{\tilde{\bp}}}{2\pi} \dfrac{1}{\delta^2+4 \sin^2\Bigr(\theta_{\tilde{\bp}}/2 \Bigr)}\bigintsss_0^\Lambda dx  \dfrac{\tanh\Bigr(\dfrac{\sqrt{x^2+|\Delta_h(\bk)|^2}}{2 T}\Bigr)}{\sqrt{x^2+\Delta_h(\bk)^2}},
 \label{int3}
\end{align}  
where $N_0$ is the density of states and $\Lambda$ is the upper pairing cutoff. Performing the angular integration over $\theta_{\tilde{\bp}}$ and defining the effective coupling $g=\dfrac{N_0\, \bar{g}}{2\, \delta}$, Eq.~\eqref{int3} reduces to
\begin{align}
1=g \, \cos^2 2\theta_{\bk} \bigintsss_0^\Lambda dx  \dfrac{\tanh\Bigr(\dfrac{\sqrt{x^2+|\Delta_h(\bk)|^2}}{2 T}\Bigr)}{\sqrt{x^2+\Delta_h(\bk)^2}}.
\label{int4}
\end{align} 
Since Eq.~\eqref{int4} depends only on the modulus of the gap $\Delta_h$, its solution does not provide any information about the phase of that gap:  $\Delta_h=|\Delta_h(\bk)| e^{i \phi(\bk)}$. Solving the linearized equation by putting $\Delta_h=0$ in Eq.~\eqref{int4}, we find an angle dependent critical temperature $\bar{T}_c(\theta_{\bk})=1.13 \, \Lambda \exp(-1/g \cos^22\theta_\bk)$. The largest critical temperature happens when $\cos^22\theta_\bk=1$ and the superconducting gap will first develop at these discrete number of points on the Fermi surface and remain zero everywhere. The set of these points is $W=\{0,\pi/2,\pi,3\pi/2\}$ and the true transition temperature of the system is  $T_c=1.1.3 \Lambda \exp(-1/g)$. At any lower temperature $T<T_c$, the gap develops around these $4$ special points($\in W$) and remains non-zero upto $\theta_\bk< |\theta_0(T)-\theta_W|$ where $\theta_0(T)$ is the solution of Eq.~\eqref{int4} with $\Delta_h=0$. \begin{align}
\label{theta1}
&\dfrac{1}{g \, \cos^2 2\theta_0(T)}=\bigintsss_0^\Lambda dx  \dfrac{\tanh\Bigr(x/2 T\Bigr)}{x}=\log\dfrac{1.13 \, \Lambda}{T}=\dfrac{1}{g}+\log\dfrac{T_c}{T}\\
&\rightarrow \theta_0(T)=\dfrac{1}{2}\arctan\sqrt{g\log\dfrac{T_c}{T}}.
\label{theta0}
\end{align} We plot $\theta_0(T)$ in Fig.\ref{gap plots}b as a function of the reduced temperature $t=T/T_c$. For $\theta_\bk>|\theta_0(T)-\theta_W|$, the gap vanishes and the Fermi surface remains intact with the width $\Delta \theta(T)=\pi/2-2\, \theta_0(T)$ where $\theta_0(T)$ is defined in Eq.~\eqref{theta0}. At zero temperature, $\Delta \theta$ reduces to zero letting the gap open everywhere except on the cold spots(red dots in Fig.\ref{gap plots}a): $\theta_c=\pi/4+n\, \pi/2$ where $n$ is an integer. In Fig.\ref{gap plots}c, we plot the angular variation of the gap magnitude for a set of temperatures. At $T=0$, the angular variation of the gap turns out to be
\begin{align}
|\dfrac{\Delta_h(\bk)|}{T_c}\Biggr|_{T=0}\approx 1.76 \exp{-\, \tan^2 2\theta_\bk/g}.
\label{gap at 0}
\end{align} Unlike the conventional $d-$ wave superconducting gap($\propto \cos 2\theta_\bk$), where near the cold spot the gap is linearly proportional to the angular deviation($\propto (\theta-\theta_c)$), in this case the gap is exponentially suppressed near the cold spots and behaves as $\exp{-\dfrac{1}{4\, g(\theta-\theta_c)^2}}$. 
    
\section{Calculation of the Specific Heat $C_v$ at $\delta=0$}
\begin{figure}[]
  \centering
 \subfigure[]{\includegraphics[width = 0.45 \textwidth]{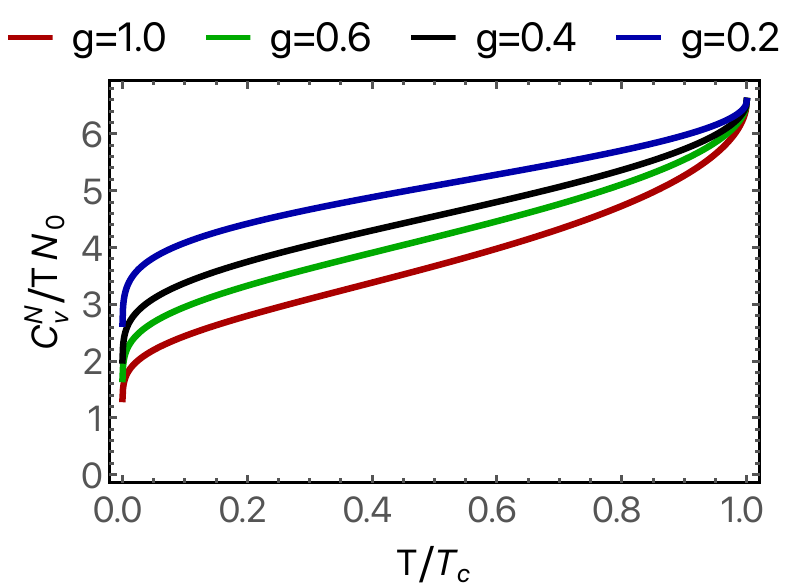}} \hspace{1 cm}
 \subfigure[]{\includegraphics[width = 0.47 \textwidth]{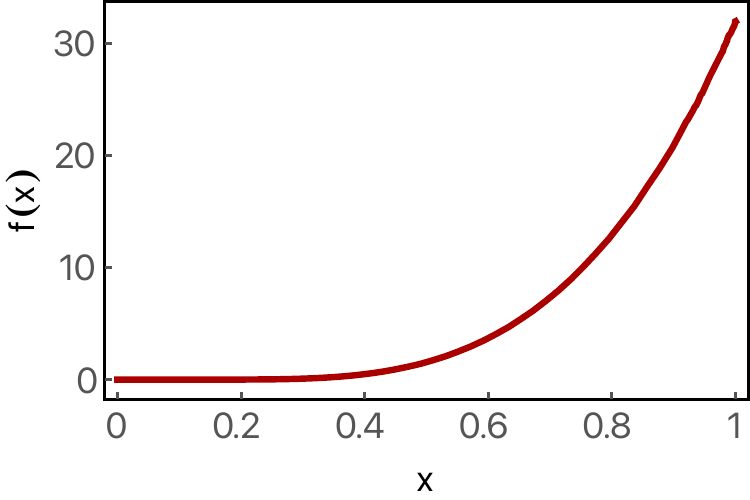}} 
 \subfigure[]{\includegraphics[width = 0.45 \textwidth]{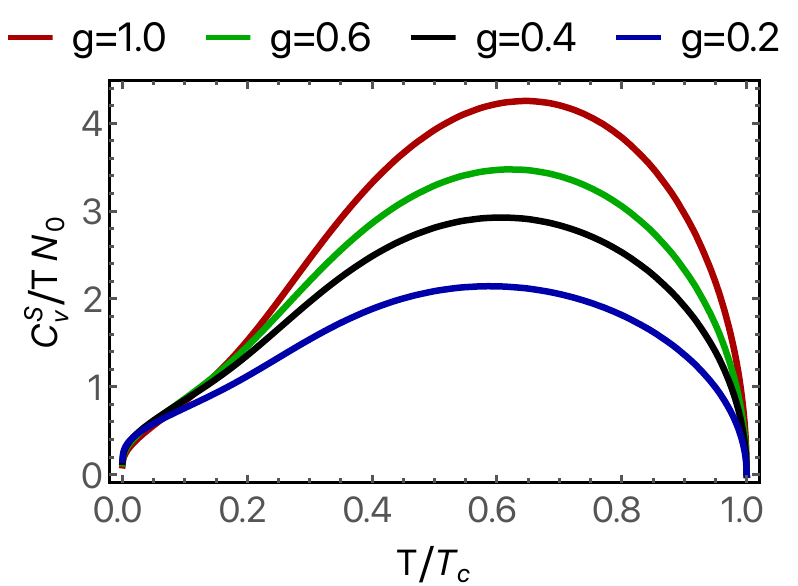}} \hspace{1 cm}
 \subfigure[]{\includegraphics[width = 0.47 \textwidth]{cvplot}} 
  \caption{We plot the temperature variation of the metallic($C_v^N$) and superconducting component($C_v^S$) of the specific heat coefficient defined in Eq.~\eqref{normal cv} and \eqref{sc cv} for a range of values of coupling constant $g$ in (a) and (c). In (b), we plot the scaling function $f(x)$ defined in Eq.~\eqref{scaling f} as a function of its argument $x$. In (d), we add the contribution from part (a) and (c) and plot the total specific heat coefficient as a function of the reduced temperature $T/T_c$ where $T_c$ is the critical temperature for $g=1$ in the main frame and $g=0.6,0.4$ and $0.2$ in the inset.}
\label{specific heat}
 \end{figure}

 The specific heat $C_V(T)$ for the hole band is given by the expression
\begin{align}
C_v(T)&=\dfrac{2}{T}\bigintss \dfrac{d^2 \bk}{(2\pi)^2}\Bigr(-\dfrac{\partial\,  n_\bk}{\partial E_h(\bk)}\Bigr) \left[E_h(\bk)^2-\dfrac{T}{2} \dfrac{\partial |\Delta_h(\bk)|^2}{\partial T}\right]
\label{sp heat 1}
\end{align} 
where $n_\bk$ is the Fermi function $1/[\exp(\beta E_h)+1]$ and $E_h(\bk)=\sqrt{\xi_h(\bk)^2+|\Delta_h(\bk)|^2}$ is the excitation energy. Since $\dfrac{\partial \, n_\bk}{\partial\, E_h(\bk)}=-\dfrac{1}{T}\sech^2\bigr(\beta \, E_h(\bk)/2\bigr) $ is peaked near the Fermi surface at low temperature, we convert the momentum integration into the energy and angle variables and approximate the energy integral by keeping the density of state at the Fermi surface which gives
\begin{align}
 C_v(T)=\dfrac{N_0}{2\, T^2} \bigintss_{-\infty}^\infty dx \bigintss_0^{2\pi} \dfrac{d\theta}{2\pi} \sech^2\Bigr(\dfrac{\sqrt{x^2+|\Delta_h(\theta)|^2}}{2\,T}\Bigr) \Bigr[x^2+|\Delta_h(\theta)|^2-\dfrac{T}{2} \dfrac{\partial |\Delta_h(\theta)|^2}{\partial T}\Bigr].
\label{scaled sp heat 1}
\end{align} Here, $N_0$ is the density of states at the fermi surface. In Sec.\ref{gap solution}, we solve the non-linear gap equation, Eq.~\eqref{int4} and use the solution to compute the specific heat in this section. Since the gap function $\Delta_h(\theta)$ vanishes for $\theta\geq \theta_0(T)$ at the temperature $T$ for $\theta \in (0,\pi/4)$(See Fig.\ref{gap plots}a), we can split the specific heat into two parts, one for the metallic part called $C_v^N$ for which the gap vanishes and another for the superconducting part called $C_v^S$ for which the gap exist: $C_v(T)=C_v^N(T)+ C_v^S(T)$ defined as
\begin{align}
\label{sc cv}
C_v^S(T)&=4\dfrac{N_0}{ T^2}\bigintss_0^{\theta_0(T)} \dfrac{d\theta}{2\pi}  \bigintss_{-\infty}^\infty dx \sech^2\Bigr(\dfrac{\sqrt{x^2+|\Delta_h(\theta)|^2}}{2\,T}\Bigr) \Bigr[x^2+|\Delta_h(\theta)|^2-\dfrac{T}{2} \dfrac{\partial |\Delta_h(\theta)|^2}{\partial T}\Bigr],\\
C_v^N(T)&=4\dfrac{N_0}{ T^2} \bigintss_{\theta_0(T)}^{\pi/4} \dfrac{d\theta}{2\pi}\bigintss_{-\infty}^\infty dx \sech^2\Bigr( \dfrac{x}{2\,T}\Bigr) \,  x^2
\label{normal cv}
\end{align}
  A straightforward calculation gives the normal part of the specific heat
\begin{align}
C_v^N(T)=\dfrac{8\,N_0\, \pi\, T}{3} \Big[\dfrac{\pi}{4}-\theta_0(T)\Bigr].
\end{align}
We plot the specific heat coeffcient for the metalic part, $C_v^N/T$ as a function of the reduced temperature $t=T/T_c$ in Fig.\ref{specific heat}a. At $t=1$, $C_v^N/T$ reduces to the normal metallic contribution $2 \pi^2\,N_0/3$. On the other hand, at low temperature, it behaves as \begin{align}
\dfrac{C_v^N(T)}{T}\Biggr|_{T \rightarrow 0}= \dfrac{4 \,N_0 \,\pi}{3\, \sqrt{|g \, \log t|}}+0((\log\,t)^{-3/2}).
\end{align} To compute the superconducting part of the specific heat, we scale the temperature, gap and energy integration variable by the critical temperature $\bar{T}_c(\theta)$ inside the angular integration of Eq.~\eqref{sc cv} and write 
\begin{align}
C_v^S(T)=4\dfrac{N_0\, T_c}{ t^2}\bigintss_0^{\theta_0(T)} \dfrac{d\theta}{2\pi} \Biggr(\dfrac{\bar{T}_c(\theta)}{T_c}\Biggr)^3  \bigintss_{-\infty}^\infty d\tilde{x} \sech^2\Bigr(\dfrac{\sqrt{\tilde{x}^2+|\tilde{\Delta}_h(\theta)|^2}}{2 t_\theta}\Bigr) \Bigr[\tilde{x}^2+|\tilde{\Delta}_h(\theta)|^2-\dfrac{t_\theta}{2} \dfrac{\partial |\tilde{\Delta}_h(\theta)|^2}{\partial t_\theta}\Bigr],
\label{sc cv 2}
\end{align}
where  $t_\theta=T/\bar{T}_c(\theta),\,  \tilde{\Delta}_h(\theta)= \Delta_h(\theta)/T_c(\theta)$ and  $\tilde{x}=x/T_c(\theta)$. We define the integration under $\tilde{x}$ variable in Eq.~\eqref{sc cv 2} as
\begin{align}
f(t_\theta)=\bigintss_{-\infty}^\infty d\tilde{x} \sech^2\Bigr(\dfrac{\sqrt{\tilde{x}^2+|\tilde{\Delta}_h(\theta)|^2}}{2 t_\theta}\Bigr) \Bigr[\tilde{x}^2+|\tilde{\Delta}_h(\theta)|^2-\dfrac{t_\theta}{2} \dfrac{\partial |\tilde{\Delta}_h(\theta)|^2}{\partial t_\theta}\Bigr],
\label{scaling f}
\end{align}
such that \begin{align}
C_v^S(T)&=4\dfrac{N_0\, T_c}{ t^2}\bigintss_0^{\theta_0(T)} \dfrac{d\theta}{2\pi}\,  \Biggr(\dfrac{\bar{T}_c(\theta)}{T_c}\Biggr)^3 f(t_\theta)\\
&=4\dfrac{N_0\, T_c}{ t^2}\bigintss_0^{\theta_0(T)} \dfrac{d\theta}{2\pi}\,  e^{-3\tan^2 2\theta/g} f( e^{\tan^2 2\theta/g}\, t)\\
&=N_0 \, g\, T_c\, \bigintss_t^1 dx \dfrac{f(x)}{x^4 \bigr(1+g \log\dfrac{x}{t}\bigr) \sqrt{g\log\dfrac{x}{t}}\bigr) }.
\label{sc cv 3}
\end{align} 

Here we have used the relations $\bar{T}_c(\theta)/T_c=\exp(-\tan^2 2\theta/g)$ and $f(t_\theta)\equiv f\biggr(t\times \dfrac{T_c}{\bar{T}_c(\theta)}\biggr)$. Eq.~\eqref{sc cv 3} has a log singularity near $x=t$ which approximates the integration to
\begin{align}
\dfrac{C_v^S(T)}{T}= N_0 \, g\, \dfrac{f(t)}{t^3} \bigintss_1^{1/t} \dfrac{dy}{\sqrt{g\, \log y}}=\sqrt{\pi}\, N_0 \, \sqrt{g}\, \dfrac{f(t)}{t^3} \text{Erfi}[\sqrt{|\log t|}],
\end{align} where $\text{Erfi}(x)$ is the imaginary error function. We obtain the scaling function $f(x)$ numerically and plot in Fig.\ref{specific heat}b. We numerically compute Eq.~\eqref{sc cv 3} and present our result for the specific heat coefficient of the superconducting part, $C_v^S/T$ in Fig.\ref{specific heat}c.   
 
In Fig.\ref{specific heat}d, we plot the total specific heat coefficient, $C_v/T$ as a function of the temperature for a range of values for the coupling constant $\bar{g}$.   The dashed black line represent the normal state result for the specific heat. We find there is no specific heat jump at the onset of the superconductivity because gap opens only at the discrete number of points on the Fermi surface and remains zero everywhere. This is captured by the fact that $C_v^S/T=0$ as $T$ goes to $T_c$. Near the critical temperature $T_c$,  $C_v/T$ first goes up, attains a maximum around $t=0.8$ and then falls  with lowering the temperature. At very low temperature, specific heat declines very rapidly as it goes as $\log(t)^{-1/2}$. This is a consequence of the existence of the exponentially suppressed gap near the cold spots, captured in Eq.~\eqref{gap at 0}. 

\section{Effect of the Pair-Hopping Interaction at $\delta=0$}
So far in our analysis we have neglected the effect of the inter-pocket pairing interaction a.k.a pair hopping defined in Eq.~\eqref{interaction in band basis}. In this section, we study if the inter-pocket pairing interaction can break the degeneracy between $s-,d-$ and $p-$ wave channel at $\delta=0$ limit. For the simplicity of the calculation, we have neglected the pairing interaction within the electron pockets and keep the density of states $N_0$ same for all the pockets. In the presence of the pair-hopping interaction, the non-linear gap equation for $\Delta_h, \Delta_x$ and $\Delta_y$ become 
\begin{align}
\label{NGE hole}
\Delta_h(\theta_\bk)&=g\, \Delta_h(\theta_\bk) \cos^22\theta_\bk  \bigintss_0^\Lambda d\xi_h  \dfrac{\tanh\Bigr(\dfrac{\sqrt{\xi_h^2+|\Delta_h(\theta_\bk)|^2}}{2 T}\Bigr)}{\sqrt{\xi_h^2+|\Delta_h(\theta_\bk)|^2}}
 -\Delta_x  \, N_0 \, \left(U_s+U_d \, \cos 2\theta_\bk \right) \bigintss_0^\Lambda d\xi_x \dfrac{\tanh\Bigr(\dfrac{\sqrt{\xi_x^2+|\Delta_x|^2}}{2 T}\Bigr)}{\sqrt{\xi_x^2+|\Delta_x|^2}}\nn &-\Delta_y \, N_0 \left(U_s-U_d \, \cos2\theta_\bk \right) \bigintss_0^\Lambda d\xi_y \dfrac{\tanh\Bigr(\dfrac{\sqrt{\xi_y^2+|\Delta_y|^2}}{2 T}\Bigr)}{\sqrt{\xi_y^2+|\Delta_y|^2}}
\end{align}
\begin{align}
\Delta_x&=-N_0 \, \bigintss_0^{2\pi} \dfrac{d\theta_\bp}{2\pi} (U_s+U_d\, \cos2\theta) \, \Delta_h(\theta_\bp) \bigintss_0^\Lambda d\xi_h  \dfrac{\tanh\Bigr(\dfrac{\sqrt{\xi_h^2+|\Delta_h(\theta_\bp)|^2}}{2 T}\Bigr)}{\sqrt{\xi_h^2+|\Delta_h(\theta_\bp)|^2}}\\
\Delta_y&=-N_0\, \bigintss_0^{2\pi} \dfrac{d\theta_\bp}{2\pi} (U_s-U_d\, \cos2\theta) \, \Delta_h(\theta_\bp) \bigintss_0^\Lambda d\xi_h  \dfrac{\tanh\Bigr(\dfrac{\sqrt{\xi_h^2+|\Delta_h(\theta_\bp)|^2}}{2 T}\Bigr)}{\sqrt{\xi_h^2+|\Delta_h(\theta_\bp)|^2}}.
\label{NGE Y}
\end{align}
 Because of the pair-hopping interaction, the gap on the hole pocket will induce uniform gap on the electron pockets which, as a feedback induces non-zero gap everywhere on the hole pocket. We break the inter-pocket pairing interaction into $s-$ and $d-$ wave components and call them $U_s$ and $U_d$ respectively. In terms of the bare interactions, $U_s=(U+J)/2$ and $U_d=(U-J)/2$. We first focus on the linearized gap equations
\begin{align}
\label{L Eq 1}
\Delta_h(\theta_\bk)&=g \, \Delta_h(\theta_\bk)\, \cos^22\theta_\bk \, L-(\Delta_x+\Delta_y)\, u_s  \, L-(\Delta_x-\Delta_y)\, u_d  \, \cos2\theta_\bk \, L ,\\
\Delta_x+\Delta_y &=-2\, L\, u_s \bigintss_0^{2\pi} \dfrac{d\theta_\bk}{2\pi} \Delta_h(\theta_\bk),\label{L eq 2}\\
\Delta_x-\Delta_y &=-2\, L\, u_d \bigintss_0^{2\pi} \dfrac{d\theta_\bk}{2\pi} \Delta_h(\theta_\bk)\, \cos2\theta_\bk,
\label{L eq 3}
\end{align}
where $L=\log \dfrac{\Lambda}{T}, u_s=N_0 \, U_s$ and $u_d=N_0 \, U_d$. We plug Eqs.~(\ref{L eq 2}-\ref{L eq 3}) into Eq.~\eqref{L Eq 1} and find the effective linearized gap equation on the hole pocket as 
\begin{align}
\Delta_h(\theta_\bk)&=g \, \Delta_h(\theta_\bk)\, \cos^22\theta_\bk \, L+2\,U^2_s \, L^2 \bigintss_0^{2\pi} \dfrac{d\theta_\bk}{2\pi} \Delta_h(\theta_\bk)+2\, U^2_d \, L^2\, \cos2\theta_\bk  \bigintss_0^{2\pi} \dfrac{d\theta_\bk}{2\pi} \Delta_h(\theta_\bk)\, \cos2\theta_\bk.
\label{linearized eq with pair hopping}
\end{align} In the absence of the pair hopping, Eq.~\eqref{linearized eq with pair hopping} reduces to Eq.~\eqref{int4} and we get back the results of Section.\ref{gap solution}. In its presence, we make the following observation from the Eq.~\eqref{linearized eq with pair hopping}. First, pair-hopping only affects the $s-$ and $d-$ wave components of the gap, not the $p-$ wave. For the $s-$ wave solution, the third term vanishes while for the $d-$ wave the second term goes away. On the other hand, for the $p$ wave solution, both the second and third term will vanish. Second, just like the isolated case, the first superconducting instability will happen at $4$ points dictated by the relation $\cos^22\theta_\bk=1$ on the Fermi surface. These are the points along the $k_x$ and $k_y$ axis on the Fermi surface. Third, the critical temperature gets modified by the pair-hopping term. We argue that because the gap opens only at discrete number of points, to compute the critical temperature one needs to substitute the integration over angle in the second and third term of Eq.~\eqref{linearized eq with pair hopping} by a summation over these $4$ points. This gives
\begin{align}
\Delta_h(\theta_i)&=g \, \Delta_h(\theta_i)\, \cos^22\theta_i \, L+2\,U^2_s \, L^2 \sum_{k}\Delta_h(\theta_k)+2\, U^2_d \, L^2\, \cos2\theta_i  \sum_{k}\Delta_h(\theta_k) \cos2\theta_W,
\label{sum equation}
\end{align}
where $\theta_k =\{0,\pi/2,\pi, 3\pi/2\}$. Writing Eq.~\eqref{sum equation} in $s-,p-$ and $d-$ wave channel gives the following conditions for the instability respectively
\begin{align}
\label{cr temp in s }
 1&=g\, L+ 8\, L^2 \, u_s^2,\\
 1&=g\, L,\\
 1&=g\, L+ 8\, L^2 \, u_d^2.\label{cr temp in d}
\end{align}
  Solving Eqs.~(\ref{cr temp in s }-\ref{cr temp in d}) we find the renormalized critical temperatures  $T^*_{0,s}$( for $s-$ wave), $T^*_{0,d}$(for $d-$ wave) and $T^*_{0,p}$(for $p-$ wave)  such that \begin{align}
T_{0,s}^*\approx\Lambda\,  e^{-1/g} e^{8 u_s^2/g^3}\geq  T_{0,d}^*\approx\Lambda \, e^{-1/g} e^{8 u_d^2/g^3}>T_{0,p}^*=\Lambda e^{-1/g}.
\end{align}
Inter-pocket pairing interaction favors $s-$ and $d-$ wave channel compared to the $p-$ wave. When $u_s=u_d$, the equality sign holds and $s-$ and $d-$ wave becomes degenerate. To realize what gap symmetry at low temperature prevails for $u_s=u_d$, we solve the non-linear gap equations listed in Eq.~(\ref{NGE hole}-\ref{NGE Y}) numerically and present our result in Fig. Our results supports an $s-$ symmetry at all low temperature down to zero within the numerical accuracy. To understand if there is a possibility of an $s+i\, d$ state with a tiny $d-$ wave component not captured by the numerical result, we perform a perturbative calculation around the numerical solution. We write the gap as \begin{align}
\Delta_h(\theta)=\Delta_h^s(\theta)+i\, \Delta_h^d(\delta), \quad \Delta_{x,y}=\Delta_{x,y}^s+i\, \Delta_{x,y}^d,
\end{align} where $\Delta_h^s, \Delta_x^s$ and $\Delta_y^s$ are the numerical solution of Eq.~(\ref{NGE hole}-\ref{NGE Y})   for the hole, $X-$ and $Y-$ pocket respectively.  $\Delta_h^d, \Delta_x^d$ and $\Delta_y^d$ are the small perturbing fields. Expanding Eq.~(\ref{NGE hole}-\ref{NGE Y}) in $\Delta^d$'s, we find the linearized equation for $\Delta^d_{h,x,y}$ in the presence of the full grown $\Delta_{h,x,y}^s$ gap as
\begin{align}
\label{perturb hole}
\Delta_h^d(\theta)&=g \, \Delta_h^d(\theta)\, \cos^22\theta\,  J\bigr(|\Delta_h^s(\theta)|,T \bigr)-\Delta_x^d \, \bar{u}\, \cos^2\theta\, J\bigr( |\Delta_x^s| ,T \bigr)-\Delta_y^d \, \bar{u}\, \sin^2\theta\, J\bigr( |\Delta_y^s|,T \bigr)\\
\Delta_x^d&=-\bar{u} \bigintss_0^{2\pi}\dfrac{d\theta}{2\pi} \cos^2\theta \, \Delta_h^d(\theta) J\bigr( |\Delta_h^s(\theta)|,T \bigr) \label{perturb x}\\
\Delta_y^d&=-\bar{u} \bigintss_0^{2\pi}\dfrac{d\theta}{2\pi} \sin^2\theta \, \Delta_h^d(\theta) J\bigr( |\Delta_h^s(\theta)|,T \bigr),
\label{perturb y}
\end{align} 
where $\bar{u}=2 u_s=2 u_d$ and the function $J$ defined as 
\begin{align}
J(|x|,T)=\int_0^{\Lambda} d\xi \dfrac{\tanh\Bigr(\dfrac{\sqrt{\xi^2+|x|^2}}{2 T}\Bigr)}{\sqrt{\xi^2+|x|^2}}.
\label{BB}
\end{align}
Combining Eqs.~(\ref{perturb x}-\ref{perturb y}) into Eq.~(\ref{perturb hole}) and doing a little bit of manipulation we write the effective linearized equation for $\Delta_h^d$ 
\begin{align}
\Delta_h^d(\theta)&=g\, \Delta_h^d(\theta)\, \cos^2 2\theta\, J(|\Delta_h^s(\theta)|,T)+\dfrac{\bar{u}^2}{2} \cos2\theta \bigintss_0^{2\pi} \dfrac{d\theta'}{2\pi}\,  \Delta_h^d(\theta')\,  J(|\Delta_h^s(\theta')|,T) \, J(|\Delta_e^s|,T),\\
&=g\, \Delta_h^d(\theta)\, \cos^2 2\theta\, J(|\Delta_h^s(\theta)|,T)+\dfrac{\bar{u}^2}{2} \cos2\theta \, \Gamma(T)
\label{self consistent 1}
\end{align} 
where $\Gamma(T)=\bigintss_0^{2\pi} \dfrac{d\theta'}{2\pi}\,  \Delta_h^d(\theta')\,  J(|\Delta_h^s(\theta')|,T) \, J(|\Delta_e^s|,T)$ and we use $|\Delta_x^s|=|\Delta_y^s|=\Delta_e^s$. We convert Eq.~\eqref{self consistent 1} into an algebraic equation using the definition of $\Gamma$ and find the self-consistent equation for the existence of the $d-$ wave component as
\begin{align}
1= J(|\Delta_e^s|) \dfrac{\bar{u}^2}{2} \bigintss_0^{2\pi} \dfrac{d\theta}{2\pi} \dfrac{ \cos^2 2\theta\,  J(|\Delta_h^s(\theta)|,T)}{1-g\,  \cos^2 2\theta \, J(|\Delta_h^s(\theta)|,T)}.
\label{condition 1}
\end{align}
To complete our proof, we revisit Eq.~(\ref{NGE hole}-\ref{NGE Y}) and write the non-linear integral equation for $\Delta_h^s$
\begin{align}
\Delta_h^s(\theta)=g\, \Delta_h^s(\theta)\, \cos^2 2\theta \,J(|\Delta_h^s(\theta)|,T)+\dfrac{\bar{u}}{2}\,  J(|\Delta_e|,T) \bigintss_0^{2\pi} \dfrac{d\theta'}{2\pi}\,  \Delta_h^s(\theta')\,  J(|\Delta_h^s(\theta')|,T).
\label{s eq 1}
\end{align}
With a little bit of manipulation, Eq.~\eqref{s eq 1} can be rearranged as
\begin{align}
1-g\, \cos^2 2\theta J(|\Delta_h^s(\theta)|,T)=\dfrac{\bar{u}}{2} \, J(|\Delta_e|,T) \, \dfrac{\bigintss_0^{2\pi} \dfrac{d\theta'}{2\pi}\,  \Delta_h^s(\theta')\,  J(|\Delta_h^s(\theta')|,T)}{\Delta_h^s(\theta)}.
\label{s eq 2}
\end{align}
 Using Eq.~\eqref{s eq 2}, we further simplify the r.h.s of Eq.\eqref{condition 1} and arrive
 \begin{align}
 J(|\Delta_e^s|) \dfrac{\bar{u}^2}{2} \bigintss_0^{2\pi} \dfrac{d\theta}{2\pi} \dfrac{ \cos^2 2\theta\,  J(|\Delta_h^s(\theta)|,T)}{1-g\,  \cos^2 2\theta \, J(|\Delta_h^s(\theta)|,T)}&=\dfrac{\bigintss_0^{2\pi}\dfrac{d\theta}{2\pi} \cos^2 2\theta \, J(|\Delta_h^s(\theta)|,T)\, \Delta_h^s(\theta) }{\bigintss_0^{2\pi}\dfrac{d\theta}{2\pi}  J(|\Delta_h^s(\theta)|,T)\, \Delta_h^s(\theta) }<1. 
 \end{align}
Because of the extra $\cos^2 2\theta$ factor in the numerator, the ratio will always be less than $1$, and Eq.~\eqref{condition 1} can never be fulfilled below the critical temperature $T_{0,s}^*$. This proves that the gap symmetry remains $s-$ wave at all temperature below the critical point. 

\section{Numerical Solution of the Gap Equation, the Possible Gap Symmetry and Specific Heat away from the QCP: $\delta \neq 0$}
\begin{figure}[H]
  \centering
  \subfigure[]{\includegraphics[width = 0.45 \textwidth]{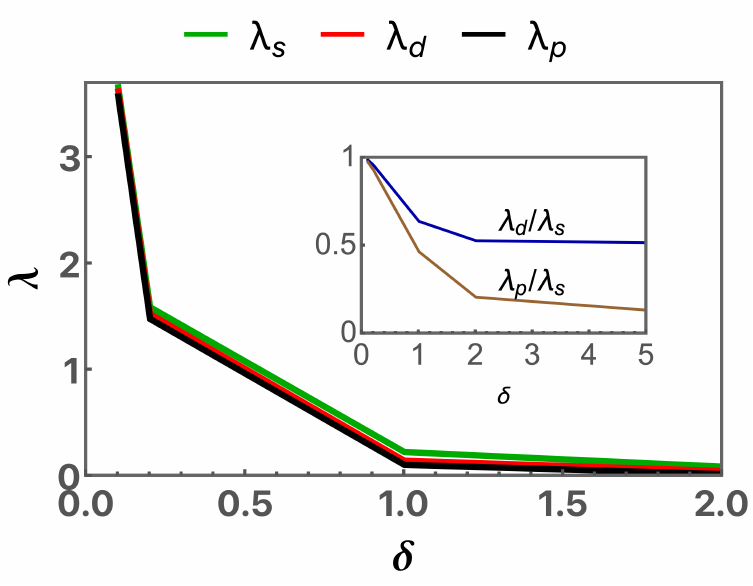}} 
 \subfigure[]{\includegraphics[width = 0.46 \textwidth]{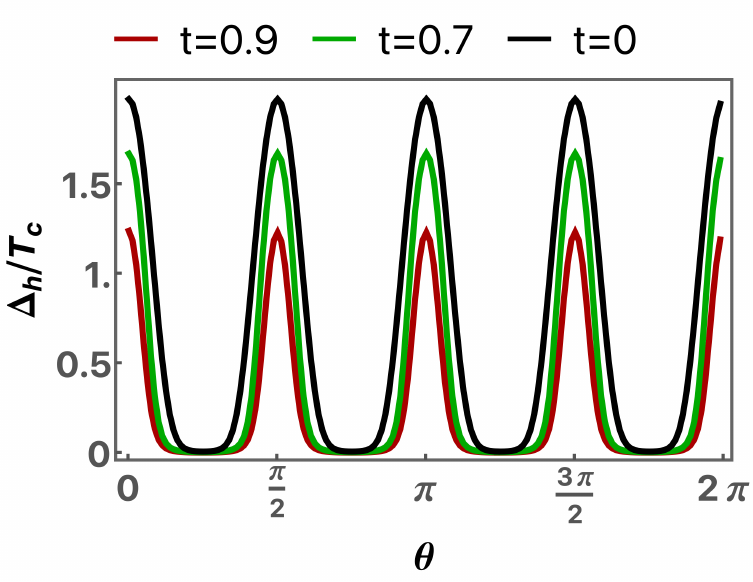}}
 \subfigure[]{\includegraphics[width = 0.47 \textwidth]{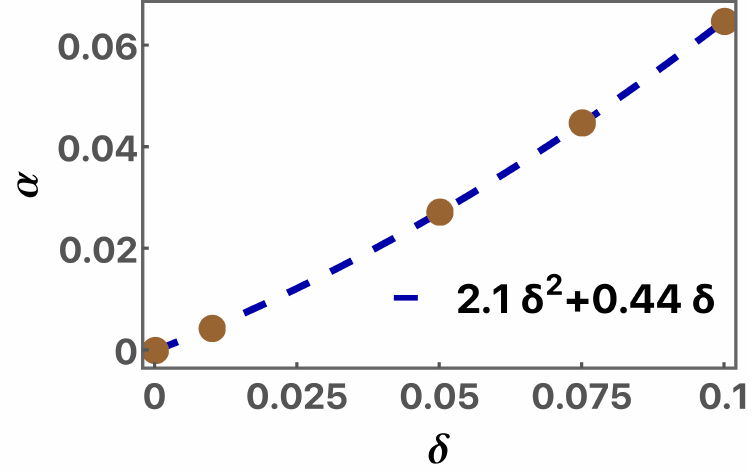}}
 \subfigure[]{\includegraphics[width = 0.45 \textwidth]{SpHeatFiniteDelta}} 
  \caption{In (a) we plot the largest eigenvalue $\lambda$ of the matrix $M$ defined in Eq.~\eqref{M linearized matrix} in different angular momentum channels labeled as $\lambda_s,\lambda_d$ and $\lambda_p$ for the $s,d$ and $p-$ wave respectively. We represent the ratio of these eigenvalues in the inset and find $\lambda_p/\lambda_s< \lambda_d/\lambda_s<1$ which indicates that $s-$ wave is the leading instability. With nematic mass parameter decreasing, the ratio of the eigenvalues becoming closer to each other and approaching unity which indicates degeneracy among different pairing channels. In (b) we plot the angular variation of the gap  $\Delta_h(\theta)$ on the hole pocket  for a set of temperatures. In (c) we plot(brown dots) the ratio of the gap along the $k_x=\pm k_y$ axis($\theta=\pi/4$) to $k_x$ axis($\theta=0$) defined as $\alpha$ in the main text as a function of $\delta$ and fit (blue line) our result upto second order in $\delta$ with the fitting parameters $\Delta(\pi/4)/\Delta(0)=2.12 \, \delta^2+0.44 \, \delta$. In (d) we plot the variation of the specific heat coefficient divided by the density of state, $N_0$ with the reduced temperature $T/T_c$ for a set of values of $\delta=0.01$(blue) and $0.1$(orange). The black dashed line represents the normal state result for the specific heat coefficient and is equal to $2\pi^2/3$. }
\label{numerical solution}
 \end{figure}

In Section.\ref{gap solution} we solve the non-linear gap equation  at the nematic quantum critical point($\delta=0$) by transforming it from an integral equation(Eq.~\eqref{int1}) to an algebraic equation(Eq.~\eqref{int4}). Even though this is right to do at $\delta= 0$, the price one has to pay is that the solution gives only the modulus of the gap function $\Delta_h(\bk)$ leading to a functional degeneracy to the obtained solution, as any function of the form $\Delta_h(k)\, e^{i\, \Phi(\bk)}$ is also another possible solution. Away from the critical point ($\delta\neq 0$), one has to solve the integral equation. In this section, we present our numerical result for the solution of the non-linear integral gap equation for non-zero values of $\delta$. We first focus on the linearized gap equation
\begin{align}
\dfrac{1}{\lambda}\Delta_h(\theta_\bk)= \bigintsss \dfrac{d \theta_\bp}{2\pi} \dfrac{\cos^2\Bigr(\theta_{\bk}+\theta_\bp\Bigr)}{\delta^2+4 \sin^2\Bigr(\theta_{\bk}-\theta_{\bp}/2 \Bigr)}\Delta_h(\theta_\bp) =  \bigintsss \dfrac{d \theta_\bp}{2\pi} \, M(\theta_\bk,\theta_\bp)\, \Delta_h(\theta_\bp),
\label{matrix in angle space}
\end{align} cast into a matrix equation in angular space with eigenvalue labeled as $\lambda$ and kernel matrix
\begin{align}
 M(\theta_\bk,\theta_\bp)=\dfrac{\cos^2\Bigr(\theta_{\bk}+\theta_\bp\Bigr)}{\delta^2+4 \sin^2\Bigr(\theta_{\bk}-\theta_{\bp}/2 \Bigr)}.
\label{M linearized matrix}
\end{align} Here, we suppress the $\delta$ dependence of the matrix $M$ for the simplicity of the representation. We break Eq.~\eqref{matrix in angle space} into different angular momentum channels and solve for the largest eigenvalue in each channel to find the leading instability. The relevant channels consistent with the pairing interaction, Eq.~\eqref{pairing interaction} are $A_{1g}$ (eigen functions are like: $1, \cos 4\theta, \cos 8\theta$ etc), $B_{1g}$ (eigen functions are like: $\cos2\theta, \cos 6\theta, \cos 10\theta$ etc), and $E_g$ (eigen functions are like: $\cos\theta, \cos 3\theta, \cos 5\theta$ etc). We will refer them as $s,d$ and $p-$ waves respectively. In Fig.\ref{numerical solution}a, we plot how the largest eigenvalue $\lambda_s,\lambda_d$ and $\lambda_p$ in $s,d,$ and $p-$ wave channel respectively vary with the nematic mass parameter $\delta$. We find the leading instability happens in the $s-$ channel followed by $d-$ and $p-$ wave respectively from the linearized gap equation. In the inset, we plot the ratio of $\lambda_d/\lambda_s$ and $\lambda_p/\lambda_s$ with varying $\delta$. As $\delta$ goes to zero, these eigen values get closer to each other and become degenerate at $\delta=0$. Below the transition point, there could be $2$ situations possible: (a) after the gap in $s-$ wave channel develops, it acts against the subleading $d$ and $p$ waves and never let them win. As a result, $s-$ wave remains down to zero temperature, or (b) at some low temperature there is a second phase transition from $s-$ to either $s+i \, d$ or $p$ wave. We solve the full non-linear gap equation Eq.~\eqref{int1} numerically for $\delta=0.01$ at different temperatures and show our result in Fig.\ref{numerical solution}b. Our results support the first case of $s-$ wave gap symmetry being intact at all the temperature below the critical temperature $T_c$. \\ \\
Even though $s-$ wave wins, we argue there is a possibility of having only $p-$ wave gap symmetry if one includes the repulsive component of the pairing interaction presented in Eq.~\eqref{regular piece}. The repulsive regular part of the intra-pocket pairing interaction does not scale with the nematic susceptibility and has only $s-$ and $d-$ wave components. As a result, the pairing strength for the $p- $ wave is larger than for $s-$ and $d-$ wave when one includes both the attractive ($V^h_{\text{singular}}$) and repulsive ($V^h_{\text{regular}}$) components of the effective pairing interaction in the solution of the non-linear gap equation. The $p-$ wave has to be of the form $p_x\pm i\, p_y$, as $|\Delta_h(\bk)|$ is same on the $k_x$ and $k_y$ axis found from the analysis of section.\ref{gap solution}. \\

 We define the gap anisotropy $\alpha=\Delta_h(\theta=\pi/4)/\Delta_h(\theta=0)$ as the ratio of the gap along the $k_x=\pm k_y$ axis($\theta=\pi/4$) to $k_x$ axis($\theta=0$)  show its variation with the nematic mass parameter $\delta$ in Fig.\ref{numerical solution}c. We fit our result upto second order in $\delta$. We find a good fit with: $\alpha(\delta)=2.12 \, \delta^2+0.44 \, \delta$. We numerically obtain the specific heat coefficient as a function of the reduced temperature $T/T_c$ for two different values of $\delta=0.01$ and $0.1$ and show the result in Fig.\ref{numerical solution}d. For small but non-zero $\delta$, there will be jump at the transition point as mentioned before. On the other hand, at the low temperature the behavior of the specific heat coefficient will be qualitatively similar to the $\delta=0$ case because of the presence of the exponentially small gap near the cold spot for small values of $\delta$. 
\end{document}